%% file: 0.main.tex
\documentclass[acmlarge]{acmart}

\usepackage[normalem]{ulem}
\usepackage{enumitem}
\usepackage{xcolor}
\definecolor{RED}{rgb}{1,0,0}
\definecolor{ORANGE}{rgb}{1, 0.65, 0}
\AtBeginDocument{%
  }

\acmJournal{TOIS}

\begin{document}

\title{A Survey of Large Language Model Empowered Agents for Recommendation and Search: Towards Next-Generation Information Retrieval}

\author{Yu Zhang}
\authornote{Both authors contributed equally to this research.}
\affiliation{%
  \institution{Tsinghua University}
  \city{Beijing}
  \country{China}}
\email{zhangyuthu14@gmail.com}

\author{Shutong Qiao}
\authornotemark[1]
\orcid{0000-0002-7368-1535}
\affiliation{
  \institution{University of Queensland}
  \city{Brisbane}
  \country{Australia}
}
\email{shutong.qiao@uq.edu.au}

\author{Jiaqi Zhang}
\orcid{0009-0003-8317-2229}
\affiliation{
  \institution{University of Queensland}
  \city{Brisbane}
  \country{Australia}
}
\email{jqzhang927@gmail.com}

\author{Tzu-Heng Lin}
\affiliation{%
  \institution{Tsinghua University}
  \city{Beijing}
  \country{China}}
\email{lzhbrian@gmail.com}

\author{Chen Gao}
\orcid{0000-0002-7561-5646}
\affiliation{
	\institution{Tsinghua University}
  \city{Beijing}
  \country{China}
}
\email{chgao96@gmail.com}

\author{Yong Li}
\orcid{0000-0001-5617-1659}
\affiliation{
	\institution{Tsinghua University}
	\city{Beijing}
	\country{China}
}
\email{liyong07@tsinghua.edu.cn}

\renewcommand{\shortauthors}{Yu Zhang and Shutong Qiao et.al.}

\begin{abstract}
Information technology has profoundly altered the way humans interact with information. The vast amount of content created, shared, and disseminated online has made it increasingly difficult to access relevant information. Over the past two decades,  recommender systems and search (collectively referred to as information retrieval systems) have evolved significantly to address these challenges. Recent advances in large language models (LLMs) have demonstrated capabilities that surpass human performance in various language-related tasks and exhibit general understanding, reasoning, and decision-making abilities. This paper explores the transformative potential of LLM agents in enhancing recommender and search systems. We discuss the motivations and roles of LLM agents, and establish a classification framework to elaborate on the existing research. We highlight the immense potential of LLM agents in addressing current challenges in recommendation and search, providing insights into future research directions. This paper is the first to systematically review and classify the research on LLM agents in these domains, offering a novel perspective on leveraging this advanced AI technology for information retrieval.
To help understand the existing works, we list the existing papers on LLM agent based recommendation and search at this link: \url{https://github.com/tsinghua-fib-lab/LLM-Agent-for-Recommendation-and-Search}.
\end{abstract}

  \keywords{Large language model agent; recommender system; search system; information system}

\maketitle

\section{Introduction} 
Information technology has fundamentally transformed human life, particularly in how we interact with information. The development of the information age has led to the creation, sharing, and dissemination of vast amounts of content on the internet, making it increasingly challenging for individuals to access relevant information. 
Against this backdrop, the past two decades have witnessed the evolution of information systems, including Recommender Systems (RSs)~\cite{zhang2019deep,gao2023survey} and search systems~\cite{brin1998anatomy,speretta2005personalized}.
Typically, search systems passively retrieve relevant content from vast amounts of information with users' query words, while recommender systems actively generate a list of candidates that users might be interested in based on collected user behavioral data or profile information \cite{schutze2008introduction,aggarwal2016recommender,levene2011introduction}. Some researchers consider recommendation as a specialized form of search, where the understanding of potential user needs is modeled as search terms~\cite{xu2018deep}. 

Recently, large language models (LLMs) have achieved remarkable success and are considered one of the most viable paths toward general artificial intelligence~\cite{achiam2023gpt}. LLMs have demonstrated capabilities similar to or even surpass human performance, not just in language-related tasks, exhibiting general understanding, reasoning, and decision-making abilities \cite{zhao2023survey}. Researchers have leveraged the text processing and commonsense reasoning capabilities of large models to conduct a series of recommendation and search tasks \cite{white2024advancing,zhu2023large,ai2023information}. Furthermore, to overcome the limitations of LLMs,
researchers build various LLM agents \cite{xi2023rise}. Specifically, LLM agents use LLM as the core and are equipped with functions such as memory management, workflow, input-output interfaces, external tool invocation, and large-small model collaboration, making their intelligence levels closer to human and capable of handling more complex tasks \cite{zhao2024expel,park2023generative,zhou2023language}.

For the crucial and fundamental tasks of recommendation and search, researchers increasingly recognize the high value of LLM agents.
In this paper, we make the first attempt to provide a systematic and comprehensive review of the research efforts on recommendation and search with LLM agents.
We first introduce relevant background knowledge and further discuss the motivations for utilizing LLM agents, as well as how LLM agents address those critical challenges faced by recommendation and search.
Next, we establish a taxonomy, categorizing these recent advances according to the role of LLM agents in the overall recommender or search system.
We then elaborate on these works and analyze how they construct and use LLM agents.
It is worth mentioning there have already been several works discussing the application of LLMs in the fields of IR, which either focus solely on LLMs~\cite{chen2024large,bao2024large,wu2024survey,gao2024large} or a specific IR application~\cite{peng2025surveyllmpoweredagentsrecommender}, but this survey paper significantly differs by focusing on the application of LLM agents across the entire spectrum of IR functionalities.

The contributions of this paper can be summarized as follows:
\begin{itemize}[leftmargin=*]
    \item First, this paper is the first to review and organize the research on LLM agents in the research of search and recommendation, which is a new yet fast-growing research field.
    
    \item Second, we construct a taxonomy that well organizes the existing work by answering the fundamental question of why LLM agents are needed and how LLM agents enhance recommendation and search.

    \item Finally, we conduct open discussions about the unresolved challenges and important future research directions, which can inspire the following works in this area. 
\end{itemize}

The structure of this paper is as follows: In Section~\ref{sec::background}, we introduce the background knowledge of recommender systems, search systems, LLMs, and LLM agents. In Section~\ref{sec::motivation}, we analyze why LLM agents are necessary for recommendation and search. In Section~\ref{sec::advances}, we first establish a coherent and comprehensive taxonomy system and then elaborate on the existing work.
In Section~\ref{sec::Embodied}, we extensively discuss the advancements of embodied LLM agents within the cyber environment, highlighting their potential to inspire the next generation of recommendation and search. 
In Section~\ref{sec::discussion}, we systematically analyze the current problems and significant future directions for recommendation and search based on LLM agents. 
Finally, we conclude the paper in Section~\ref{sec::conclusion}.

\input{2.background}

\input{3.Why}

\input{4.Advances}
\input{5.Embodied}
\input{6.discussions}

\section{Conclusions}\label{sec::conclusion}
In this paper, we take the pioneering step in reviewing the advanced applications of LLM agents in recommender and search systems. 
We begin the survey with a brief introduction to the concepts of recommendation and search tasks, as well as LLM agents, offering newcomers a foundational overview and critical background knowledge. 
Then, we elucidate the intrinsic nature of LLM agents, highlighting three specific aspects to underscore why LLM agents can be used for recommendation and search. 
Moreover, we provide a detailed introduction to how to apply LLM agents in each task. For each phase, we provide a detailed taxonomy to categorize the major techniques and roles, drawing connections among the existing publications. 
Additionally, we extensively discuss the potential of applying embodied agents in both tasks, delving into the rationale behind the advanced development. 
Finally, we summarize several challenges and promising directions in this field, which are expected to guide potential future directions.

\bibliographystyle{ACM-Reference-Format}
\bibliography{bibliography}

\appendix

\end{document}

%% file: 2.background.tex
\section{Background}\label{sec::background}

\subsection{Recommendation and Search}

Recommendation and search are pivotal challenges in information retrieval and have been extensively studied for decades. Despite their distinct application scenarios, they share similar core components (namely \textit{Interaction Interface}, \textit{User/Query Modeling}, \textit{Item Modeling}, and \textit{Matching, Ranking, and Re-ranking}). This section introduces the background and concepts of each core component.

\subsubsection{Interaction Interface}
The rapid development of smart devices and the mobile Internet has exposed users to numerous applications employing recommender systems and search engines.
Interacting with a traditional recommender system is simple, where users are mostly unconsciously browsing contents (\textit{e.g.}, on an e-commerce website, or a video sharing platform) provided by the recommender system \cite{resnick1997recommender}.
On the other hand, the interaction interface between users and search engines usually consists of a keyword query submission from users (\textit{e.g.}, on a search engine website or a project management software) and a one-round result retrieval from the search engines \cite{baeza1999modern}.
Conversational recommender systems and search engines have recently gained attention \cite{zhang2018towards}.
In this interface, users interact through a chat interface using natural language, iteratively providing additional information or feedback based on previous results, forming a multi-turn dialogue. This conversational interface is more closely aligned with human communication methods and achieves more precise and satisfying results.

\subsubsection{User/Query Modeling}
For recommender systems and search engines to retrieve satisfying results, they must understand the users' intentions. For example, identifying the products a user wants to buy on an e-commerce app or determining relevant websites based on a keyword query. This requires a precise modeling of user characteristics and input queries.
In recommender systems, this includes utilizing the user profile (gender, age, preferences, etc.), browsing/clicking/purchasing history, and more \cite{lops2011content}. 
In search engines, this involves text segmentation, named entity recognition, part-of-speech tagging, relation extraction, etc \cite{croft2010search}. 
Advanced systems like multi-modal and conversational search engines also incorporate image recognition, multi-turn dialogue comprehension, and other algorithms \cite{jannach2021survey, wang2016comprehensive}.

\subsubsection{Item Modeling}
Apart from user/query modeling, item modeling also plays an essential role in efficient and extensive retrieval.
It gathers information of the items, and preprocesses them into some representation (\textit{e.g.}, a feature vector) for later usage.
In recommender systems, item modeling considers item descriptions, attributes, images, buyers, and other related information \cite{lops2011content, liu2024multimodal}.
In traditional search engines, this process is known as \textit{Indexing} \cite{croft2010search}, where significant keywords or tags from the content of web pages are identified and stored beforehand for faster retrieval. Advanced search engines may also use auxiliary knowledge bases specific to certain domains.

\subsubsection{Matching, Ranking, and Re-ranking}
To effectively retrieve accurate items according to the users/queries from a huge-size candidate item set (usually millions or more), a recommender system or a search engine typically comprises of three consecutive stages: \textit{Matching}, \textit{Ranking}, and \textit{Re-ranking} \cite{hron2021component}.
\begin{itemize}
    \item \textbf{Matching:} The \textit{Matching} (or \textit{Recall})  \cite{covington2016deep} stage retrieves potentially relevant hundreds of items from the full millions of candidates.
    In this stage, the algorithm is usually coarse (\textit{e.g.}, vector inner product) and highly efficient due to the huge-size input item set. 
    \item \textbf{Ranking:} The \textit{Ranking} \cite{cheng2016wide} stage ranks the output of the \textit{Matching} stage and selects the top ten of the items.
    With a smaller item set, this stage employs more complex models (\textit{e.g.}, deep neural networks) for higher accuracy.
    \item \textbf{Re-ranking:} The \textit{Re-ranking} \cite{pei2019personalized} stage further adjusts the final item list according to some additional requirements such as diversity, fairness, freshness, and business goals, \textit{etc}.
\end{itemize}
By leveraging these stages, recommender systems and search engines ensure precise and relevant results.

\subsection{Large Language Model and LLM Agents} 
\subsubsection{Large Language Models}
Language models play a pivotal role in Natural Language Processing (NLP), primarily used for assessing sentence probabilities and generating grammatically correct text. In its early stages, the development of language models relied on statistical and probability theories such as n-grams, maximum entropy, and Hidden Markov Models, but was limited by its ability to handle long-range dependencies, known as the ``curse of dimensionality.'' With the rise of deep learning technologies, neural network language models gradually became mainstream due to their robust generalization capabilities and flexible architectural designs. Models like Recurrent Neural Networks (RNNs) \cite{sherstinsky2020fundamentals}, Long Short-Term Memory networks (LSTMs) \cite{graves2012long}, and Gated Recurrent Units (GRUs) \cite{cho2014learning} effectively mitigated the challenges of processing long sequences by utilizing internal memory units, thereby better capturing contextual dependencies within sequences and significantly enhancing language model performance.

The introduction of the Transformer model \cite{vaswani2017attention} revolutionized the NLP field by replacing recurrent structures with self-attention mechanisms, thereby improving training speed and parallelism. Concurrently inspired by the pre-training and fine-tuning paradigm from computer vision, language models began leveraging large amounts of unlabeled data for pre-training, followed by task-specific fine-tuning, which gave rise to high-performance models such as BERT \cite{devlin2018bert}. With the rapid expansion of training data and significant growth in computational resources, LLMs have established absolute dominance in the NLP. This wave not only symbolizes the comprehensive transformation of NLP from focusing on single-task optimization to multimodal perception and generation but also heralds language models reaching new heights in generation, understanding, and transformation. 
In the latest research, LLMs such as GPT-4 \cite{achiam2023gpt}, Claude \cite{claude-3}, etc., not only perform well in tasks across various NLP tasks such as text generation, multilingual translation, and sentiment analysis but also demonstrate impressive performance in tasks traditionally considered to require deep cognitive abilities like solving mathematical problems, logical reasoning, and strategic planning. 
This comprehensive performance marks the maturity of artificial intelligence technology, steadily advancing in a direction closer to Artificial General Intelligence (AGI).

\subsubsection{LLM Agents}

The concept of agent originated in the 1960s, aiming to develop software entities that can simulate human intelligence. By the 1970s and 1980s, the rise of expert systems \cite{waterman1985guide} represented the ability of intelligent programs to solve professional problems based on fixed rules and knowledge bases. Although these systems were powerful, they were limited by static knowledge and rigid logic.
In the late 1980s, researchers such as Michael Wooldridge redefined agents, emphasizing their autonomy, responsiveness to external events, and interaction with the environment and other agents. Minsky elaborated on the concept of agents in his work \cite{minsky1988society}, describing them as individuals that can participate in social interactions and exhibit intelligent behavior. Minsky's idea that many relatively simple components (i.e., agents) in a complex system can solve complex problems by collaborating with each other later became the basis for research in Multi-Agent Systems (MAS) \cite{dorri2018multi} and distributed artificial intelligence \cite{bond2014readings}.
In the 1990s, with the popularity of the Internet and distributed computing, the research focus turned to the cooperation of agents in an open, dynamic environment. This required them to have intelligent strategies and interaction capabilities, promoting complex intelligent ecosystems' design.

At the beginning of the 21st century, breakthroughs in machine learning, particularly represented by the rise of deep learning, enabled autonomous learning from massive amounts of data rather than relying solely on pre-set programming rules. This marked a shift in agent research from rule-based methods to data-driven approaches. Deep neural networks significantly enhance the capabilities of intelligent agents in areas such as image recognition, speech processing, and natural language understanding, thus ushering in a new era of Artificial Intelligence (AI) agents.
Recently, GPT series \cite{chatgpt, achiam2023gpt} and other major language models have made significant progress in NLP, greatly improving the performance of artificial intelligence agents and demonstrating their profound potential.

\begin{figure}
    \centering
    \includegraphics[width=0.9\linewidth]{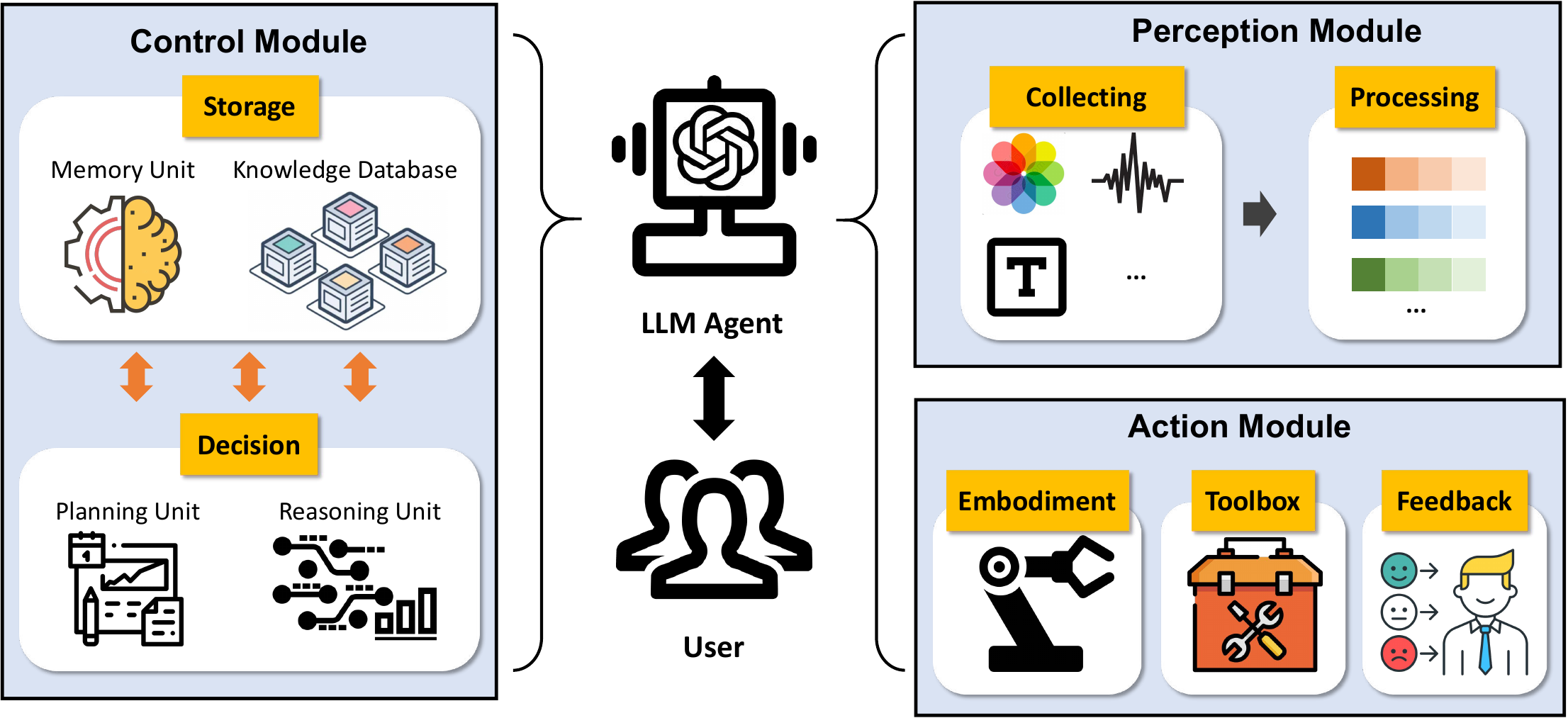}
    \caption{Schematic diagram of the three core modules of agent}
    \label{fig:Agent Modules}
\end{figure}

An agent is an autonomous computing entity capable of perceiving its environment, making decisions, and taking actions. Figure \ref{fig:Agent Modules} shows the three core modules of the LLM agent, each detailed as follows:
\begin{itemize}
    \item \textbf{Perception Module: }The perception module is equivalent to the agent's \textbf{``senses''}, ensuring that the agent can obtain the information needed for decision-making on time. Unlike conventional single-channel information processing systems, LLM agents demonstrate superior capability in seamlessly integrating multimodal information sources, including textual, auditory, visual, and even video data, thereby establishing a more robust and comprehensive perceptual framework \cite{schumann2024velma}. 
    \item \textbf{Control Module: }The control module plays the role of a proxy \textbf{``brain''}, analyzing the collected information, evaluating possible action plans, and selecting the best strategy. In LLM agents, an LLM serves as the core of the control module, enhancing the agent's understanding and generation ability of language contexts and strengthening its situational adaptability, creativity, and long-term memory functions. 
    \item \textbf{Action Module:} The action module acts as the agent’s \textbf{``limbs''}, executing decisions to alter environmental states or achieve goals. In the LLM agent, this module is not limited to simple execution actions. Still, it is endowed with deeper functions, which can realize complex tasks such as language generation, tool use, and even concrete actions, greatly expanding the application scope and influence of the agent.
\end{itemize}

Currently, LLM agents have emerged in many fields. They can not only handle basic transactional work but also involve complex tasks that require deep understanding and creative thinking. For example, in game development, LLM agents can give non-player characters (NPCs) vivid and natural language capabilities to create a more immersive gaming environment \cite{park2023generative, hu2024survey}; in the field of education, LLM agents can provide personalized tutoring for students as classroom assistants and help teachers correct homework \cite{lagakis2024evaai}. In the field of scientific research, LLM agents such as CellAgent can perform complex tasks such as single-cell RNA sequencing data analysis that require the combination of specific domain knowledge \cite{xiao2024cellagent}. LLM agents are constantly expanding the boundaries of human capabilities. As the technology matures, more innovative application scenarios will continue to emerge. LLM agents will drive human civilization in a more intelligent and sustainable direction.

%% file: 3.Why.tex
\section{Why LLM agents can be used for recommendation and search}\label{sec::motivation}
As discussed in previous sections, LLM agents introduce new capabilities compared to simpler models and have diverse applications, such as game NPCs, education, RNA design, etc. In this section, we will demonstrate why these new capabilities provide LLM agents with distinct advantages in addressing IR problems.

\subsection{LLM Agents Can Do Deep Thinking and Task Decomposition}
In an era of information overload, user queries are often complex and diverse, making it challenging for traditional information retrieval systems to grasp the users' true needs accurately. However, LLM agents, with their advanced language comprehension abilities, can thoroughly analyze user queries, identifying key information, implicit intents, and contextual relationships. 
Two key capabilities underpin LLM agents' advances: deep thinking and extended information input. Firstly, Deep thinking enables the agent to perform a more in-depth analysis of a given issue, typically achieved through techniques of Chain-of-Thought (CoT) \cite{wei2022chain,feng2024towards}. Secondly, the use of extended information inputs enables the agent to remember more content, including user preferences, context information, and item profiles. It is usually facilitated by technologies such as a long context window that supports millions of token inputs. \cite{ding2024longrope,jin2024llm}.
Together, these capabilities enable the agent to decompose a complex IR task into multiple sub-tasks for more effective execution \cite{joko2024doing,aliannejadi2024trec}. For example, when a user plans a trip, an LLM agent can decompose the process into selecting a destination, planning the itinerary, booking flights and hotels, and calculating the budget, ultimately presenting the user with a complete report \cite{xie2024human}, which can be quite challenging for traditional IR systems. 

\subsection{LLM Agents can Interact with Environment and Integrate Information}
In the field of information retrieval, another significant advantage of LLM agents is their ability to interact with environments and integrate results \cite{he2025pasa,zhou2023agents}. IR often involves multiple data sources and complex retrieval scenarios where traditional retrieval systems may struggle to collect information effectively from different sources. LLM agents with action modules can use various tools, browse web pages, operate mobile apps, and even independently search the internet to find needed information \cite{schick2024toolformer,qin2023toolllm}. Additionally, LLM agents can go a step further by filtering and summarizing the collected content, presenting users with a clear and concise report instead of delivering raw, unprocessed results like traditional IR systems.
For example, when conducting academic literature searches, an LLM agent can interact with multiple academic databases to obtain research outcomes from different fields. It then analyzes and consolidates these results to provide users with a comprehensive literature review. Furthermore, LLM agents can interact with users, continuously adjusting retrieval strategies based on user feedback, thereby enhancing the quality of the retrieval results.

\subsection{LLM Agents Can Serve as User Simulators to Generate Feedback to Information System}
IR systems rely on real user feedback for improvement, but conducting experiments on actual users can be costly and negatively impact user experience. The various modules within LLM agents enable them to simulate human perception, memory, and actions, making them well-suited for collecting feedback in place of real users.
LLM agents can simulate real user responses when using a product and identifying potential improvements \cite{zhang2024usimagent}. They can also be used to explore the evolution of users' mindsets and interests, aiding researchers in better understanding user behavior \cite{park2023generative,takata2024spontaneous}. 
LLM agents can simulate users' preferences and interact with recommender and search systems. These generated interaction data can be used to train more robust IR models~\cite{sekulic2024reliable,zhang2024llm}.

In summary, LLM agents bring several critical capabilities to the table that make them highly suitable for recommendation and search tasks. Their ability to think and plan, interface with various tools, and simulate user behavior makes a significant difference in the user experience of IR in the new era.

%% file: 4.Advances.tex
\section{Recent Advances of LLM Agents for Recommendation and Search}\label{sec::advances}
\begin{figure}
    \centering
    \includegraphics[width=0.9\linewidth]{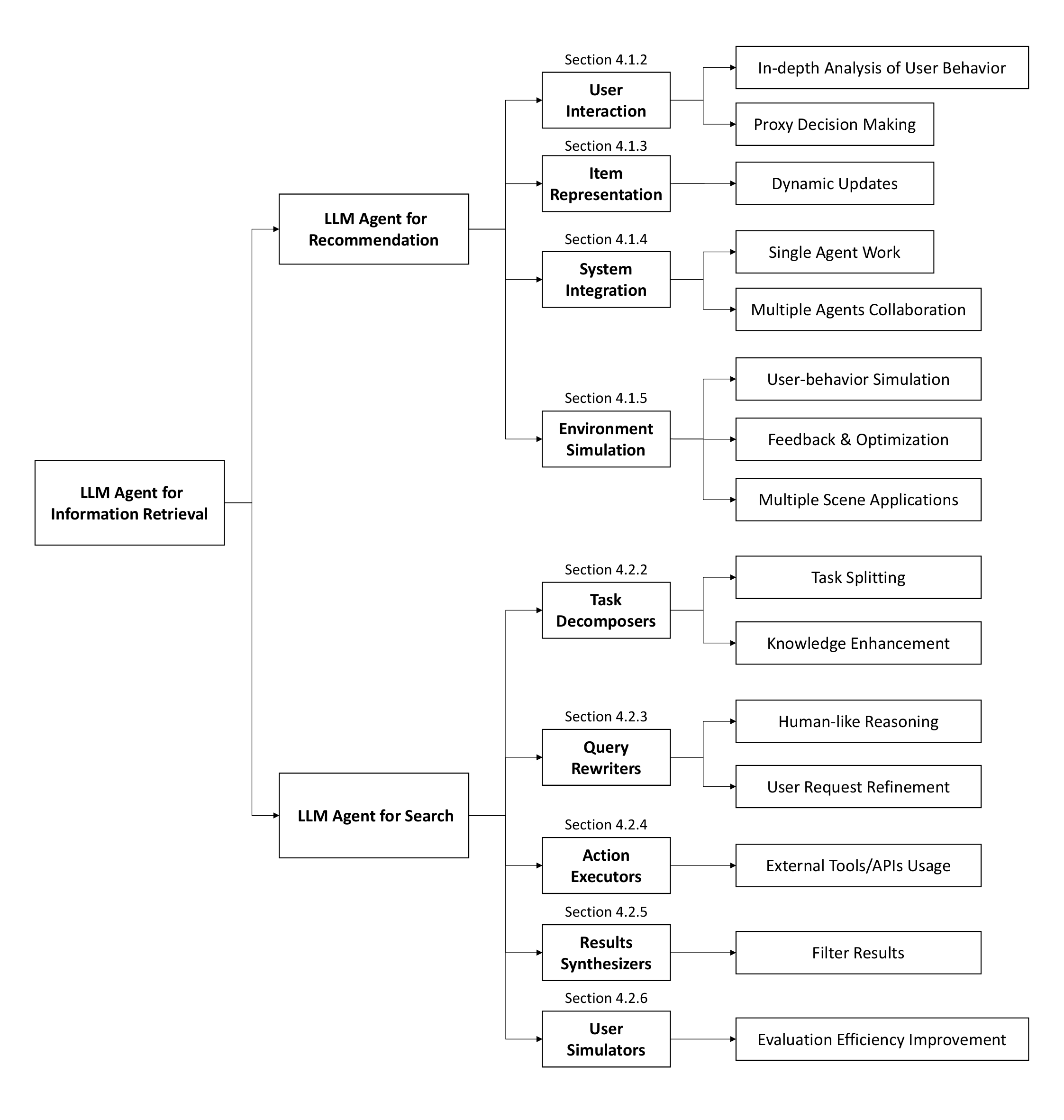}
    \caption{Illustration of existing work of LLM agents for recommendation and search.}
    \label{fig:2}
\end{figure}

Recently, significant advancements have been made in the research on combining LLM agents with traditional recommendation and search algorithms.
This section provides a comprehensive overview of these developments, categorized into the contributions and innovations made in three primary areas: the overall domain and taxonomy, the role of agents, and recent significant papers.
The classification of existing work in LLM agents for Recommendation and Search is shown in Figure \ref{fig:2}.
\subsection{LLM Agents for Recommendation}
\subsubsection{Taxonomy Introduction}

Deploying LLM agents in recommendation and search domains has given rise to new frameworks and methods. We redefine the taxonomy of these domains, identifying the key domains of LLM agents for RSs as user interaction, representation optimization, system integration, and environment simulation. 

\begin{figure}
    \centering
    \includegraphics[width=0.7\linewidth]{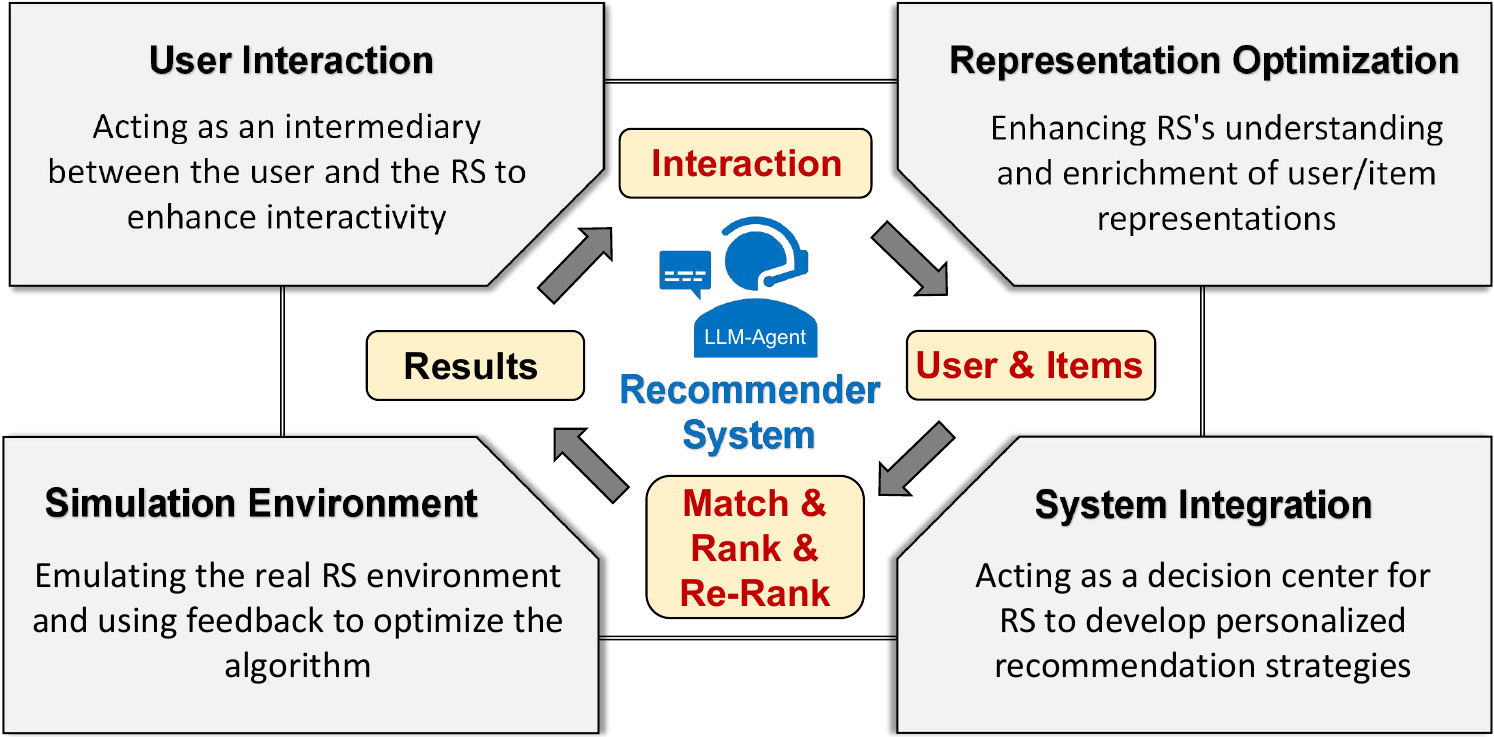}
    \caption{Four domains of LLM agent's role in recommendation tasks}
    \label{fig:Agent4Rec}
\end{figure}
Figure \ref{fig:Agent4Rec} shows the four domains used by LLM agents for recommendation tasks, with detailed descriptions of each domain as follows:
\begin{itemize}
    \item \textbf{User Interaction:} LLM agents serve as intermediaries between users and systems, enhancing interaction through natural language understanding and generation.
    \item \textbf{Representation Optimization:} The innovation focuses on better understanding and representation of users/items through LLM agents, making the recommendation process more precise.
    \item \textbf{System Integration:} LLM agent acts as the brain of RS, helping the RS analyze and make decisions for more effective results.
    \item \textbf{Environment Simulation:} LLM agent is used as a user simulator to build a simulation test environment for the RS and optimize feedback to achieve continuous improvement.
\end{itemize}
We categorize research work according to the descriptions of the domains as mentioned above and list typical works related to the four domains in Table \ref{tab:LLM4RS}.

\begin{table}[t!]
	\centering
	\footnotesize
	\caption{A list of representative works of LLM agents for Recommendation.}
        \label{tab:LLM4RS}
	\resizebox{1.0\textwidth}{!}{
        \begin{tabular}{c|c|l}
		\toprule
	    \bf Domain & \bf Paper & \bf What agents can do (ability) \\ \midrule
	Interaction & RAH \cite{shu2024rah} & Assists users in receiving customized recommendations and provide feedback \\
        Interaction & ToolRec \cite{zhao2024let} & Uses tools for specific recommendation tasks \\ 
        Interaction & RecAI \cite{lian2024recai} & Utilizes LLMs as an interface for traditional recommendation tools \\
        Interaction & AutoConcierge \cite{zeng2024automated}  & Conducts real conversations with users \\
        Interaction & FLOW \cite{cai2024flow} & Introduces a feedback loop to enable collaboration between the recommendation agent and the user agent \\
        \midrule
        
        Representation & AgentCF \cite{zhang2024agentcf} & Facilitates collaborative learning between user and item agents \\
        Representation & Rec4Agentverse \cite{zhang2024prospect} & Controls the collaboration between the Intelligent Agent items and the Agent Recommenders \\ 
        Representation & KGLA \cite{guo2024knowledge} & Improves user agent memory \\
        \midrule

         System & RecMind \cite{wang2023recmind} & Introduces a self-inspiring algorithm for decision-making \\
        System & InteRec \cite{huang2023recommender} & Integrates LLMs and RSs for interactive recommendations \\
        System & MACRec \cite{wang2024multi} & Develops a multi-agent collaboration framework for RSs \\
        System & BiLLP \cite{shi2024enhancing} & Emphasizes long-term user retention using LLM-planned RL algorithms \\
        System & MACRS \cite{fang2024multi} & Tackles dialog control and user feedback integration with multi-agent framework \\
        System & PMS  \cite{thakkar2024personalized} & Uses multimodal, autonomous, multi-agent systems \\
        System & CORE \cite{jin2023lending} & Combines conversational agents and RSs for better interaction \\ 
        System &  Hybrid-MACRS \cite{nie2024hybrid} & Combines LLM agent and search engine to optimize conversational recommendation \\
        \midrule
        
        Simulation & Agent4Rec \cite{zhang2023generative} & Trains LLM agents to simulate real users for evaluation \\
        Simulation & RecAgent \cite{wang2023recagent} & Simulates user behaviors related to the RS \\
        Simulation & Yoon et al. \cite{yoon2024evaluating} & Uses LLMs to simulate users for conversational recommendation tasks \\
        Simulation & SUBER \cite{coreccosuber} & Develops an RL environment using LLM to simulate user feedback \\
        Simulation & CSHI \cite{zhu2024llm} & Proposes a framework for LLM-based user simulators in conversational RSs \\
        Simulation & iEvaLM \cite{wang2023rethinking} & Suggests new evaluation methods using LLMs \\
        Simulation & Zhu et al. \cite{zhu2024reliable} & Examines reliability and limitations of current LLM-based simulators \\
        Simulation & OS-1 \cite{xu2024can} & Develops an LLM-based eyewear system with conversational common ground \\
        Simulation & CheatAgent \cite{ning2024cheatagent} & Uses LLM agent to attack LLM-driven RSs \\
        Simulation & Zhang et al. \cite{zhang2024llm} & Improves the training efficiency and effectiveness of RSs based on reinforcement learning \\
        \bottomrule

	\end{tabular}
    }
\end{table}

\subsubsection{User Interaction}
Traditional RSs rely on static data (such as user historical behavior) to model user preferences, but this approach is difficult to cope with scenarios where user preferences change rapidly, new users or new products (cold start problem), and niche product recommendations (long tail effect). In addition, although the dialogue recommendation algorithm in the traditional recommendation algorithm can dynamically capture user intentions through natural language interaction, its implementation faces challenges such as how to efficiently understand complex human language, how to maintain coherence and goal orientation in the dialogue, and how to deal with ambiguity in the dialogue. 
LLM agent supports users to interact in natural language and provides highly personalized recommendations by analyzing user preferences, historical behaviors, conversation context, etc. The LLM agent can also flexibly call external tools to further optimize recommendation results according to specific needs, providing users with a richer and more personalized interactive experience.

Specifically, AutoConcierge \cite{zeng2024automated} uses LLM to convert user questions into logical predicates, checks the consistency of information through the Answer Set Programming (ASP) system and updates the system status, and automatically generates questions to complete missing information when necessary. After obtaining all user preferences, the system searches for matching restaurants in its knowledge base and uses LLM to convert the results into natural language recommendations. If the user changes his mind or requests different suggestions, AutoConcierge can dynamically adjust its strategy, re-evaluate, and provide new recommendations. This process ensures efficient and accurate responses to user needs.
ToolRec \cite{zhao2024let} framework uses LLM as a surrogate user to evaluate the degree of match between the user's preferences and the current scenario, aiming to simulate the real user decision-making process. Then, external tools (ranking tools and retrieval tools) are called according to the user's attribute instructions to explore different parts of the project pool. RecAI \cite{lian2024recai} is a practical toolkit that enhances or innovates RSs through the capabilities of LLMs. The LLM agent first develops a comprehensive execution plan based on the intent in the user conversation, then calls the tools, tracks the output of each tool, and ultimately generates a response for the user. 
In the FLOW \cite{cai2024flow} framework, the recommendation agent uses LLM, combined with the memory module and the recommendation module, to generate initial recommendations and optimize the recommendation results based on user feedback; the user agent is responsible for simulating user behavior and more accurately capturing the user's potential interests by analyzing the interaction history with the recommendation agent. This iterative refinement process progressively enhances the cognitive capabilities of both the recommendation agent and the user simulation agent, consequently enabling more nuanced recommendation generation and higher-fidelity user behavior emulation.
The RAH framework \cite{shu2024rah} combines RSs, assistants, and humans, using LLM agents to perceive, learn, act, criticize, and reflect. These agents collaborate through a learning-act-critic cycle to continuously improve their understanding of user personality. For example, when a user interacts or gives feedback, the learning agent extracts preliminary personality features, the action agent predicts user behavior based on this, and the criticism agent evaluates the accuracy of the prediction. If the prediction is inaccurate, the criticism agent will analyze the reasons in depth and make suggestions for improvement, and the learning agent will adjust the personality features accordingly until the prediction is consistent with the user's actual behavior. 

In summary, LLM agents can not only provide more accurate and personalized recommendation services but also provide users with a richer and more efficient interactive experience through continuous self-optimization and calling external resources. This marks the transformation of the RS from a simple information provider to an intelligent interactive partner, significantly improving the user experience and service quality.

\subsubsection{Representation Optimization}

In the traditional RS framework, the information records of each user and item are independent and static, including everything from basic information to detailed attributes, and most of these data updates rely on manual updates. However, user interests change dynamically, and different users have significantly different interests in the same item, and even the same user's interests change at different times. Traditional representation methods based on static data have difficulty capturing these personalized needs and changes over time, which poses a challenge to providing accurate and effective recommendation services.
The LLM agent can generate detailed representations through deep semantic analysis, multimodal data fusion, and external world knowledge, and it can keenly perceive context changes and user feedback, promoting dynamic learning and updating of user and item representations.

Specifically, AgentCF\cite{zhang2024agentcf} treats both users and items as agents, each with its memory module to maintain the preferences and tastes of potential adopters. In addition, the study introduces a new collaborative learning method for simultaneously optimizing user agents and item agents and designs a collaborative reflection mechanism that enables agents to adjust their memories based on differences in real-world interaction records to more accurately reflect user behavior. 
KGLA \cite{guo2024knowledge} converts the path information in the knowledge graph into natural language descriptions, enhances the language agent's understanding of user preferences, and dynamically updates user memory based on the interaction between users and items during the simulation phase. At the same time, it uses information such as brands, categories, and product features in the knowledge graph to generate detailed item descriptions, and more accurately characterizes item features by analyzing the path relationship between users and items, thereby improving the personalization and accuracy of the recommendation.
In the Rec4Agentverse \cite{zhang2024prospect} framework, items are converted into interactive, intelligent, and proactive LLM agents that can dynamically acquire user preferences and continuously update their own characteristics through multiple rounds of dialogue. Unlike traditional static items, LLM agents can provide multi-dimensional knowledge representation and enhance service content through collaboration with other agents. For example, a travel agent can not only recommend itineraries based on user interests but also cooperate with other agents to provide more comprehensive services. In addition, agent items can accurately model user preferences and improve the accuracy of personalized recommendations and user experience.

These works effectively enrich the representation of users and items by leveraging individual agent autonomy and inter-agent collaboration, giving them more initiative.
This shift makes recommendation algorithms more situational and dynamic and can capture subtle changes in user preferences and the potential value of product attributes. The domain is still in its infancy, with less relevant work, and more excellent results are expected to emerge in the future.

\subsubsection{System Integration}

Traditional RSs often use a batch processing architecture to regularly update models and recommendation results. For example, models can be retrained every day or every week to reflect the latest user preferences. Data cleaning, feature engineering, model selection, and so on need to be considered separately during system integration. In contrast, LLM agents significantly reduce the need for manual intervention by learning and dynamically adjusting recommendation strategies in real-time and supporting an instant feedback mechanism, which not only improves the efficiency of the recommendation system but also enhances the user experience. In addition, LLM agents also support a team collaboration framework where each agent can focus on different data sources or tasks. Through multi-level and multi-dimensional collaboration and information sharing between these agents, the system can more comprehensively understand user preferences and provide highly personalized and context-related recommendation content.
From the perspective of framework structure, these studies can be divided into single-agent work \cite{wang2023recmind, huang2023recommender,jin2023lending, shi2024enhancing}, multi-agent collaboration \cite{fang2024multi,wang2024multi, nie2024hybrid}. 

In the \textbf{single-agent work category}, RecMind \cite{wang2023recmind} introduces a self-inspiration algorithm for the LLM agent which is designed for RS to retain the status of all historical paths and use this historical information to optimize planning decisions, with the aim of addressing the limitations of existing RSs in generalizing their ability to perform new tasks and effectively utilizing external knowledge.
InteRecAgent \cite{huang2023recommender} takes LLM as its core, processing instruction understanding, common sense reasoning, and human-computer interaction, while the recommendation model serves as a tool for domain knowledge and user behavior patterns. Through memory components and dynamic examples to enhance task planning and reflection mechanisms, it upgrades traditional RSs, such as ID-based matrix decomposition to systems that support natural language interaction.
The CORE framework \cite{jin2023lending} adopts an offline training and online checking model, where the RS model acts as an offline relevance score estimator, while the LLM conversational agent checks these scores online to reduce uncertainty by minimizing the sum of unchecked relevance scores. The LLM agent in BiLLP \cite{shi2024enhancing} is designed as a high-level planner that divides the learning process into two levels: macro-learning and micro-learning. Macro learning is responsible for acquiring high-level guiding principles, while micro-learning focuses on learning personalized recommendation strategies. This dual-level approach aims to achieve long-term recommendation strategies in RSs. 

In the \textbf{multi-agent work category}, MACRS \cite{fang2024multi} is a multi-agent conversational RS that controls dialogue flow, generates responses through multiple LLM agents, learns from user feedback to refine dialogue strategies, prevents errors, and interprets implicit user semantics.
MACRec \cite{wang2024multi} is a multi-agent framework that includes specialized agents like managers, analysts, reflectors, searchers, and task interpreters to address diverse recommendation tasks collaboratively. PAS \cite{thakkar2024personalized} consists of three agents: The first agent recommends products suitable for answering a given question; the second agent asks follow-up questions based on images belonging to these recommended products; and the third agent then conducts the autonomous search. It also features real-time data extraction, recommendations based on user preferences, and adaptive learning.
Hybrid-MACRS \cite{nie2024hybrid} consists of a central agent and a search agent. The central agent is driven by LLM and is responsible for user interaction, preference identification, search query generation, recommendation generation, etc. The search agent consists of a search engine and a relative search module, which is responsible for processing the search queries generated by the Central Agent, returning a list of matching products, and optimizing search results using personalized sorting based on user historical behavior.

These investigations collectively strive to transcend the constraints inherent in conventional recommendation systems by harnessing the advanced capabilities of LLM agents across multiple dimensions, including natural language comprehension, personalized preference modeling, and context-aware adaptive decision-making.

\subsubsection{Environment Simulation}

In the past, collecting user feedback to optimize recommendation algorithms was a time-consuming and resource-intensive process. However, with the advancement of technology, a LLM agent has emerged as an advanced simulator that can simulate user behaviors and preferences and support real-time response and large-scale concurrent testing.
By leveraging the highly simulated environment provided by LLM agents, development teams can now test and optimize recommendation algorithms more efficiently. This simulation not only greatly shortens the testing cycle, but also reduces the reliance on real user data, while ensuring user privacy and data security. Most of the studies \cite{wang2023rethinking,wang2023recagent, zhu2024llm, zhang2024generative, yoon2024evaluating, zhu2024reliable} focus on using agents in conversational recommendation scenarios to simulate user behaviors and evaluate their effectiveness.
In addition, some studies \cite{coreccosuber, zhang2024llm, ning2024cheatagent} have explored in depth the new challenges and risks that LLM agents may bring in more practical applications.

Specifically, iEvaLM \cite{wang2023rethinking} represents an innovative interactive evaluation framework leveraging LLMs, incorporating an LLM-based user simulator to comprehensively model diverse system-user interaction scenarios. Extensive experimental evaluations conducted on two publicly available Conversational Recommender System (CRS) datasets demonstrate that iEvaLM achieves substantial performance enhancements over conventional evaluation methodologies. Notably, this framework introduces a novel dimension by incorporating the assessment of recommendation interpretability, thereby addressing a critical aspect often overlooked in traditional evaluation paradigms.
Agent4Rec \cite{zhang2024generative} generates 1,000 LLM-enabled agents with different social traits and preferences by initializing from the MovieLens-1M dataset. The agents interact with the RS on a page-by-page basis, performing operations such as watching, rating, evaluating, exiting, and interviewing. Experimental results show that the generative agent can effectively identify and respond to items that match user preferences, and the feedback provided by the agent can be used as enhanced data for iterative training and improvement of the recommendation strategy, thus forming a dual-track approach for comprehensive evaluation of recommendation algorithms.
Yoon et al. \cite{yoon2024evaluating} introduced a novel evaluation framework designed to quantitatively assess the behavioral alignment between LLMs and human users in CRS. This comprehensive protocol encompasses five distinct evaluation tasks: (1) item selection for discussion, (2) binary preference expression, (3) open-ended preference articulation, (4) recommendation solicitation, and (5) feedback provision. Through systematic experimentation, their study not only identifies significant behavioral discrepancies between LLM-generated responses and human interactions but also provides empirical insights into mitigating these deviations through optimized model selection criteria and advanced prompting techniques.
Zhu et al. \cite{zhu2024reliable} conducted an analysis of the inherent limitations in current LLM-based user simulation approaches, particularly focusing on the issue of inadvertent data leakage. To address these challenges, they proposed SimpleUserSim, an innovative framework that implements a controlled information disclosure mechanism. This novel simulator employs a direct conversational strategy to steer dialogue topics toward target items while maintaining strict information constraints - specifically, it restricts access to target item titles and operates solely on item attribute information until successful recommendation completion.
The RecAgent \cite{wang2023recagent} framework has a detailed definition of LLM agent simulation users, including the Profile Module (ensuring that the agent has a unique personality), Memory Module (remember past behaviors and evolve dynamically in the environment), and Action Module (including RS-related searches, browsing, clicks, and page turns, as well as some social behaviors), so that it can effectively simulate real human behavior. 
The CSHI framework \cite{zhu2024llm} introduces control mechanisms, enhances the scalability of simulation, and incorporates elements of human-computer collaboration, enabling agents to flexibly adjust their behavior based on real-time context and historical data, thereby providing a more accurate personalized experience in conversational RSs.
Beyond acting as user simulators in conversational recommendation scenarios, LLM agents have further broadened their application areas, playing a vital role in reinforcement learning and attacks on RS.
In the latest research, the SUBER framework \cite{coreccosuber} ingeniously utilizes LLM agents to simulate human behavior, thereby constructing a synthetic environment specifically for the training and evaluation of reinforcement learning-based RSs. This innovative approach effectively mitigates the high training costs associated with the scarcity of online data and addresses the challenges in evaluating RS models, particularly in accurately measuring model performance without directly impacting the real user experience.
Zhang et al. \cite{zhang2024llm} addresses the shortcomings of current user simulators such as SUBER and Agent4Rec (including high computational cost, susceptibility to ``hallucinations'', and failure to fully capture the complex user-system interaction dynamics) by explicitly modeling user preferences, combining the advantages of logical reasoning and statistical learning, and designing an effective evaluation framework to improve the reliability and efficiency of the simulators.
CheatAgent \cite{ning2024cheatagent} first analyzes the parts of the text that have the most significant impact on recommendation results. Then, it utilizes an LLM agent to generate adversarial perturbations that can mislead the RS and optimize the attack strategy through prompt tuning. Finally, it fine-tunes the LLM's attack strategy based on feedback from the target system to enhance the attack's effectiveness.

In summary, these studies range from simple dialogue interactions to complex reinforcement learning environments, collectively representing the advancing application landscape of LLM agents in simulating user behavior.

\subsection{LLM Agents for Search}
\subsubsection{Taxonomy Introduction}
Typically, the interaction between users and search engines can be divided into four key steps: user intent, query writing, search execution, and result understanding \cite{white2024advancing}.
Figure \ref{fig:Agent4Search} shows the five domains utilizing LLM agents in search tasks, which can correspond to the typical IR interaction process mentioned \ref{sec::background}. Detailed descriptions of each domain are as follows:
\begin{figure}
    \centering
    \includegraphics[width=0.7\linewidth]{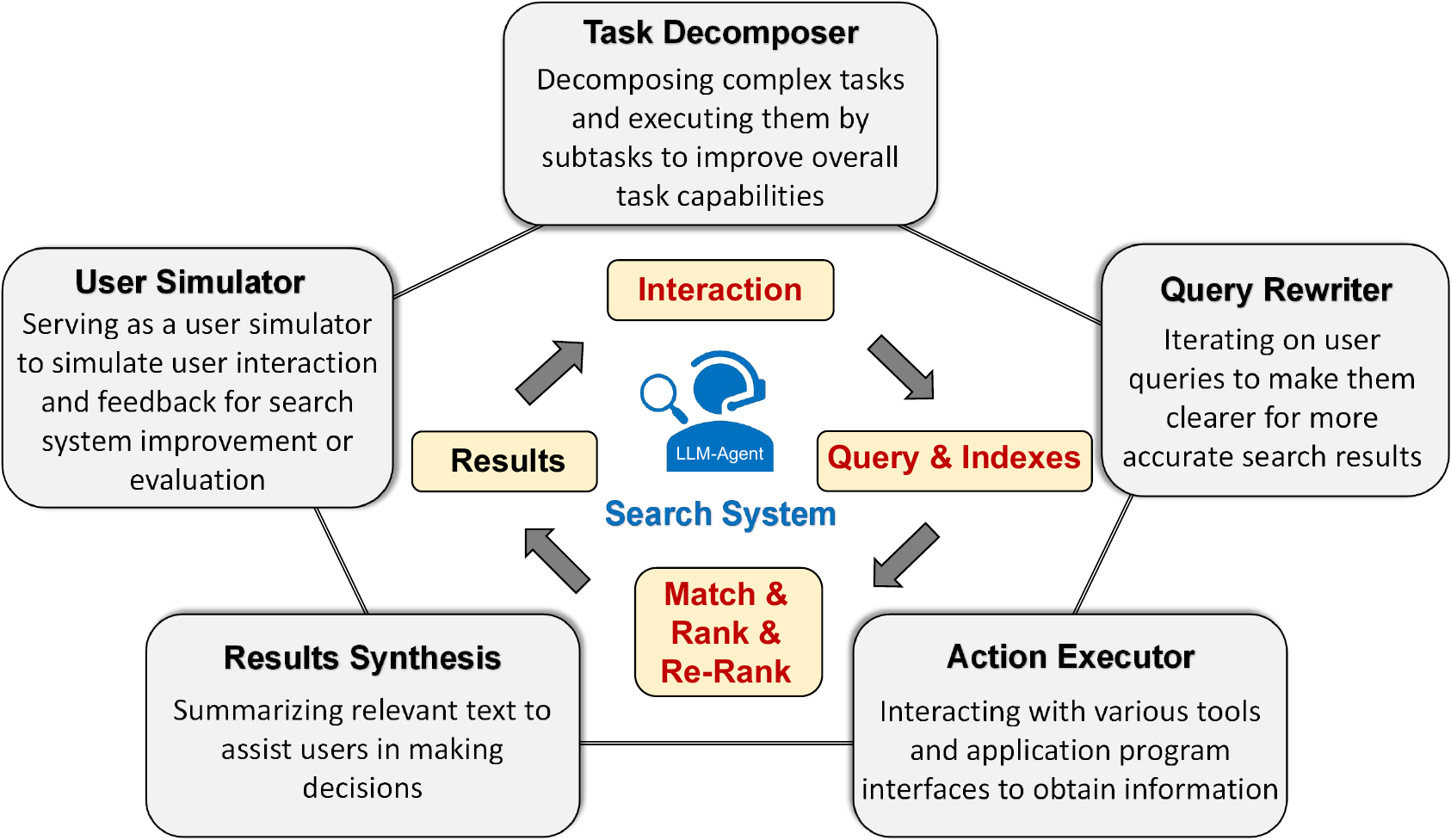}
    \caption{Five domains of LLM agent's role in search tasks}
    \label{fig:Agent4Search}
\end{figure}
\begin{itemize}
    \item \textbf{Task Decomposers:} LLM agents can break down complex tasks into smaller, manageable components to align with user intent and enhance overall task execution. In this case, LLM agents serve as the leading interfaces of search engines.
    \item \textbf{Query Rewriters:} LLM agents are good at refining user queries to make them clearer and more specific, leading to more accurate search results. In this case, LLM agents serve as the query modeling modules.
    \item \textbf{Action Executors:} LLM agents can interact with various tools and APIs to gather necessary information on behalf of the users. In this case, LLM agents serve as the matching and ranking modules.
    \item \textbf{Results Synthesizers:} LLM agents are suitable for summarizing large amounts of search results and helping users quickly grasp the essential details to make decisions. In this case, LLM agents bring new abilities into the traditional searching process.
    \item \textbf{User Simulators:} LLMs can serve as user simulators to mimic user action and interaction for search systems improvement or evaluation. In this case, LLM agents give feedback to search engines.
\end{itemize}
We categorize research work related to search tasks according to the descriptions in the domains mentioned above and list typical works in the five domains in Table \ref{tab:LLM4Search}.

\begin{table}[t!]
	\centering
	\footnotesize
	\caption{A list of representative works of LLM agents in search.}
	\resizebox{1.0\textwidth}{!}{
        \begin{tabular}{c|c|l}
            \toprule
	    \bf Role of agent & \bf Paper & \bf What agents can do (ability) \\ \midrule
	    Decomposer & Laser \cite{ma2023laser} & Uses state-space exploration for web navigation tasks \\
	    Decomposer & Knowagent \cite{zhu2024knowagent} & Integrates knowledge base for task decomposition and logical action execution \\
	    Decomposer & Deng et al. \cite{deng2024multi} & Utilizes self-reflection memory enhancement planning for web navigation tasks \\
	    Decomposer & Gur et al. \cite{gur2023real} & Learns from experience to complete tasks and divide complex instructions \\
	    Decomposer & SteP \cite{sodhi2023step} & Introduces dynamic strategy combination through task decomposition \\
	    Decomposer & Koh et al. \cite{koh2024tree} & Enhances web navigation using tree search algorithms \\
            Decomposer & Agent Q \cite{putta2024agent} & Integrates MCTS-guided search with self-critique for multi-step reasoning \\
	    \midrule
	    Rewriter & CoSearchAgent \cite{gong2024cosearchagent} & Enables collaborative search through plug-ins that understand and refine queries \\
	    Rewriter & Joko et al. \cite{joko2024doing} & Assists in constructing personalized dialogue datasets to enhance query quality \\
	    Rewriter & Aliannejadi et al. \cite{aliannejadi2024trec} & Utilizes internal knowledge of LLMs for better retrieval and response generation \\
            Rewriter & Jain et al. \cite{jain2024llm} & Proposes RAG-powered agents with multi-stream ensemble for semantic code search \\
	    \midrule
	    Executor & AVATAR \cite{wu2024avatar} & Utilizes a comparator LLM to teach the agent how to use tools \\
	    Executor & EASYTOOL \cite{yuan2024easytool} & extracts key information from tool documentation and designs a unified interface \\
            Executor & CodeAct \cite{wang2024executable} & Integrates LLm agents with a Python interpreter in order to execute code actions \\
            Executor & CodeNav \cite{gupta2024codenav} & Proposes code-as-tool paradigm through semantic code search engines \\
	    \midrule
            Synthesizer & PersonaRAG \cite{zerhoudi2024personarag} & Uses real-time personalized data to enhance the relevance of the returned results. \\
	    Synthesizer & ChatCite \cite{li2024chatcite} & Mimics human methods to extract key points and write summaries for literature reviews. \\
            Synthesizer & PaSa \cite{he2025pasa} & Utilizes a selector to determine whether search results should be included or not. \\
	    \midrule
	    Simulator & Sekuli et al. \cite{sekulic2024analysing} & Explores user emulators in conversation search systems for multi-round clarification \\
            Simulator & Usimagent \cite{zhang2024usimagent} & Simulates users' query, click, and stop behavior in search tasks \\     
            Simulator & BASES \cite{ren2024bases} & Establishes a parameterized user profiling system with validation framework \\
            Simulator & Chatshop \cite{chen2024chatshop} & Introduces LLM-simulated shoppers to evaluate agents' multi-turn interaction \\
            \bottomrule
	\end{tabular}
        \label{tab:LLM4Search}
    }
\end{table}

\subsubsection{Task Decomposers}
The existing search technologies typically focus on learning the relevance between user queries and documents without any thoughtful processing. This limitation is critical for complex search tasks. For example, when a user searches for “where to go on vacation?”, his/her expectation might be a comparison of various travel destinations, details on travel itineraries, and hotel/flight prices. As discussed in Section~\ref{sec::motivation}, LLM agents possess strong reasoning and task decomposition abilities, enabling them to collect and aggregate information step by step and provide comprehensive results to the user.

Laser \cite{ma2023laser} introduces a method that views web navigation tasks as state-space exploration. LLM agents decompose tasks into a predefined set of states to handle unfamiliar scenarios and flexible backtracking. 
Deng et al. \cite{deng2024multi} propose a new task called Conversational Web Navigation, which requires complex multi-turn interactions with both users and the environment. In cases where tasks involve intricate actions, the agents will initially decompose the task into smaller sub-tasks. The decomposition results will later be verified and corrected by human annotators.
Gur et al. \cite{gur2023real} introduce a LLM agent named WebAgent. WebAgent has the capability to decompose instructions into standardized sub-instructions and plan ahead, which enables it to better understand and execute complex tasks. For instance, in a complex web operation task, it can first identify the main steps that need to be executed and then carry out these steps sequentially. 
SteP \cite{sodhi2023step} proposes a policy stack to better coordinate different actions and accomplish complex tasks. By decomposing tasks into various policy layers, each policy can focus on specific aspects of the task, thereby improving efficiency and accuracy in task execution. For example, in handling web tasks, one policy might be responsible for page navigation, while another policy might handle form filling. Together, these policies form a policy stack that collaboratively completes the entire task.
Koh et al. \cite{koh2024tree} propose an inference-time search algorithm where agents do a tree search to decide which action to take according to the current state. This method helps LLM agents better perform multi-step planning in interactive web environments.
Knowagent \cite{zhu2024knowagent} utilizes an action knowledge base and a knowledge-driven self-learning approach to guide the action paths during the planning phase. The action knowledge base furnishes the agent with extensive action knowledge, allowing for more rational action selection during planning. Meanwhile, the knowledge-driven self-learning approach enables the agent to continuously learn and refine its planning strategies, enhancing overall planning performance.
Xie et al. \cite{xie2024human} introduce a human-like reasoning framework to improve the planning ability that can help LLM agents when solving multi-phase problems like travel planning.
Agent Q \cite{putta2024agent} proposes a framework that integrates guided Monte Carlo Tree Search (MCTS) with iterative fine-tuning for agents. This approach enables LLM agents to learn effectively from all trajectories, including both successful and unsuccessful ones, thereby enhancing their ability to generalize in complex, multi-step reasoning tasks.

Overall, the task decomposition capability of LLM agents significantly enhances the efficiency of users in obtaining information for complex search tasks. Some of the existing approaches predefine the action space and do search progress to decide the next move. These approaches offer strong control but are limited to solving specific problems. In contrast, the others allow the model to autonomously plan, offering greater flexibility, while the decomposition results may be in turn.

\subsubsection{Query Rewriters}
In traditional search scenarios, users must decide on the search terms themselves and master complex syntax to emphasize which keywords should necessarily be included or excluded from the search, and the length and context of these queries are usually limited. These limitations can prevent users from fully expressing their true intentions. However, these obstacles can be solved through LLM agents.  LLM agents allow users to input longer context to fully express their intentions and then formulate more effective queries by themselves, thereby enhancing search efficiency.

CoSearchAgent \cite{gong2024cosearchagent} is a lightweight collaborative search agent powered
by LLMs and serves as a search plugin on the instant message platform Slack. This agent can autonomously extract the user's search intent based on the conversation between users, formulate it into a query, and then invoke a search engine to return the results. The entire process does not require the user to explicitly input a query, thereby avoiding disruption to the conversational experience.
Joko et al. \cite{joko2024doing} propose a scheme called LAPS (LLM-Augmented Personalized Self-Dialogue) to gather personalized dialogues, which can be used to train preference extraction and personalized response generation. As more conversational data is collected, LLM agents can gradually learn about users' various preferences. For instance, when a vegetarian user inquires about recipes, the agents can account for vegetarian needs and directly search for vegetarian-related rather than normal recipes.
Aliannejadi et al. \cite{aliannejadi2024trec} organize a competition on how to utilize users' prior interactions and present context and observed two main pipelines, namely, multi-turn retrieve-then-generate (R→G) and generate-retrieve-then-generate (G→R→G). In this context, ``G'' includes conversational query expansion and conversational rewriting.
Jain et al. \cite{jain2024llm} propose an approach utilizing Retrieval Augmented Generation (RAG) to enable LLM agents to enrich user queries with additional information. Specifically, by leveraging RAG, the agents can augment user queries with pertinent details sourced from GitHub repositories.

In summary, LLM agents can significantly enhance users' ability to utilize search systems. They can eliminate the need for users to learn search syntax, obviate the necessity of explicitly writing search queries, and even supplement information based on historical interactions or external data to improve search accuracy.

\subsubsection{Action Executors}
The searching process often involves the use of numerous tools. For example, in travel planning, users need to gather various types of information, such as geographical locations, hotels, flights, and weather. Specially designed LLM agents can learn how to judiciously select the appropriate tools and execute these actions on behalf of the user.

AVATAR \cite{wu2024avatar} is a novel automatic framework to optimize the performance of LLM agents when using external tools and knowledge to complete complex tasks. It consists of an actor LLM and a comparator LLM. During optimization, the actor generates actions to answer queries by utilizing the provided tools. The comparator then evaluates a set of well-performing (positive) and poorly-performing (negative) queries, teaching the actor more effective retrieval strategies and tool usage.
EASYTOOL \cite{yuan2024easytool} extracts key information from extensive tool documentation from various sources and meticulously designs a unified interface called tool instructions. This interface provides LLM-based agents with standardized tool descriptions and functionalities. By doing so, it effectively reduces the cognitive load on LLMs when understanding tool functionalities, thereby enhancing the efficiency and accuracy of tool usage.
CodeAct \cite{wang2024executable} proposes consolidating LLM agents' actions into a unified action space using executable Python code with a Python interpreter integrated. In multi-turn interaction scenarios, CodeAct can execute code actions, dynamically revise previous actions, or generate new actions based on new observations.
CodeNav \cite{gupta2024codenav} points out that tool-using LLM agents typically require manual registration of all relevant tools in the LLM context, which presents limitations when dealing with complex real-world codebases. CodeNav addresses this issue by automatically indexing and searching code fragments within the target codebase, eliminating the need for manual tool registration. It can locate relevant code snippets and incorporate them to iteratively develop solutions. Compared to traditional tool-using LLM agents, CodeNav is more flexible and efficient, capable of adapting to different codebase structures and requirements.

In summary, recent researches have focused on two key areas. Firstly, how to equip LLMs agents with a broader range of tools, such as code execution and web retrieval. Secondly, how to make LLM agents better select the most appropriate tools from various options.

\subsubsection{Results Synthesizers}
In traditional search processes, after the search engine returns results, users must review each document individually, determine their relevance, and synthesize conclusions on their own. However, this laborious process can be entirely replaced by LLM agents, which can significantly enhance search efficiency and user experience.

PersonaRAG \cite{zerhoudi2024personarag} is an innovative framework that incorporates user-centric agents to select retrieval and generation based on real user preference. This framework works through three steps: retrieval, user interaction analysis, and cognitive dynamic adaptation. PersonaRAG outperforms baseline models, delivering personalized answers that better address user needs on various question-answering datasets.
ChatCite \cite{li2024chatcite} is an LLM agent guided by human workflows for comparative literature summarization. The agent mimics the human workflow to extract key elements from relevant literature and generate summaries through a reflective incremental mechanism. This approach enables ChatCite to achieve a deeper understanding of the literature content and produce more targeted and comparative summaries.
PaSa \cite{he2025pasa} introduces a hierarchical architecture comprising Crawler and Selector agents to address complex academic search challenges. The Crawler agent dynamically generates search queries and expands citation networks through tool invocation, while the Selector agent evaluates paper relevance via fine-grained scoring. To optimize this process, a novel conversational PPO algorithm is developed to handle sparse rewards in long-horizon search tasks.

In conclusion, large model agents can assist in summarizing search results from multiple perspectives. On one hand, they can consider personalized user information to filter and highlight content relevant to the user. On the other hand, they can be equipped with better mechanisms for filtering information, such as mimicking human expertise or employing specialized selectors.

\subsubsection{User Simulators}
Evaluating search effectiveness has always been a challenge. Existing solutions typically measure algorithm performance by conducting A/B tests and collecting the metrics difference between A/B groups, such as users' click rate and stay time. However, these statistical measures cannot directly reflect user satisfaction, and conducting A/B tests may disrupt real user experiences when the test strategy is suboptimal. LLM agents offer a solution to these issues by simulating user behavior. As previously mentioned, with their independent reasoning and action capabilities, LLM agents can mimic user decision-making processes by learning from real user data. This allows them not only to provide simple feedback like clicks but also to articulate specific reasons for their preferences or dislikes.

Usimagent \cite{zhang2024usimagent} is a search user behavior emulator based on LLMs. USimAgent can simulate users' query, click, and stop behaviors in search tasks, generating complete search sessions. Through empirical research on actual user behavior datasets, researchers found that USimAgent outperforms existing methods in generating queries and performs similarly to traditional methods in predicting user click and stop behaviors.
Sekuli \cite{sekulic2024analysing} explores the use cases of user simulators in conversation search systems. User simulators that can automatically answer clarification questions in multiple rounds of conversations and evaluate the rationality of system search results. Additionally, user simulators can generate large amounts of training data, thereby enhancing the performance of conversational systems.
BASES \cite{ren2024bases} is an innovative user simulation framework based on LLM agents, designed to simulate web search user behavior comprehensively. It is capable of generating unique user profiles on a large scale, leading to diverse search behaviors. Specifically, through learning from vast amounts of web knowledge, the LLM can achieve human-like intelligence and generalization capabilities.
Chen et al. \cite{chen2024chatshop} proposed an online shopping task named ChatShop, where an agent needs to interact with the shopper to understand their needs and preferences gradually. They used LLM agents simulated shoppers to test the capabilities of the agent and found that the simulated shopper made the task more realistic, increasing its complexity and challenge like real human shoppers. 

On the whole, the use of LLM agents as user simulators in conversational search systems shows great promise. They not only enhance evaluation efficiency but also offer valuable insights for improving these systems. Future research could explore further refinements and applications of such simulators to address the challenges in evaluating conversational search results.

\subsection{Benchmark and Datasets}
Existing benchmarks on LLM primarily focus on language understanding and generation. However, these datasets and benchmarks are inadequate for comprehensively assessing the agents on their decision-making and interaction capabilities in complex, dynamic environments. In this subsection, we will elaborate on the research approaches and progress in this area.

Furuta et al. \cite{furuta2024exposing} introduced a benchmark, CompWoB, to highlight the limitations of language model agents in sequential task compositions and to emphasize the need for robust and generalized LLM agents for task combinations. CompWoB is a benchmark consisting of 50 new composite Web Automation tasks that reflect more realistic scenarios. Experimental results show that existing language model agents (such as gpt-3.5-turbo or gpt-4) achieve an average success rate of 94.0\% on basic tasks, but their success rate drops to 24.9\% on composite tasks.
Agentbench \cite{liu2023agentbench} includes diverse agent-task scenarios and assesses LLM agent capabilities through task performance data. This benchmark not only pinpoints the strengths and weaknesses of LLMs in agent tasks but also provides a unified standard for comparison.
WebArena \cite{zhou2023webarena} and MIND2WEB \cite{deng2024mind2web} provide environments that replicate realistic web browsing experiences, incorporating fully functional websites from diverse domains, such as e-commerce and social forums, with URLs, open tabs, and keyboard and mouse interactions.
Cocktail \cite{dai2024cocktail} is a benchmark providing a comprehensive tool for evaluating information retrieval models in the LLM era. Cocktail comprises 16 different datasets covering various text retrieval tasks and domains, including a mix of human-written and LLM-generated corpora. The diversity of these datasets enables researchers to assess the performance of information retrieval models across different scenarios. To avoid potential biases from datasets previously included in LLM training, the authors introduce a new dataset named NQ-UTD. This dataset features queries originating from recent events, ensuring the fairness and validity of the evaluation.

Overall, researchers have introduced various datasets and benchmarks to evaluate the performance of LLM agents in IR scenarios. Unlike conventional LLM evaluation datasets that primarily focus on textual information, these datasets often include real online browsing tasks and user interaction behaviors.

%% file: 5.Embodied.tex
\section{Embodied LLM Agents: Towards Next Generation Recommendation and Search}
\label{sec::Embodied}
Although LLM agents enhance traditional recommendation and search models with comprehensive capabilities, they are often limited to analyzing static user-item interactions, such as pre-training from fixed-size and limited-domain data. This limitation hinders their ability to respond to emerging interactions in changing environments such as smart devices, and provide proactive services from both user and environmental perspectives. Recent embodied agents take a forward step over the LLM agent above by actively perceiving and interacting with environments. Their applications in the cyber world, such as autonomous exploration and online content retrieval in web services, naturally encompass a myriad of information retrieval processes that closely align with the task of recommendation and search. In this section, we discuss the latest development of embodied agents in the cyber environment, their potential applications in recommendation and search, limitations, and promising directions.

\subsection{Recent Advance of Embodied Agents in Cyber Environment}
\textbf{Embodied agents.} LLMs have been recognized with notable reasoning abilities and capacities to utilize external tools and knowledge. Building on this foundation, researchers have integrated these capabilities into unified LLM agents with diverse domain profiles and knowledge of specific targets. However, these LLM agents remain constrained by their dependence on passive learning from static data of images, videos, and text, which are pre-sampled from real-world information systems, like the Internet.

In recent years, researchers have been exploring the next piece of the puzzle of realizing AGI. Reinforcement Learning (RL) and Embodied AI have emerged in response to the thriving research of LLM agents. Consequently, there has been a shift from ``Internet agents'' that focus on learning from constant datasets collected from the Internet, towards ``Embodied agents'' which enable LLM agents to learn through interactions with their surroundings \cite{duan2022survey_embodied_ai}. Just as its name implies, the term \textit{embodied} grants the agent a physical substrate, manifesting in the real world as using robotic arms to measure and interact with the environment or in cyber environments as the ``driving software'' of information systems. Web agents serve as one of the key representations of embodied agents within the cyber environment. These agents must not only learn to navigate websites by interacting with GUI (Graphical User Interface) functions but also respond effectively to feedback from both web systems and human users. Such advancement inherently involves information retrieval tasks, requiring agents to perform complex multi-step searching and multi-modal retrievals. As a result, embodied agents are evolving into a new testbed and paradigm for research in recommendation and search areas.

\textbf{How embodied agents interact with cyber environment.} The current enthusiasm for building web agents is mainly divided into two directions; they consider static and dynamic environments respectively \cite{rawles2024AndroidWorld, zhou2023webarena, gur2023HTMLT5}. Most existing approaches follow the static environments by comparing an agent’s action trajectory to a pre-collected human demonstration. For example, AiTW \cite{rawles2023AiTW}, the most commonly used dataset, comprises over 700K sequential samples with screenshot-action pairs. This lightweight environment offers greater flexibility for designing model architectures and promotes the adoption of more complex technology stacks, such as Chain-of-Thought (CoT) \cite{deng2024mobilebench}. However, agents trained within this constrained framework often follow a predetermined path to complete tasks, which can result in a limited understanding of real-world scenarios and difficulty in responding to new targets. To achieve more realistic evaluations, recent research has developed dynamic environments where agents learn interactively from their mistakes, enabling them to test the boundaries of these systems. AndroidWorld \cite{rawles2024AndroidWorld} is the representative benchmark where a simulated Android system is well-exhibited without restricted action trajectories. This introduces complexity and instability of the real environment, posing a huge challenge to the agent's robustness and ability to cope with complex tasks. In summary, both of these directions entail extensive multi-modal retrieval tasks, which could represent either novel intersection fields or evaluation environments for recommendation and search research.

\textbf{Applications.} Along with these approaches, embodied agents have been developed in diverse fields related to recommendation and search areas. In website search, web agents have been used to instantiate search engine models into autonomous agents, equipped with external web-browsing tools \cite{koh2024visualwebarena, zheng2024SEEACT, he2024webvoyager, kim2024RCIPrompt}. In general GUI control, GUI agents have been supported by fine GUI element grounding capability to retrieve potential interactive objects \cite{hong2023CogAgent, zhang2024AutoGUI, ma2024coco_agent, cheng2024SeeClick}. Other domains, such as games \cite{zhao2024See_and_Think} and 3D environments \cite{huang2024LEO}, have also witnessed the flourishing application of embodied agents.

\subsection{Why Embodied Agents Can Be Used for Recommendation and Search}
The widespread application of embodied agents is based on two key factors: the inherent human-like language reasoning ability of LLM and the interactive learning of embodied agents. The following discussion delves into the intrinsic qualities behind these factors that make embodied agents viable tools for recommendation and search tasks, by reviewing their capabilities as lifelong learners, one-for-all (One4all) models, and personal assistants, respectively.

\begin{itemize}[leftmargin=*]
\item \textbf{Lifelong learner:} One of the core strengths of embodied agents is their ability for continuous and even lifelong learning \cite{wang2023voyager}. Unlike traditional machine learning models that are typically trained on a fixed dataset, embodied agents can persistently expand their knowledge and skills through ongoing interaction with environments. In the recommendation field, previous lifelong works \cite{yuan2021Conure, Lin2023ReLLaRL, li2023STAN} continually focus on identifying long-term behavior dependencies and incorporating emerging intentions, while these traditional task-specific models constantly suffer from catastrophic forgetting. The nature of embodied agents as lifelong learners could be a novel solution. 

\item \textbf{One4all model:} Another crucial advantage of embodied agents is their ability to achieve the One4all model, i.e., generalist problem solver. Differing from learning lifelong, the capability of one4all modeling concentrates on cross-domain tasks, where one4all agents are expected to efficiently adapt to unseen domains at minimal cost (in terms of adaptation time, fine-tuning data, loss of performance, etc.). Existing recommendation work \cite{fu2024iisan, Fu2024Adapter, Li2023OneforAll, cheng2024image} has made great efforts on this topic thanks to techniques such as large-scale pre-training and efficient adaptation. However, a universal representation that seamlessly spans various domains is still out of reach \cite{shin2021one4all, li2023exploring, zhang2024NineRec}. Embodied agents could contribute massive supportive information to knowledge transfer through rapid adaptation to new scenarios.
    
\item \textbf{Personal assistant:} Embodied agents can be omnipotent personal assistants. The characteristics of learning lifelong and being flexible as the one4all model could make the embodied agent a competent assistant for perceiving the rapidly changing personalized needs in the spatial-temporal dimension. In recommendation and search tasks, the possibility of condensing well-trained models into cloud-edge personal services gradually becomes a hot spot in both academics and industries. Prior research \cite{wang2024poisoning, yuan2023federated, yin2024device, rawassizadeh2023odsearch} have delved into the path of achieving lightweight on-device recommender systems and search tools in privacy-preserving and cybersecure approaches, where embodied agents could be the next generation of solutions. 
\end{itemize}

\textbf{Embodied Agent Designing for Recommendation and Search.} 
Existing embodied agents within cyber environment primarily focused on interaction with GUI. These agents concentrate on GUI operations and task-solving on a wide range of information systems like websites, desktops, and smartphones, which are typical scenarios for recommendation and search tasks. This close relationship could inspire innovative approaches and solutions for the next generation of recommendation and search. WorkArena \cite{drouin2024WorkArena} introduces the first environment that supports chat-based agent-user interactions for the development and evaluation of web agents. Through a chat interface, the real human user can exchange messages with web agents. This allows for information retrieval tasks where a specific answer is expected from the agent, but also more practical tasks where user instructions change over time. 
This advancement may inspire further research and discussion around recommendation and search tasks, as agents would need to complete information retrieval tasks and output specific actions or plans. SmartAgent \cite{zhang2024smartagent} defines a new task known as embodied personalized learning, where embodied cognition and item recommendation are interlaced as in many real-world personal assistance scenarios. To address this, a novel reasoning paradigm called Chain-of-User-Thought (COUT) is proposed to align user feedback with embodied agent actions under the progressive thoughts from basic GUI navigation to explicit and implicit user requirements. Leveraging the COUT paradigm, SmartAgent demonstrates the first full-stage embodied personalized capabilities in collaboration with a new dataset called SmartSpot, which for the first time supports a range of diverse embodied personalized environments.

In summary, the combination of embodied agents with recommender and search systems could lead to the creation of intelligent systems offering more personalized services, but may also realize truly bionic personal assistants. This could be a potentially important direction for the next generation of recommendation and search research.

\subsection{Limitation and Promising Directions}
\textbf{Limitation.} Embodied agents have shown promise in the fields of recommendation and search, but they are also challenged by several key limitations. One major challenge is the limited generalization capability of current agents. Most existing agents are primarily designed for static environments, resulting in suboptimal performance when faced with dynamic real-world scenarios. Additionally, these static environments typically represent one-sided functions of specific systems, which further complicates the training of general embodied agents across diverse situations. 
This further leads to the second issue, where currently embodied agents are still not ready to effectively handle information retrieval tasks involved in complex systems. This is mainly due to their limited understanding of environments, as they are still stuck on learning systems' functionality. 
However, real-world application scenarios frequently involve personalized needs, which are often expressed through non-standard paths that extend beyond the training samples. Goal-completion-oriented embodied agents may struggle to address these personalized requirements underlying, resulting in lower user engagement and stickiness.
Finally, there are still other critical technical bottlenecks that have not been fully addressed. For example, executing advanced cross-app operations on smartphone devices and multi-web page tasks on desktop environments remains challenging.

\noindent \textbf{Promising directions.} These limitations also spur new research directions. One of them is leveraging the one-for-all capability of embodied agents for transferable recommendation or search. Currently, large-scale pre-trained recommendation models require extremely high costs to transfer to new scenarios. In some new scenarios where the gap is relatively large, the current pre-training and fine-tuning (PEFT) paradigm finds it difficult to achieve better domain adaptation. The ability of embodied agents to understand dynamic changes in the environment may be able to help with this. Another promising direction is the use of lightweight embodied agents for on-device recommendation. LLMs are often too resource-intensive to deploy on mobile devices, hindering the implementation of personalized services. Embodied agents, with their smaller footprint, may provide a solution to this problem. Additionally, the demand for zero-shot services on end-user devices also poses new requirements for embodied agent capabilities, which could drive more research. Finally, as embodied agents extend the capabilities of LLMs to recommendation and search areas within personal settings, the issue of privacy preservation becomes more crucial. Balancing user behavior modeling and individual privacy protection while providing intelligent services will be a critical and urgent problem. 

%% file: 6.discussions.tex
\section{Open Problems and Future Directions}\label{sec::discussion}
Even though LLM agents have shown great potential to bring promising advancement to IR, there are still many challenges to address and numerous directions to explore. In this section, we will discuss these topics.

\textbf{Hallucinations}. Hallucinations in LLMs refer to situations where the content generated by LLMs is inconsistent with facts or cannot be verified. For LLM agents in the IR field, hallucinations can result in the retrieval of incorrect information, thereby degrading the user experience. Hallucinations can be divided into factuality hallucinations and faithfulness hallucinations \cite{huang2023survey}. The former refers to the inconsistency between the content generated by the LLM and the verifiable facts of the real world, while the latter refers to the inconsistency between the content generated by the model and the user's instructions. Researchers have proposed many methods to alleviate these two types of hallucinations. For factual hallucinations, solutions such as RAG \cite{gao2023retrieval,xu2024tad,ram2023context}, CoT \cite{wei2022chain}, and consistency enhancing \cite{wang2022self,shi2023trusting,o2023contrastive} can constrain LLM to obtain data from given sources. For command-type hallucinations, prompt engineering, RLHF, and other alignment technologies \cite{sharma2023towards,perez2022discovering,wei2023simple} can strengthen the consistency of execution. 

\textbf{Bias}. Bias is a classic issue in the IR field, typically arising from improper data collection or training methods, leading to a significant discrepancy between the model results and real data, thereby degrading the accuracy of information retrieval. The capabilities of LLMs are largely constrained by their training data, making them more likely to encounter bias problems. Studies \cite{dai2023llms,dai2024unifying,ma2023large,zheng2023large} indicate that LLMs can introduce various biases, such as gender discrimination \cite{dhingra2023queer}, political extremism \cite{dong2024building}, and echo chamber \cite{sharma2024generative}. Dai et al. \cite{dai2023llms} find that LLM-based retrievers tend to prioritize content generated by LLMs over human-written text and introduce a bias-oriented loss to alleviate this trend. 

\textbf{Deployment Cost}. LLM agents are typically composed of multiple LLM modules, resulting in a higher number of invocations and model parameters compared to general LLM applications. So, deploying LLM agents requires substantial resources, resulting in longer inference latency, which imposes higher hardware requirements on the end devices and may impact users' experience. Researchers have put a lot of effort into this issue. Wilkins et al. \cite{wilkins2024hybrid} optimize LLM efficiency by proposing a cost-based scheduling framework, while Avatar \cite{wu2025avatar} and QueryAgent \cite{huang2024queryagent} utilize tools or environmental feedback to reduce inference time.

\textbf{Multi-Modal Agent}. Multi-modal LLM agents can integrate data from different modalities, such as images and text, to more comprehensively understand user queries. For instance, when querying about a product, a user might provide both a textual description and an image of the product. A multi-modal LLM agent can analyze both the text and image information simultaneously, better understand the product's features, and thereby provide more accurate retrieval results \cite{xie2024large,zhang2024multimodal}. Recently, numerous studies have begun exploring how to leverage multi-modal capabilities to enhance LLM agents' ability to operate on web pages \cite{zheng2024gpt,furuta2023multimodal} and mobile devices \cite{nong2024mobileflow,wang2024Mobileagent}. These studies typically utilize visual perception capabilities to identify visual and textual elements of mobile applications accurately.

\textbf{Domain Specific Agent}. 
In specialized domains with substantial domain knowledge, general LLM agents often perform poorly due to their reliance on common knowledge and data sources. However, data in these specialized domains is relatively sparse, making it insufficient to fully train LLMs. Therefore, efficiently and economically transforming general LLM agents into domain-specific LLM agents is a crucial direction for future research. Currently, related research has already been applied to relatively niche but popular categories such as clinical \cite{van2024adapted,li2023beginner,russe2024improving,yan2024clinicallab,jin2024agentmd} and legal \cite{sun2024lawluo,li2024legalagentbench}. In the future, as LLM agents progressively improve their ability to utilize sparse data, there may be opportunities for enhancements in niche areas such as psychological counseling and job seeking.

\textbf{Multi Agent Interaction}. In multi-agent systems, individual agents can solve complex problems through information sharing and collaboration. Different agents may possess distinct knowledge and skills, and by cooperating, they can leverage their strengths to enhance overall performance information retrieval \cite{choi1998multi,wondergem1998agents}.
Recently, studies \cite{park2023generative,park2024generative} have explored the interaction and evolution processes among multiple LLM agents, and several multi-agent frameworks \cite{chen2023agentverse,qian2023communicative,team2023xagent} have been proposed. Infogent \cite{reddy2024infogent} utilized multi-agent systems to simulate human cognition for gathering information from the web. Chateval \cite{chan2023chateval} and MAD \cite{liang2023encouraging} improve agents' generating quality through multi-agent debate.

\textbf{Personalization}. Different users often have varying needs and preferences when performing information retrieval. For instance, the search term ``Apple'' might refer to Apple Inc. for a tech investor, whereas it might mean the fruit for others. By leveraging the memory module and long context processing capabilities of LLM agents, personalized information retrieval tailored to individual users can be achieved \cite{li2024personal}. One approach to do so is to treat the user's interaction history as context and input it into the model \cite{mo2024leverage,richardson2023integrating,aknouche2012integrating}. 
Mo et al. \cite{mo2024leverage} incorporate conversational and personalized elements with LLM agents to fulfill user needs through multi-turn interactions. They indicate that Personal Textual Knowledge Bases (PTKBs) can effectively enhance conversational information retrieval, allowing retrieval results to align more closely with the user's background.

%% file: 0.main.bbl

\begin{thebibliography}{191}

%%% the Web address.

\ifx \showCODEN    \undefined \def \showCODEN     #1{\unskip}     \fi
\ifx \showDOI      \undefined \def \showDOI       #1{#1}\fi
\ifx \showISBNx    \undefined \def \showISBNx     #1{\unskip}     \fi
\ifx \showISBNxiii \undefined \def \showISBNxiii  #1{\unskip}     \fi
\ifx \showISSN     \undefined \def \showISSN      #1{\unskip}     \fi
\ifx \showLCCN     \undefined \def \showLCCN      #1{\unskip}     \fi
\ifx \shownote     \undefined \def \shownote      #1{#1}          \fi
\ifx \showarticletitle \undefined \def \showarticletitle #1{#1}   \fi
\ifx \showURL      \undefined \def \showURL       {\relax}        \fi
% invisible to TeX
\providecommand\bibfield[2]{#2}
\providecommand\bibinfo[2]{#2}
\providecommand\natexlab[1]{#1}
\providecommand\showeprint[2][]{arXiv:#2}

\bibitem[Achiam et~al\mbox{.}(2023)]%
        {achiam2023gpt}
\bibfield{author}{\bibinfo{person}{Josh Achiam}, \bibinfo{person}{Steven Adler}, \bibinfo{person}{Sandhini Agarwal}, \bibinfo{person}{Lama Ahmad}, \bibinfo{person}{Ilge Akkaya}, \bibinfo{person}{Florencia~Leoni Aleman}, \bibinfo{person}{Diogo Almeida}, \bibinfo{person}{Janko Altenschmidt}, \bibinfo{person}{Sam Altman}, \bibinfo{person}{Shyamal Anadkat}, {et~al\mbox{.}}} \bibinfo{year}{2023}\natexlab{}.
\newblock \showarticletitle{Gpt-4 technical report}.
\newblock \bibinfo{journal}{\emph{arXiv preprint arXiv:2303.08774}} (\bibinfo{year}{2023}).
\newblock


\bibitem[Aggarwal et~al\mbox{.}(2016)]%
        {aggarwal2016recommender}
\bibfield{author}{\bibinfo{person}{Charu~C Aggarwal} {et~al\mbox{.}}} \bibinfo{year}{2016}\natexlab{}.
\newblock \bibinfo{booktitle}{\emph{Recommender systems}}. Vol.~\bibinfo{volume}{1}.
\newblock \bibinfo{publisher}{Springer}.
\newblock


\bibitem[Ai et~al\mbox{.}(2023)]%
        {ai2023information}
\bibfield{author}{\bibinfo{person}{Qingyao Ai}, \bibinfo{person}{Ting Bai}, \bibinfo{person}{Zhao Cao}, \bibinfo{person}{Yi Chang}, \bibinfo{person}{Jiawei Chen}, \bibinfo{person}{Zhumin Chen}, \bibinfo{person}{Zhiyong Cheng}, \bibinfo{person}{Shoubin Dong}, \bibinfo{person}{Zhicheng Dou}, \bibinfo{person}{Fuli Feng}, {et~al\mbox{.}}} \bibinfo{year}{2023}\natexlab{}.
\newblock \showarticletitle{Information retrieval meets large language models: a strategic report from chinese ir community}.
\newblock \bibinfo{journal}{\emph{AI Open}}  \bibinfo{volume}{4} (\bibinfo{year}{2023}), \bibinfo{pages}{80--90}.
\newblock


\bibitem[Aknouche et~al\mbox{.}(2012)]%
        {aknouche2012integrating}
\bibfield{author}{\bibinfo{person}{Rachid Aknouche}, \bibinfo{person}{Ounas Asfari}, \bibinfo{person}{Fadila Bentayeb}, {and} \bibinfo{person}{Omar Boussaid}.} \bibinfo{year}{2012}\natexlab{}.
\newblock \showarticletitle{Integrating query context and user context in an information retrieval model based on expanded language modeling}. In \bibinfo{booktitle}{\emph{International Conference on Availability, Reliability, and Security}}. Springer, \bibinfo{pages}{244--258}.
\newblock


\bibitem[Aliannejadi et~al\mbox{.}(2024)]%
        {aliannejadi2024trec}
\bibfield{author}{\bibinfo{person}{Mohammad Aliannejadi}, \bibinfo{person}{Zahra Abbasiantaeb}, \bibinfo{person}{Shubham Chatterjee}, \bibinfo{person}{Jeffery Dalton}, {and} \bibinfo{person}{Leif Azzopardi}.} \bibinfo{year}{2024}\natexlab{}.
\newblock \showarticletitle{Trec ikat 2023: The interactive knowledge assistance track overview}.
\newblock \bibinfo{journal}{\emph{arXiv preprint arXiv:2401.01330}} (\bibinfo{year}{2024}).
\newblock


\bibitem[Anthropic(2024)]%
        {claude-3}
\bibfield{author}{\bibinfo{person}{Anthropic}.} \bibinfo{year}{2024}\natexlab{}.
\newblock \bibinfo{title}{Claude 3 haiku: our fastest model yet.}
\newblock \bibinfo{howpublished}{\url{https://www.anthropic.com/news/claude-3-haiku}}.
\newblock


\bibitem[Baeza-Yates et~al\mbox{.}(1999)]%
        {baeza1999modern}
\bibfield{author}{\bibinfo{person}{Ricardo Baeza-Yates}, \bibinfo{person}{Berthier Ribeiro-Neto}, {et~al\mbox{.}}} \bibinfo{year}{1999}\natexlab{}.
\newblock \bibinfo{booktitle}{\emph{Modern information retrieval}}. Vol.~\bibinfo{volume}{463}.
\newblock \bibinfo{publisher}{ACM press New York}.
\newblock


\bibitem[Bao et~al\mbox{.}(2024)]%
        {bao2024large}
\bibfield{author}{\bibinfo{person}{Keqin Bao}, \bibinfo{person}{Jizhi Zhang}, \bibinfo{person}{Xinyu Lin}, \bibinfo{person}{Yang Zhang}, \bibinfo{person}{Wenjie Wang}, {and} \bibinfo{person}{Fuli Feng}.} \bibinfo{year}{2024}\natexlab{}.
\newblock \showarticletitle{Large Language Models for Recommendation: Past, Present, and Future}. In \bibinfo{booktitle}{\emph{Proceedings of the 47th International ACM SIGIR Conference on Research and Development in Information Retrieval}}. \bibinfo{pages}{2993--2996}.
\newblock


\bibitem[Bond and Gasser(2014)]%
        {bond2014readings}
\bibfield{author}{\bibinfo{person}{Alan~H Bond} {and} \bibinfo{person}{Les Gasser}.} \bibinfo{year}{2014}\natexlab{}.
\newblock \bibinfo{booktitle}{\emph{Readings in distributed artificial intelligence}}.
\newblock \bibinfo{publisher}{Morgan Kaufmann}.
\newblock


\bibitem[Brin and Page(1998)]%
        {brin1998anatomy}
\bibfield{author}{\bibinfo{person}{Sergey Brin} {and} \bibinfo{person}{Lawrence Page}.} \bibinfo{year}{1998}\natexlab{}.
\newblock \showarticletitle{The anatomy of a large-scale hypertextual web search engine}.
\newblock \bibinfo{journal}{\emph{Computer networks and ISDN systems}} \bibinfo{volume}{30}, \bibinfo{number}{1-7} (\bibinfo{year}{1998}), \bibinfo{pages}{107--117}.
\newblock


\bibitem[Cai et~al\mbox{.}(2024)]%
        {cai2024flow}
\bibfield{author}{\bibinfo{person}{Shihao Cai}, \bibinfo{person}{Jizhi Zhang}, \bibinfo{person}{Keqin Bao}, \bibinfo{person}{Chongming Gao}, {and} \bibinfo{person}{Fuli Feng}.} \bibinfo{year}{2024}\natexlab{}.
\newblock \showarticletitle{FLOW: A Feedback LOop FrameWork for Simultaneously Enhancing Recommendation and User Agents}.
\newblock \bibinfo{journal}{\emph{arXiv preprint arXiv:2410.20027}} (\bibinfo{year}{2024}).
\newblock


\bibitem[Chan et~al\mbox{.}(2023)]%
        {chan2023chateval}
\bibfield{author}{\bibinfo{person}{Chi-Min Chan}, \bibinfo{person}{Weize Chen}, \bibinfo{person}{Yusheng Su}, \bibinfo{person}{Jianxuan Yu}, \bibinfo{person}{Wei Xue}, \bibinfo{person}{Shanghang Zhang}, \bibinfo{person}{Jie Fu}, {and} \bibinfo{person}{Zhiyuan Liu}.} \bibinfo{year}{2023}\natexlab{}.
\newblock \showarticletitle{Chateval: Towards better llm-based evaluators through multi-agent debate}.
\newblock \bibinfo{journal}{\emph{arXiv preprint arXiv:2308.07201}} (\bibinfo{year}{2023}).
\newblock


\bibitem[Chen et~al\mbox{.}(2024a)]%
        {chen2024large}
\bibfield{author}{\bibinfo{person}{Jin Chen}, \bibinfo{person}{Zheng Liu}, \bibinfo{person}{Xu Huang}, \bibinfo{person}{Chenwang Wu}, \bibinfo{person}{Qi Liu}, \bibinfo{person}{Gangwei Jiang}, \bibinfo{person}{Yuanhao Pu}, \bibinfo{person}{Yuxuan Lei}, \bibinfo{person}{Xiaolong Chen}, \bibinfo{person}{Xingmei Wang}, {et~al\mbox{.}}} \bibinfo{year}{2024}\natexlab{a}.
\newblock \showarticletitle{When large language models meet personalization: Perspectives of challenges and opportunities}.
\newblock \bibinfo{journal}{\emph{World Wide Web}} \bibinfo{volume}{27}, \bibinfo{number}{4} (\bibinfo{year}{2024}), \bibinfo{pages}{42}.
\newblock


\bibitem[Chen et~al\mbox{.}(2024b)]%
        {chen2024chatshop}
\bibfield{author}{\bibinfo{person}{Sanxing Chen}, \bibinfo{person}{Sam Wiseman}, {and} \bibinfo{person}{Bhuwan Dhingra}.} \bibinfo{year}{2024}\natexlab{b}.
\newblock \showarticletitle{ChatShop: Interactive Information Seeking with Language Agents}.
\newblock \bibinfo{journal}{\emph{arXiv preprint arXiv:2404.09911}} (\bibinfo{year}{2024}).
\newblock


\bibitem[Chen et~al\mbox{.}(2023)]%
        {chen2023agentverse}
\bibfield{author}{\bibinfo{person}{Weize Chen}, \bibinfo{person}{Yusheng Su}, \bibinfo{person}{Jingwei Zuo}, \bibinfo{person}{Cheng Yang}, \bibinfo{person}{Chenfei Yuan}, \bibinfo{person}{Chen Qian}, \bibinfo{person}{Chi-Min Chan}, \bibinfo{person}{Yujia Qin}, \bibinfo{person}{Yaxi Lu}, \bibinfo{person}{Ruobing Xie}, {et~al\mbox{.}}} \bibinfo{year}{2023}\natexlab{}.
\newblock \showarticletitle{Agentverse: Facilitating multi-agent collaboration and exploring emergent behaviors in agents}.
\newblock \bibinfo{journal}{\emph{arXiv preprint arXiv:2308.10848}} (\bibinfo{year}{2023}).
\newblock


\bibitem[Cheng et~al\mbox{.}(2016)]%
        {cheng2016wide}
\bibfield{author}{\bibinfo{person}{Heng-Tze Cheng}, \bibinfo{person}{Levent Koc}, \bibinfo{person}{Jeremiah Harmsen}, \bibinfo{person}{Tal Shaked}, \bibinfo{person}{Tushar Chandra}, \bibinfo{person}{Hrishi Aradhye}, \bibinfo{person}{Glen Anderson}, \bibinfo{person}{Greg Corrado}, \bibinfo{person}{Wei Chai}, \bibinfo{person}{Mustafa Ispir}, {et~al\mbox{.}}} \bibinfo{year}{2016}\natexlab{}.
\newblock \showarticletitle{Wide \& deep learning for recommender systems}. In \bibinfo{booktitle}{\emph{Proceedings of the 1st workshop on deep learning for recommender systems}}. \bibinfo{pages}{7--10}.
\newblock


\bibitem[Cheng et~al\mbox{.}(2024b)]%
        {cheng2024SeeClick}
\bibfield{author}{\bibinfo{person}{Kanzhi Cheng}, \bibinfo{person}{Qiushi Sun}, \bibinfo{person}{Yougang Chu}, \bibinfo{person}{Fangzhi Xu}, \bibinfo{person}{Yantao Li}, \bibinfo{person}{Jianbing Zhang}, {and} \bibinfo{person}{Zhiyong Wu}.} \bibinfo{year}{2024}\natexlab{b}.
\newblock \showarticletitle{Seeclick: Harnessing gui grounding for advanced visual gui agents}.
\newblock \bibinfo{journal}{\emph{arXiv preprint arXiv:2401.10935}} (\bibinfo{year}{2024}).
\newblock


\bibitem[Cheng et~al\mbox{.}(2024a)]%
        {cheng2024image}
\bibfield{author}{\bibinfo{person}{Yu Cheng}, \bibinfo{person}{Yunzhu Pan}, \bibinfo{person}{Jiaqi Zhang}, \bibinfo{person}{Yongxin Ni}, \bibinfo{person}{Aixin Sun}, {and} \bibinfo{person}{Fajie Yuan}.} \bibinfo{year}{2024}\natexlab{a}.
\newblock \showarticletitle{An Image Dataset for Benchmarking Recommender Systems with Raw Pixels}. In \bibinfo{booktitle}{\emph{Proceedings of the 2024 SIAM International Conference on Data Mining (SDM)}}. SIAM, \bibinfo{pages}{418--426}.
\newblock


\bibitem[Cho et~al\mbox{.}(2014)]%
        {cho2014learning}
\bibfield{author}{\bibinfo{person}{Kyunghyun Cho}, \bibinfo{person}{Bart Van~Merri{\"e}nboer}, \bibinfo{person}{Caglar Gulcehre}, \bibinfo{person}{Dzmitry Bahdanau}, \bibinfo{person}{Fethi Bougares}, \bibinfo{person}{Holger Schwenk}, {and} \bibinfo{person}{Yoshua Bengio}.} \bibinfo{year}{2014}\natexlab{}.
\newblock \showarticletitle{Learning phrase representations using RNN encoder-decoder for statistical machine translation}.
\newblock \bibinfo{journal}{\emph{arXiv preprint arXiv:1406.1078}} (\bibinfo{year}{2014}).
\newblock


\bibitem[Choi and Yoo(1998)]%
        {choi1998multi}
\bibfield{author}{\bibinfo{person}{Yong~S Choi} {and} \bibinfo{person}{Suk~I Yoo}.} \bibinfo{year}{1998}\natexlab{}.
\newblock \showarticletitle{Multi-agent learning approach to www information retrieval using neural network}. In \bibinfo{booktitle}{\emph{Proceedings of the 4th international conference on Intelligent user interfaces}}. \bibinfo{pages}{23--30}.
\newblock


\bibitem[Corecco et~al\mbox{.}({[n.\,d.]})]%
        {coreccosuber}
\bibfield{author}{\bibinfo{person}{Nathan Corecco}, \bibinfo{person}{Giorgio Piatti}, \bibinfo{person}{Luca~A Lanzend{\"o}rfer}, \bibinfo{person}{Flint~Xiaofeng Fan}, {and} \bibinfo{person}{Roger Wattenhofer}.} \bibinfo{year}{[n.\,d.]}\natexlab{}.
\newblock \showarticletitle{SUBER: An RL Environment with Simulated Human Behavior for Recommender Systems}.
\newblock  (\bibinfo{year}{[n.\,d.]}).
\newblock


\bibitem[Covington et~al\mbox{.}(2016)]%
        {covington2016deep}
\bibfield{author}{\bibinfo{person}{Paul Covington}, \bibinfo{person}{Jay Adams}, {and} \bibinfo{person}{Emre Sargin}.} \bibinfo{year}{2016}\natexlab{}.
\newblock \showarticletitle{Deep neural networks for youtube recommendations}. In \bibinfo{booktitle}{\emph{Proceedings of the 10th ACM conference on recommender systems}}. \bibinfo{pages}{191--198}.
\newblock


\bibitem[Croft et~al\mbox{.}(2010)]%
        {croft2010search}
\bibfield{author}{\bibinfo{person}{W~Bruce Croft}, \bibinfo{person}{Donald Metzler}, {and} \bibinfo{person}{Trevor Strohman}.} \bibinfo{year}{2010}\natexlab{}.
\newblock \bibinfo{booktitle}{\emph{Search engines: Information retrieval in practice}}. Vol.~\bibinfo{volume}{520}.
\newblock \bibinfo{publisher}{Addison-Wesley Reading}.
\newblock


\bibitem[Dai et~al\mbox{.}(2024a)]%
        {dai2024cocktail}
\bibfield{author}{\bibinfo{person}{Sunhao Dai}, \bibinfo{person}{Weihao Liu}, \bibinfo{person}{Yuqi Zhou}, \bibinfo{person}{Liang Pang}, \bibinfo{person}{Rongju Ruan}, \bibinfo{person}{Gang Wang}, \bibinfo{person}{Zhenhua Dong}, \bibinfo{person}{Jun Xu}, {and} \bibinfo{person}{Ji-Rong Wen}.} \bibinfo{year}{2024}\natexlab{a}.
\newblock \showarticletitle{Cocktail: A Comprehensive Information Retrieval Benchmark with LLM-Generated Documents Integration}.
\newblock \bibinfo{journal}{\emph{arXiv preprint arXiv:2405.16546}} (\bibinfo{year}{2024}).
\newblock


\bibitem[Dai et~al\mbox{.}(2024b)]%
        {dai2024unifying}
\bibfield{author}{\bibinfo{person}{Sunhao Dai}, \bibinfo{person}{Chen Xu}, \bibinfo{person}{Shicheng Xu}, \bibinfo{person}{Liang Pang}, \bibinfo{person}{Zhenhua Dong}, {and} \bibinfo{person}{Jun Xu}.} \bibinfo{year}{2024}\natexlab{b}.
\newblock \showarticletitle{Unifying Bias and Unfairness in Information Retrieval: A Survey of Challenges and Opportunities with Large Language Models}.
\newblock \bibinfo{journal}{\emph{arXiv preprint arXiv:2404.11457}} (\bibinfo{year}{2024}).
\newblock


\bibitem[Dai et~al\mbox{.}(2023)]%
        {dai2023llms}
\bibfield{author}{\bibinfo{person}{Sunhao Dai}, \bibinfo{person}{Yuqi Zhou}, \bibinfo{person}{Liang Pang}, \bibinfo{person}{Weihao Liu}, \bibinfo{person}{Xiaolin Hu}, \bibinfo{person}{Yong Liu}, \bibinfo{person}{Xiao Zhang}, {and} \bibinfo{person}{Jun Xu}.} \bibinfo{year}{2023}\natexlab{}.
\newblock \showarticletitle{Llms may dominate information access: Neural retrievers are biased towards llm-generated texts}.
\newblock \bibinfo{journal}{\emph{arXiv preprint arXiv:2310.20501}} (\bibinfo{year}{2023}).
\newblock


\bibitem[Deng et~al\mbox{.}(2024b)]%
        {deng2024mobilebench}
\bibfield{author}{\bibinfo{person}{Shihan Deng}, \bibinfo{person}{Weikai Xu}, \bibinfo{person}{Hongda Sun}, \bibinfo{person}{Wei Liu}, \bibinfo{person}{Tao Tan}, \bibinfo{person}{Jianfeng Liu}, \bibinfo{person}{Ang Li}, \bibinfo{person}{Jian Luan}, \bibinfo{person}{Bin Wang}, \bibinfo{person}{Rui Yan}, {et~al\mbox{.}}} \bibinfo{year}{2024}\natexlab{b}.
\newblock \showarticletitle{Mobile-Bench: An Evaluation Benchmark for LLM-based Mobile Agents}.
\newblock \bibinfo{journal}{\emph{arXiv preprint arXiv:2407.00993}} (\bibinfo{year}{2024}).
\newblock


\bibitem[Deng et~al\mbox{.}(2024a)]%
        {deng2024mind2web}
\bibfield{author}{\bibinfo{person}{Xiang Deng}, \bibinfo{person}{Yu Gu}, \bibinfo{person}{Boyuan Zheng}, \bibinfo{person}{Shijie Chen}, \bibinfo{person}{Sam Stevens}, \bibinfo{person}{Boshi Wang}, \bibinfo{person}{Huan Sun}, {and} \bibinfo{person}{Yu Su}.} \bibinfo{year}{2024}\natexlab{a}.
\newblock \showarticletitle{Mind2web: Towards a generalist agent for the web}.
\newblock \bibinfo{journal}{\emph{Advances in Neural Information Processing Systems}}  \bibinfo{volume}{36} (\bibinfo{year}{2024}).
\newblock


\bibitem[Deng et~al\mbox{.}(2024c)]%
        {deng2024multi}
\bibfield{author}{\bibinfo{person}{Yang Deng}, \bibinfo{person}{Xuan Zhang}, \bibinfo{person}{Wenxuan Zhang}, \bibinfo{person}{Yifei Yuan}, \bibinfo{person}{See-Kiong Ng}, {and} \bibinfo{person}{Tat-Seng Chua}.} \bibinfo{year}{2024}\natexlab{c}.
\newblock \showarticletitle{On the Multi-turn Instruction Following for Conversational Web Agents}.
\newblock \bibinfo{journal}{\emph{arXiv preprint arXiv:2402.15057}} (\bibinfo{year}{2024}).
\newblock


\bibitem[Devlin et~al\mbox{.}(2018)]%
        {devlin2018bert}
\bibfield{author}{\bibinfo{person}{Jacob Devlin}, \bibinfo{person}{Ming-Wei Chang}, \bibinfo{person}{Kenton Lee}, {and} \bibinfo{person}{Kristina Toutanova}.} \bibinfo{year}{2018}\natexlab{}.
\newblock \showarticletitle{Bert: Pre-training of deep bidirectional transformers for language understanding}.
\newblock \bibinfo{journal}{\emph{arXiv preprint arXiv:1810.04805}} (\bibinfo{year}{2018}).
\newblock


\bibitem[Dhingra et~al\mbox{.}(2023)]%
        {dhingra2023queer}
\bibfield{author}{\bibinfo{person}{Harnoor Dhingra}, \bibinfo{person}{Preetiha Jayashanker}, \bibinfo{person}{Sayali Moghe}, {and} \bibinfo{person}{Emma Strubell}.} \bibinfo{year}{2023}\natexlab{}.
\newblock \showarticletitle{Queer people are people first: Deconstructing sexual identity stereotypes in large language models}.
\newblock \bibinfo{journal}{\emph{arXiv preprint arXiv:2307.00101}} (\bibinfo{year}{2023}).
\newblock


\bibitem[Ding et~al\mbox{.}(2024)]%
        {ding2024longrope}
\bibfield{author}{\bibinfo{person}{Yiran Ding}, \bibinfo{person}{Li~Lyna Zhang}, \bibinfo{person}{Chengruidong Zhang}, \bibinfo{person}{Yuanyuan Xu}, \bibinfo{person}{Ning Shang}, \bibinfo{person}{Jiahang Xu}, \bibinfo{person}{Fan Yang}, {and} \bibinfo{person}{Mao Yang}.} \bibinfo{year}{2024}\natexlab{}.
\newblock \showarticletitle{Longrope: Extending llm context window beyond 2 million tokens}.
\newblock \bibinfo{journal}{\emph{arXiv preprint arXiv:2402.13753}} (\bibinfo{year}{2024}).
\newblock


\bibitem[Dong et~al\mbox{.}(2024)]%
        {dong2024building}
\bibfield{author}{\bibinfo{person}{Yi Dong}, \bibinfo{person}{Ronghui Mu}, \bibinfo{person}{Gaojie Jin}, \bibinfo{person}{Yi Qi}, \bibinfo{person}{Jinwei Hu}, \bibinfo{person}{Xingyu Zhao}, \bibinfo{person}{Jie Meng}, \bibinfo{person}{Wenjie Ruan}, {and} \bibinfo{person}{Xiaowei Huang}.} \bibinfo{year}{2024}\natexlab{}.
\newblock \showarticletitle{Building guardrails for large language models}.
\newblock \bibinfo{journal}{\emph{arXiv preprint arXiv:2402.01822}} (\bibinfo{year}{2024}).
\newblock


\bibitem[Dorri et~al\mbox{.}(2018)]%
        {dorri2018multi}
\bibfield{author}{\bibinfo{person}{Ali Dorri}, \bibinfo{person}{Salil~S Kanhere}, {and} \bibinfo{person}{Raja Jurdak}.} \bibinfo{year}{2018}\natexlab{}.
\newblock \showarticletitle{Multi-agent systems: A survey}.
\newblock \bibinfo{journal}{\emph{Ieee Access}}  \bibinfo{volume}{6} (\bibinfo{year}{2018}), \bibinfo{pages}{28573--28593}.
\newblock


\bibitem[Drouin et~al\mbox{.}(2024)]%
        {drouin2024WorkArena}
\bibfield{author}{\bibinfo{person}{Alexandre Drouin}, \bibinfo{person}{Maxime Gasse}, \bibinfo{person}{Massimo Caccia}, \bibinfo{person}{Issam~H Laradji}, \bibinfo{person}{Manuel Del~Verme}, \bibinfo{person}{Tom Marty}, \bibinfo{person}{L{\'e}o Boisvert}, \bibinfo{person}{Megh Thakkar}, \bibinfo{person}{Quentin Cappart}, \bibinfo{person}{David Vazquez}, {et~al\mbox{.}}} \bibinfo{year}{2024}\natexlab{}.
\newblock \showarticletitle{WorkArena: How Capable are Web Agents at Solving Common Knowledge Work Tasks?}
\newblock \bibinfo{journal}{\emph{arXiv preprint arXiv:2403.07718}} (\bibinfo{year}{2024}).
\newblock


\bibitem[Duan et~al\mbox{.}(2022)]%
        {duan2022survey_embodied_ai}
\bibfield{author}{\bibinfo{person}{Jiafei Duan}, \bibinfo{person}{Samson Yu}, \bibinfo{person}{Hui~Li Tan}, \bibinfo{person}{Hongyuan Zhu}, {and} \bibinfo{person}{Cheston Tan}.} \bibinfo{year}{2022}\natexlab{}.
\newblock \showarticletitle{A survey of embodied ai: From simulators to research tasks}.
\newblock \bibinfo{journal}{\emph{IEEE Transactions on Emerging Topics in Computational Intelligence}} \bibinfo{volume}{6}, \bibinfo{number}{2} (\bibinfo{year}{2022}), \bibinfo{pages}{230--244}.
\newblock


\bibitem[Fang et~al\mbox{.}(2024)]%
        {fang2024multi}
\bibfield{author}{\bibinfo{person}{Jiabao Fang}, \bibinfo{person}{Shen Gao}, \bibinfo{person}{Pengjie Ren}, \bibinfo{person}{Xiuying Chen}, \bibinfo{person}{Suzan Verberne}, {and} \bibinfo{person}{Zhaochun Ren}.} \bibinfo{year}{2024}\natexlab{}.
\newblock \showarticletitle{A multi-agent conversational recommender system}.
\newblock \bibinfo{journal}{\emph{arXiv preprint arXiv:2402.01135}} (\bibinfo{year}{2024}).
\newblock


\bibitem[Feng et~al\mbox{.}(2024)]%
        {feng2024towards}
\bibfield{author}{\bibinfo{person}{Guhao Feng}, \bibinfo{person}{Bohang Zhang}, \bibinfo{person}{Yuntian Gu}, \bibinfo{person}{Haotian Ye}, \bibinfo{person}{Di He}, {and} \bibinfo{person}{Liwei Wang}.} \bibinfo{year}{2024}\natexlab{}.
\newblock \showarticletitle{Towards revealing the mystery behind chain of thought: a theoretical perspective}.
\newblock \bibinfo{journal}{\emph{Advances in Neural Information Processing Systems}}  \bibinfo{volume}{36} (\bibinfo{year}{2024}).
\newblock


\bibitem[Fu et~al\mbox{.}(2024a)]%
        {fu2024iisan}
\bibfield{author}{\bibinfo{person}{Junchen Fu}, \bibinfo{person}{Xuri Ge}, \bibinfo{person}{Xin Xin}, \bibinfo{person}{Alexandros Karatzoglou}, \bibinfo{person}{Ioannis Arapakis}, \bibinfo{person}{Jie Wang}, {and} \bibinfo{person}{Joemon~M Jose}.} \bibinfo{year}{2024}\natexlab{a}.
\newblock \showarticletitle{IISAN: Efficiently adapting multimodal representation for sequential recommendation with decoupled PEFT}. In \bibinfo{booktitle}{\emph{Proceedings of the 47th International ACM SIGIR Conference on Research and Development in Information Retrieval}}. \bibinfo{pages}{687--697}.
\newblock


\bibitem[Fu et~al\mbox{.}(2024b)]%
        {Fu2024Adapter}
\bibfield{author}{\bibinfo{person}{Junchen Fu}, \bibinfo{person}{Fajie Yuan}, \bibinfo{person}{Yu Song}, \bibinfo{person}{Zheng Yuan}, \bibinfo{person}{Mingyue Cheng}, \bibinfo{person}{Shenghui Cheng}, \bibinfo{person}{Jiaqi Zhang}, \bibinfo{person}{Jie Wang}, {and} \bibinfo{person}{Yunzhu Pan}.} \bibinfo{year}{2024}\natexlab{b}.
\newblock \showarticletitle{Exploring Adapter-based Transfer Learning for Recommender Systems: Empirical Studies and Practical Insights}. In \bibinfo{booktitle}{\emph{Proceedings of the 17th ACM International Conference on Web Search and Data Mining}} \emph{(\bibinfo{series}{WSDM ’24})}. \bibinfo{publisher}{ACM}.
\newblock
\urldef\tempurl%
\url{https://doi.org/10.1145/3616855.3635805}
\showDOI{\tempurl}


\bibitem[Furuta et~al\mbox{.}(2023)]%
        {furuta2023multimodal}
\bibfield{author}{\bibinfo{person}{Hiroki Furuta}, \bibinfo{person}{Kuang-Huei Lee}, \bibinfo{person}{Ofir Nachum}, \bibinfo{person}{Yutaka Matsuo}, \bibinfo{person}{Aleksandra Faust}, \bibinfo{person}{Shixiang~Shane Gu}, {and} \bibinfo{person}{Izzeddin Gur}.} \bibinfo{year}{2023}\natexlab{}.
\newblock \showarticletitle{Multimodal web navigation with instruction-finetuned foundation models}.
\newblock \bibinfo{journal}{\emph{arXiv preprint arXiv:2305.11854}} (\bibinfo{year}{2023}).
\newblock


\bibitem[Furuta et~al\mbox{.}(2024)]%
        {furuta2024exposing}
\bibfield{author}{\bibinfo{person}{Hiroki Furuta}, \bibinfo{person}{Yutaka Matsuo}, \bibinfo{person}{Aleksandra Faust}, {and} \bibinfo{person}{Izzeddin Gur}.} \bibinfo{year}{2024}\natexlab{}.
\newblock \showarticletitle{Exposing Limitations of Language Model Agents in Sequential-Task Compositions on the Web}. In \bibinfo{booktitle}{\emph{ICLR 2024 Workshop on Large Language Model (LLM) Agents}}.
\newblock


\bibitem[Gao et~al\mbox{.}(2024)]%
        {gao2024large}
\bibfield{author}{\bibinfo{person}{Chen Gao}, \bibinfo{person}{Xiaochong Lan}, \bibinfo{person}{Nian Li}, \bibinfo{person}{Yuan Yuan}, \bibinfo{person}{Jingtao Ding}, \bibinfo{person}{Zhilun Zhou}, \bibinfo{person}{Fengli Xu}, {and} \bibinfo{person}{Yong Li}.} \bibinfo{year}{2024}\natexlab{}.
\newblock \showarticletitle{Large language models empowered agent-based modeling and simulation: A survey and perspectives}.
\newblock \bibinfo{journal}{\emph{Humanities and Social Sciences Communications}} \bibinfo{volume}{11}, \bibinfo{number}{1} (\bibinfo{year}{2024}), \bibinfo{pages}{1--24}.
\newblock


\bibitem[Gao et~al\mbox{.}(2023b)]%
        {gao2023survey}
\bibfield{author}{\bibinfo{person}{Chen Gao}, \bibinfo{person}{Yu Zheng}, \bibinfo{person}{Nian Li}, \bibinfo{person}{Yinfeng Li}, \bibinfo{person}{Yingrong Qin}, \bibinfo{person}{Jinghua Piao}, \bibinfo{person}{Yuhan Quan}, \bibinfo{person}{Jianxin Chang}, \bibinfo{person}{Depeng Jin}, \bibinfo{person}{Xiangnan He}, {et~al\mbox{.}}} \bibinfo{year}{2023}\natexlab{b}.
\newblock \showarticletitle{A survey of graph neural networks for recommender systems: Challenges, methods, and directions}.
\newblock \bibinfo{journal}{\emph{ACM Transactions on Recommender Systems}} \bibinfo{volume}{1}, \bibinfo{number}{1} (\bibinfo{year}{2023}), \bibinfo{pages}{1--51}.
\newblock


\bibitem[Gao et~al\mbox{.}(2023a)]%
        {gao2023retrieval}
\bibfield{author}{\bibinfo{person}{Yunfan Gao}, \bibinfo{person}{Yun Xiong}, \bibinfo{person}{Xinyu Gao}, \bibinfo{person}{Kangxiang Jia}, \bibinfo{person}{Jinliu Pan}, \bibinfo{person}{Yuxi Bi}, \bibinfo{person}{Yi Dai}, \bibinfo{person}{Jiawei Sun}, {and} \bibinfo{person}{Haofen Wang}.} \bibinfo{year}{2023}\natexlab{a}.
\newblock \showarticletitle{Retrieval-augmented generation for large language models: A survey}.
\newblock \bibinfo{journal}{\emph{arXiv preprint arXiv:2312.10997}} (\bibinfo{year}{2023}).
\newblock


\bibitem[Gong et~al\mbox{.}(2024)]%
        {gong2024cosearchagent}
\bibfield{author}{\bibinfo{person}{Peiyuan Gong}, \bibinfo{person}{Jiamian Li}, {and} \bibinfo{person}{Jiaxin Mao}.} \bibinfo{year}{2024}\natexlab{}.
\newblock \showarticletitle{CoSearchAgent: A Lightweight Collaborative Search Agent with Large Language Models}. In \bibinfo{booktitle}{\emph{Proceedings of the 47th International ACM SIGIR Conference on Research and Development in Information Retrieval}}. \bibinfo{pages}{2729--2733}.
\newblock


\bibitem[Graves and Graves(2012)]%
        {graves2012long}
\bibfield{author}{\bibinfo{person}{Alex Graves} {and} \bibinfo{person}{Alex Graves}.} \bibinfo{year}{2012}\natexlab{}.
\newblock \showarticletitle{Long short-term memory}.
\newblock \bibinfo{journal}{\emph{Supervised sequence labelling with recurrent neural networks}} (\bibinfo{year}{2012}), \bibinfo{pages}{37--45}.
\newblock


\bibitem[Guo et~al\mbox{.}(2024)]%
        {guo2024knowledge}
\bibfield{author}{\bibinfo{person}{Taicheng Guo}, \bibinfo{person}{Chaochun Liu}, \bibinfo{person}{Hai Wang}, \bibinfo{person}{Varun Mannam}, \bibinfo{person}{Fang Wang}, \bibinfo{person}{Xin Chen}, \bibinfo{person}{Xiangliang Zhang}, {and} \bibinfo{person}{Chandan~K Reddy}.} \bibinfo{year}{2024}\natexlab{}.
\newblock \showarticletitle{Knowledge Graph Enhanced Language Agents for Recommendation}.
\newblock \bibinfo{journal}{\emph{arXiv preprint arXiv:2410.19627}} (\bibinfo{year}{2024}).
\newblock


\bibitem[Gupta et~al\mbox{.}(2024)]%
        {gupta2024codenav}
\bibfield{author}{\bibinfo{person}{Tanmay Gupta}, \bibinfo{person}{Luca Weihs}, {and} \bibinfo{person}{Aniruddha Kembhavi}.} \bibinfo{year}{2024}\natexlab{}.
\newblock \showarticletitle{CodeNav: Beyond tool-use to using real-world codebases with LLM agents}.
\newblock \bibinfo{journal}{\emph{arXiv preprint arXiv:2406.12276}} (\bibinfo{year}{2024}).
\newblock


\bibitem[Gur et~al\mbox{.}(2023a)]%
        {gur2023real}
\bibfield{author}{\bibinfo{person}{Izzeddin Gur}, \bibinfo{person}{Hiroki Furuta}, \bibinfo{person}{Austin Huang}, \bibinfo{person}{Mustafa Safdari}, \bibinfo{person}{Yutaka Matsuo}, \bibinfo{person}{Douglas Eck}, {and} \bibinfo{person}{Aleksandra Faust}.} \bibinfo{year}{2023}\natexlab{a}.
\newblock \showarticletitle{A real-world webagent with planning, long context understanding, and program synthesis}.
\newblock \bibinfo{journal}{\emph{arXiv preprint arXiv:2307.12856}} (\bibinfo{year}{2023}).
\newblock


\bibitem[Gur et~al\mbox{.}(2023b)]%
        {gur2023HTMLT5}
\bibfield{author}{\bibinfo{person}{Izzeddin Gur}, \bibinfo{person}{Hiroki Furuta}, \bibinfo{person}{Austin Huang}, \bibinfo{person}{Mustafa Safdari}, \bibinfo{person}{Yutaka Matsuo}, \bibinfo{person}{Douglas Eck}, {and} \bibinfo{person}{Aleksandra Faust}.} \bibinfo{year}{2023}\natexlab{b}.
\newblock \showarticletitle{A real-world webagent with planning, long context understanding, and program synthesis}.
\newblock \bibinfo{journal}{\emph{arXiv preprint arXiv:2307.12856}} (\bibinfo{year}{2023}).
\newblock


\bibitem[He et~al\mbox{.}(2024)]%
        {he2024webvoyager}
\bibfield{author}{\bibinfo{person}{Hongliang He}, \bibinfo{person}{Wenlin Yao}, \bibinfo{person}{Kaixin Ma}, \bibinfo{person}{Wenhao Yu}, \bibinfo{person}{Yong Dai}, \bibinfo{person}{Hongming Zhang}, \bibinfo{person}{Zhenzhong Lan}, {and} \bibinfo{person}{Dong Yu}.} \bibinfo{year}{2024}\natexlab{}.
\newblock \showarticletitle{WebVoyager: Building an End-to-End Web Agent with Large Multimodal Models}.
\newblock \bibinfo{journal}{\emph{arXiv preprint arXiv:2401.13919}} (\bibinfo{year}{2024}).
\newblock


\bibitem[He et~al\mbox{.}(2025)]%
        {he2025pasa}
\bibfield{author}{\bibinfo{person}{Yichen He}, \bibinfo{person}{Guanhua Huang}, \bibinfo{person}{Peiyuan Feng}, \bibinfo{person}{Yuan Lin}, \bibinfo{person}{Yuchen Zhang}, \bibinfo{person}{Hang Li}, {et~al\mbox{.}}} \bibinfo{year}{2025}\natexlab{}.
\newblock \showarticletitle{PaSa: An LLM Agent for Comprehensive Academic Paper Search}.
\newblock \bibinfo{journal}{\emph{arXiv preprint arXiv:2501.10120}} (\bibinfo{year}{2025}).
\newblock


\bibitem[Hong et~al\mbox{.}(2024)]%
        {hong2023CogAgent}
\bibfield{author}{\bibinfo{person}{Wenyi Hong}, \bibinfo{person}{Weihan Wang}, \bibinfo{person}{Qingsong Lv}, \bibinfo{person}{Jiazheng Xu}, \bibinfo{person}{Wenmeng Yu}, \bibinfo{person}{Junhui Ji}, \bibinfo{person}{Yan Wang}, \bibinfo{person}{Zihan Wang}, \bibinfo{person}{Yuxiao Dong}, \bibinfo{person}{Ming Ding}, {et~al\mbox{.}}} \bibinfo{year}{2024}\natexlab{}.
\newblock \showarticletitle{Cogagent: A visual language model for gui agents}. In \bibinfo{booktitle}{\emph{Proceedings of the IEEE/CVF Conference on Computer Vision and Pattern Recognition}}. \bibinfo{pages}{14281--14290}.
\newblock


\bibitem[Hron et~al\mbox{.}(2021)]%
        {hron2021component}
\bibfield{author}{\bibinfo{person}{Jiri Hron}, \bibinfo{person}{Karl Krauth}, \bibinfo{person}{Michael Jordan}, {and} \bibinfo{person}{Niki Kilbertus}.} \bibinfo{year}{2021}\natexlab{}.
\newblock \showarticletitle{On component interactions in two-stage recommender systems}.
\newblock \bibinfo{journal}{\emph{Advances in neural information processing systems}}  \bibinfo{volume}{34} (\bibinfo{year}{2021}), \bibinfo{pages}{2744--2757}.
\newblock


\bibitem[Hu et~al\mbox{.}(2024)]%
        {hu2024survey}
\bibfield{author}{\bibinfo{person}{Sihao Hu}, \bibinfo{person}{Tiansheng Huang}, \bibinfo{person}{Fatih Ilhan}, \bibinfo{person}{Selim Tekin}, \bibinfo{person}{Gaowen Liu}, \bibinfo{person}{Ramana Kompella}, {and} \bibinfo{person}{Ling Liu}.} \bibinfo{year}{2024}\natexlab{}.
\newblock \showarticletitle{A survey on large language model-based game agents}.
\newblock \bibinfo{journal}{\emph{arXiv preprint arXiv:2404.02039}} (\bibinfo{year}{2024}).
\newblock


\bibitem[Huang et~al\mbox{.}(2023b)]%
        {huang2024LEO}
\bibfield{author}{\bibinfo{person}{Jiangyong Huang}, \bibinfo{person}{Silong Yong}, \bibinfo{person}{Xiaojian Ma}, \bibinfo{person}{Xiongkun Linghu}, \bibinfo{person}{Puhao Li}, \bibinfo{person}{Yan Wang}, \bibinfo{person}{Qing Li}, \bibinfo{person}{Song-Chun Zhu}, \bibinfo{person}{Baoxiong Jia}, {and} \bibinfo{person}{Siyuan Huang}.} \bibinfo{year}{2023}\natexlab{b}.
\newblock \showarticletitle{An embodied generalist agent in 3d world}.
\newblock \bibinfo{journal}{\emph{arXiv preprint arXiv:2311.12871}} (\bibinfo{year}{2023}).
\newblock


\bibitem[Huang et~al\mbox{.}(2023c)]%
        {huang2023survey}
\bibfield{author}{\bibinfo{person}{Lei Huang}, \bibinfo{person}{Weijiang Yu}, \bibinfo{person}{Weitao Ma}, \bibinfo{person}{Weihong Zhong}, \bibinfo{person}{Zhangyin Feng}, \bibinfo{person}{Haotian Wang}, \bibinfo{person}{Qianglong Chen}, \bibinfo{person}{Weihua Peng}, \bibinfo{person}{Xiaocheng Feng}, \bibinfo{person}{Bing Qin}, {et~al\mbox{.}}} \bibinfo{year}{2023}\natexlab{c}.
\newblock \showarticletitle{A survey on hallucination in large language models: Principles, taxonomy, challenges, and open questions}.
\newblock \bibinfo{journal}{\emph{arXiv preprint arXiv:2311.05232}} (\bibinfo{year}{2023}).
\newblock


\bibitem[Huang et~al\mbox{.}(2024)]%
        {huang2024queryagent}
\bibfield{author}{\bibinfo{person}{Xiang Huang}, \bibinfo{person}{Sitao Cheng}, \bibinfo{person}{Shanshan Huang}, \bibinfo{person}{Jiayu Shen}, \bibinfo{person}{Yong Xu}, \bibinfo{person}{Chaoyun Zhang}, {and} \bibinfo{person}{Yuzhong Qu}.} \bibinfo{year}{2024}\natexlab{}.
\newblock \showarticletitle{QueryAgent: A Reliable and Efficient Reasoning Framework with Environmental Feedback based Self-Correction}.
\newblock \bibinfo{journal}{\emph{arXiv preprint arXiv:2403.11886}} (\bibinfo{year}{2024}).
\newblock


\bibitem[Huang et~al\mbox{.}(2023a)]%
        {huang2023recommender}
\bibfield{author}{\bibinfo{person}{Xu Huang}, \bibinfo{person}{Jianxun Lian}, \bibinfo{person}{Yuxuan Lei}, \bibinfo{person}{Jing Yao}, \bibinfo{person}{Defu Lian}, {and} \bibinfo{person}{Xing Xie}.} \bibinfo{year}{2023}\natexlab{a}.
\newblock \showarticletitle{Recommender ai agent: Integrating large language models for interactive recommendations}.
\newblock \bibinfo{journal}{\emph{arXiv preprint arXiv:2308.16505}} (\bibinfo{year}{2023}).
\newblock


\bibitem[Jain et~al\mbox{.}(2024)]%
        {jain2024llm}
\bibfield{author}{\bibinfo{person}{Sarthak Jain}, \bibinfo{person}{Aditya Dora}, \bibinfo{person}{Ka~Seng Sam}, {and} \bibinfo{person}{Prabhat Singh}.} \bibinfo{year}{2024}\natexlab{}.
\newblock \showarticletitle{Llm agents improve semantic code search}.
\newblock \bibinfo{journal}{\emph{arXiv preprint arXiv:2408.11058}} (\bibinfo{year}{2024}).
\newblock


\bibitem[Jannach et~al\mbox{.}(2021)]%
        {jannach2021survey}
\bibfield{author}{\bibinfo{person}{Dietmar Jannach}, \bibinfo{person}{Ahtsham Manzoor}, \bibinfo{person}{Wanling Cai}, {and} \bibinfo{person}{Li Chen}.} \bibinfo{year}{2021}\natexlab{}.
\newblock \showarticletitle{A survey on conversational recommender systems}.
\newblock \bibinfo{journal}{\emph{ACM Computing Surveys (CSUR)}} \bibinfo{volume}{54}, \bibinfo{number}{5} (\bibinfo{year}{2021}), \bibinfo{pages}{1--36}.
\newblock


\bibitem[Jin et~al\mbox{.}(2024a)]%
        {jin2024llm}
\bibfield{author}{\bibinfo{person}{Hongye Jin}, \bibinfo{person}{Xiaotian Han}, \bibinfo{person}{Jingfeng Yang}, \bibinfo{person}{Zhimeng Jiang}, \bibinfo{person}{Zirui Liu}, \bibinfo{person}{Chia-Yuan Chang}, \bibinfo{person}{Huiyuan Chen}, {and} \bibinfo{person}{Xia Hu}.} \bibinfo{year}{2024}\natexlab{a}.
\newblock \showarticletitle{Llm maybe longlm: Self-extend llm context window without tuning}.
\newblock \bibinfo{journal}{\emph{arXiv preprint arXiv:2401.01325}} (\bibinfo{year}{2024}).
\newblock


\bibitem[Jin et~al\mbox{.}(2023)]%
        {jin2023lending}
\bibfield{author}{\bibinfo{person}{Jiarui Jin}, \bibinfo{person}{Xianyu Chen}, \bibinfo{person}{Fanghua Ye}, \bibinfo{person}{Mengyue Yang}, \bibinfo{person}{Yue Feng}, \bibinfo{person}{Weinan Zhang}, \bibinfo{person}{Yong Yu}, {and} \bibinfo{person}{Jun Wang}.} \bibinfo{year}{2023}\natexlab{}.
\newblock \showarticletitle{Lending interaction wings to recommender systems with conversational agents}.
\newblock \bibinfo{journal}{\emph{Advances in Neural Information Processing Systems}}  \bibinfo{volume}{36} (\bibinfo{year}{2023}), \bibinfo{pages}{27951--27979}.
\newblock


\bibitem[Jin et~al\mbox{.}(2024b)]%
        {jin2024agentmd}
\bibfield{author}{\bibinfo{person}{Qiao Jin}, \bibinfo{person}{Zhizheng Wang}, \bibinfo{person}{Yifan Yang}, \bibinfo{person}{Qingqing Zhu}, \bibinfo{person}{Donald Wright}, \bibinfo{person}{Thomas Huang}, \bibinfo{person}{W~John Wilbur}, \bibinfo{person}{Zhe He}, \bibinfo{person}{Andrew Taylor}, \bibinfo{person}{Qingyu Chen}, {et~al\mbox{.}}} \bibinfo{year}{2024}\natexlab{b}.
\newblock \showarticletitle{AgentMD: Empowering Language Agents for Risk Prediction with Large-Scale Clinical Tool Learning}.
\newblock \bibinfo{journal}{\emph{arXiv preprint arXiv:2402.13225}} (\bibinfo{year}{2024}).
\newblock


\bibitem[Joko et~al\mbox{.}(2024)]%
        {joko2024doing}
\bibfield{author}{\bibinfo{person}{Hideaki Joko}, \bibinfo{person}{Shubham Chatterjee}, \bibinfo{person}{Andrew Ramsay}, \bibinfo{person}{Arjen~P de Vries}, \bibinfo{person}{Jeff Dalton}, {and} \bibinfo{person}{Faegheh Hasibi}.} \bibinfo{year}{2024}\natexlab{}.
\newblock \showarticletitle{Doing Personal LAPS: LLM-Augmented Dialogue Construction for Personalized Multi-Session Conversational Search}. In \bibinfo{booktitle}{\emph{Proceedings of the 47th International ACM SIGIR Conference on Research and Development in Information Retrieval}}. \bibinfo{pages}{796--806}.
\newblock


\bibitem[Kim et~al\mbox{.}(2024)]%
        {kim2024RCIPrompt}
\bibfield{author}{\bibinfo{person}{Geunwoo Kim}, \bibinfo{person}{Pierre Baldi}, {and} \bibinfo{person}{Stephen McAleer}.} \bibinfo{year}{2024}\natexlab{}.
\newblock \showarticletitle{Language models can solve computer tasks}.
\newblock \bibinfo{journal}{\emph{Advances in Neural Information Processing Systems}}  \bibinfo{volume}{36} (\bibinfo{year}{2024}).
\newblock


\bibitem[Koh et~al\mbox{.}(2024a)]%
        {koh2024visualwebarena}
\bibfield{author}{\bibinfo{person}{Jing~Yu Koh}, \bibinfo{person}{Robert Lo}, \bibinfo{person}{Lawrence Jang}, \bibinfo{person}{Vikram Duvvur}, \bibinfo{person}{Ming~Chong Lim}, \bibinfo{person}{Po-Yu Huang}, \bibinfo{person}{Graham Neubig}, \bibinfo{person}{Shuyan Zhou}, \bibinfo{person}{Ruslan Salakhutdinov}, {and} \bibinfo{person}{Daniel Fried}.} \bibinfo{year}{2024}\natexlab{a}.
\newblock \showarticletitle{Visualwebarena: Evaluating multimodal agents on realistic visual web tasks}.
\newblock \bibinfo{journal}{\emph{arXiv preprint arXiv:2401.13649}} (\bibinfo{year}{2024}).
\newblock


\bibitem[Koh et~al\mbox{.}(2024b)]%
        {koh2024tree}
\bibfield{author}{\bibinfo{person}{Jing~Yu Koh}, \bibinfo{person}{Stephen McAleer}, \bibinfo{person}{Daniel Fried}, {and} \bibinfo{person}{Ruslan Salakhutdinov}.} \bibinfo{year}{2024}\natexlab{b}.
\newblock \showarticletitle{Tree Search for Language Model Agents}.
\newblock \bibinfo{journal}{\emph{arXiv preprint arXiv:2407.01476}} (\bibinfo{year}{2024}).
\newblock


\bibitem[Lagakis and Demetriadis(2024)]%
        {lagakis2024evaai}
\bibfield{author}{\bibinfo{person}{Paraskevas Lagakis} {and} \bibinfo{person}{Stavros Demetriadis}.} \bibinfo{year}{2024}\natexlab{}.
\newblock \showarticletitle{EvaAI: A Multi-agent Framework Leveraging Large Language Models for Enhanced Automated Grading}. In \bibinfo{booktitle}{\emph{International Conference on Intelligent Tutoring Systems}}. Springer, \bibinfo{pages}{378--385}.
\newblock


\bibitem[Levene(2011)]%
        {levene2011introduction}
\bibfield{author}{\bibinfo{person}{Mark Levene}.} \bibinfo{year}{2011}\natexlab{}.
\newblock \bibinfo{booktitle}{\emph{An introduction to search engines and web navigation}}.
\newblock \bibinfo{publisher}{John Wiley \& Sons}.
\newblock


\bibitem[Li et~al\mbox{.}(2023b)]%
        {Li2023OneforAll}
\bibfield{author}{\bibinfo{person}{Chenglin Li}, \bibinfo{person}{Yuanzhen Xie}, \bibinfo{person}{Chenyun Yu}, \bibinfo{person}{Bo Hu}, \bibinfo{person}{Zang Li}, \bibinfo{person}{Guoqiang Shu}, \bibinfo{person}{Xiaohu Qie}, {and} \bibinfo{person}{Di Niu}.} \bibinfo{year}{2023}\natexlab{b}.
\newblock \showarticletitle{One for All, All for One: Learning and Transferring User Embeddings for Cross-Domain Recommendation}. In \bibinfo{booktitle}{\emph{Proceedings of the Sixteenth ACM International Conference on Web Search and Data Mining}} \emph{(\bibinfo{series}{WSDM ’23})}. \bibinfo{publisher}{ACM}.
\newblock
\urldef\tempurl%
\url{https://doi.org/10.1145/3539597.3570379}
\showDOI{\tempurl}


\bibitem[Li et~al\mbox{.}(2024b)]%
        {li2024legalagentbench}
\bibfield{author}{\bibinfo{person}{Haitao Li}, \bibinfo{person}{Junjie Chen}, \bibinfo{person}{Jingli Yang}, \bibinfo{person}{Qingyao Ai}, \bibinfo{person}{Wei Jia}, \bibinfo{person}{Youfeng Liu}, \bibinfo{person}{Kai Lin}, \bibinfo{person}{Yueyue Wu}, \bibinfo{person}{Guozhi Yuan}, \bibinfo{person}{Yiran Hu}, {et~al\mbox{.}}} \bibinfo{year}{2024}\natexlab{b}.
\newblock \showarticletitle{LegalAgentBench: Evaluating LLM Agents in Legal Domain}.
\newblock \bibinfo{journal}{\emph{arXiv preprint arXiv:2412.17259}} (\bibinfo{year}{2024}).
\newblock


\bibitem[Li et~al\mbox{.}(2023c)]%
        {li2023beginner}
\bibfield{author}{\bibinfo{person}{Qiang Li}, \bibinfo{person}{Xiaoyan Yang}, \bibinfo{person}{Haowen Wang}, \bibinfo{person}{Qin Wang}, \bibinfo{person}{Lei Liu}, \bibinfo{person}{Junjie Wang}, \bibinfo{person}{Yang Zhang}, \bibinfo{person}{Mingyuan Chu}, \bibinfo{person}{Sen Hu}, \bibinfo{person}{Yicheng Chen}, {et~al\mbox{.}}} \bibinfo{year}{2023}\natexlab{c}.
\newblock \showarticletitle{From beginner to expert: Modeling medical knowledge into general llms}.
\newblock \bibinfo{journal}{\emph{arXiv preprint arXiv:2312.01040}} (\bibinfo{year}{2023}).
\newblock


\bibitem[Li et~al\mbox{.}(2023a)]%
        {li2023exploring}
\bibfield{author}{\bibinfo{person}{Ruyu Li}, \bibinfo{person}{Wenhao Deng}, \bibinfo{person}{Yu Cheng}, \bibinfo{person}{Zheng Yuan}, \bibinfo{person}{Jiaqi Zhang}, {and} \bibinfo{person}{Fajie Yuan}.} \bibinfo{year}{2023}\natexlab{a}.
\newblock \showarticletitle{Exploring the upper limits of text-based collaborative filtering using large language models: Discoveries and insights}.
\newblock \bibinfo{journal}{\emph{arXiv preprint arXiv:2305.11700}} (\bibinfo{year}{2023}).
\newblock


\bibitem[Li et~al\mbox{.}(2023d)]%
        {li2023STAN}
\bibfield{author}{\bibinfo{person}{Wanda Li}, \bibinfo{person}{Wenhao Zheng}, \bibinfo{person}{Xuanji Xiao}, {and} \bibinfo{person}{Suhang Wang}.} \bibinfo{year}{2023}\natexlab{d}.
\newblock \showarticletitle{Stan: stage-adaptive network for multi-task recommendation by learning user lifecycle-based representation}. In \bibinfo{booktitle}{\emph{Proceedings of the 17th ACM Conference on Recommender Systems}}. \bibinfo{pages}{602--612}.
\newblock


\bibitem[Li et~al\mbox{.}(2024a)]%
        {li2024chatcite}
\bibfield{author}{\bibinfo{person}{Yutong Li}, \bibinfo{person}{Lu Chen}, \bibinfo{person}{Aiwei Liu}, \bibinfo{person}{Kai Yu}, {and} \bibinfo{person}{Lijie Wen}.} \bibinfo{year}{2024}\natexlab{a}.
\newblock \showarticletitle{ChatCite: LLM agent with human workflow guidance for comparative literature summary}.
\newblock \bibinfo{journal}{\emph{arXiv preprint arXiv:2403.02574}} (\bibinfo{year}{2024}).
\newblock


\bibitem[Li et~al\mbox{.}(2024c)]%
        {li2024personal}
\bibfield{author}{\bibinfo{person}{Yuanchun Li}, \bibinfo{person}{Hao Wen}, \bibinfo{person}{Weijun Wang}, \bibinfo{person}{Xiangyu Li}, \bibinfo{person}{Yizhen Yuan}, \bibinfo{person}{Guohong Liu}, \bibinfo{person}{Jiacheng Liu}, \bibinfo{person}{Wenxing Xu}, \bibinfo{person}{Xiang Wang}, \bibinfo{person}{Yi Sun}, {et~al\mbox{.}}} \bibinfo{year}{2024}\natexlab{c}.
\newblock \showarticletitle{Personal llm agents: Insights and survey about the capability, efficiency and security}.
\newblock \bibinfo{journal}{\emph{arXiv preprint arXiv:2401.05459}} (\bibinfo{year}{2024}).
\newblock


\bibitem[Lian et~al\mbox{.}(2024)]%
        {lian2024recai}
\bibfield{author}{\bibinfo{person}{Jianxun Lian}, \bibinfo{person}{Yuxuan Lei}, \bibinfo{person}{Xu Huang}, \bibinfo{person}{Jing Yao}, \bibinfo{person}{Wei Xu}, {and} \bibinfo{person}{Xing Xie}.} \bibinfo{year}{2024}\natexlab{}.
\newblock \showarticletitle{RecAI: Leveraging Large Language Models for Next-Generation Recommender Systems}. In \bibinfo{booktitle}{\emph{Companion Proceedings of the ACM on Web Conference 2024}}. \bibinfo{pages}{1031--1034}.
\newblock


\bibitem[Liang et~al\mbox{.}(2023)]%
        {liang2023encouraging}
\bibfield{author}{\bibinfo{person}{Tian Liang}, \bibinfo{person}{Zhiwei He}, \bibinfo{person}{Wenxiang Jiao}, \bibinfo{person}{Xing Wang}, \bibinfo{person}{Yan Wang}, \bibinfo{person}{Rui Wang}, \bibinfo{person}{Yujiu Yang}, \bibinfo{person}{Zhaopeng Tu}, {and} \bibinfo{person}{Shuming Shi}.} \bibinfo{year}{2023}\natexlab{}.
\newblock \showarticletitle{Encouraging Divergent Thinking in Large Language Models through Multi-Agent Debate}.
\newblock \bibinfo{journal}{\emph{arXiv preprint arXiv:2305.19118}} (\bibinfo{year}{2023}).
\newblock


\bibitem[Lin et~al\mbox{.}(2023)]%
        {Lin2023ReLLaRL}
\bibfield{author}{\bibinfo{person}{Jianghao Lin}, \bibinfo{person}{Rongjie Shan}, \bibinfo{person}{Chenxu Zhu}, \bibinfo{person}{Kounianhua Du}, \bibinfo{person}{Bo Chen}, \bibinfo{person}{Shigang Quan}, \bibinfo{person}{Ruiming Tang}, \bibinfo{person}{Yong Yu}, {and} \bibinfo{person}{Weinan Zhang}.} \bibinfo{year}{2023}\natexlab{}.
\newblock \showarticletitle{ReLLa: Retrieval-enhanced Large Language Models for Lifelong Sequential Behavior Comprehension in Recommendation}.
\newblock \bibinfo{journal}{\emph{Proceedings of the ACM on Web Conference 2024}} (\bibinfo{year}{2023}).
\newblock
\urldef\tempurl%
\url{https://api.semanticscholar.org/CorpusID:261065228}
\showURL{%
\tempurl}


\bibitem[Liu et~al\mbox{.}(2024)]%
        {liu2024multimodal}
\bibfield{author}{\bibinfo{person}{Qidong Liu}, \bibinfo{person}{Jiaxi Hu}, \bibinfo{person}{Yutian Xiao}, \bibinfo{person}{Xiangyu Zhao}, \bibinfo{person}{Jingtong Gao}, \bibinfo{person}{Wanyu Wang}, \bibinfo{person}{Qing Li}, {and} \bibinfo{person}{Jiliang Tang}.} \bibinfo{year}{2024}\natexlab{}.
\newblock \showarticletitle{Multimodal recommender systems: A survey}.
\newblock \bibinfo{journal}{\emph{Comput. Surveys}} \bibinfo{volume}{57}, \bibinfo{number}{2} (\bibinfo{year}{2024}), \bibinfo{pages}{1--17}.
\newblock


\bibitem[Liu et~al\mbox{.}(2023)]%
        {liu2023agentbench}
\bibfield{author}{\bibinfo{person}{Xiao Liu}, \bibinfo{person}{Hao Yu}, \bibinfo{person}{Hanchen Zhang}, \bibinfo{person}{Yifan Xu}, \bibinfo{person}{Xuanyu Lei}, \bibinfo{person}{Hanyu Lai}, \bibinfo{person}{Yu Gu}, \bibinfo{person}{Hangliang Ding}, \bibinfo{person}{Kaiwen Men}, \bibinfo{person}{Kejuan Yang}, {et~al\mbox{.}}} \bibinfo{year}{2023}\natexlab{}.
\newblock \showarticletitle{Agentbench: Evaluating llms as agents}.
\newblock \bibinfo{journal}{\emph{arXiv preprint arXiv:2308.03688}} (\bibinfo{year}{2023}).
\newblock


\bibitem[Lops et~al\mbox{.}(2011)]%
        {lops2011content}
\bibfield{author}{\bibinfo{person}{Pasquale Lops}, \bibinfo{person}{Marco De~Gemmis}, {and} \bibinfo{person}{Giovanni Semeraro}.} \bibinfo{year}{2011}\natexlab{}.
\newblock \showarticletitle{Content-based recommender systems: State of the art and trends}.
\newblock \bibinfo{journal}{\emph{Recommender systems handbook}} (\bibinfo{year}{2011}), \bibinfo{pages}{73--105}.
\newblock


\bibitem[Ma et~al\mbox{.}(2023b)]%
        {ma2023laser}
\bibfield{author}{\bibinfo{person}{Kaixin Ma}, \bibinfo{person}{Hongming Zhang}, \bibinfo{person}{Hongwei Wang}, \bibinfo{person}{Xiaoman Pan}, {and} \bibinfo{person}{Dong Yu}.} \bibinfo{year}{2023}\natexlab{b}.
\newblock \showarticletitle{Laser: Llm agent with state-space exploration for web navigation}.
\newblock \bibinfo{journal}{\emph{arXiv preprint arXiv:2309.08172}} (\bibinfo{year}{2023}).
\newblock


\bibitem[Ma et~al\mbox{.}(2023a)]%
        {ma2023large}
\bibfield{author}{\bibinfo{person}{Tianhui Ma}, \bibinfo{person}{Yuan Cheng}, \bibinfo{person}{Hengshu Zhu}, {and} \bibinfo{person}{Hui Xiong}.} \bibinfo{year}{2023}\natexlab{a}.
\newblock \showarticletitle{Large language models are not stable recommender systems}.
\newblock \bibinfo{journal}{\emph{arXiv preprint arXiv:2312.15746}} (\bibinfo{year}{2023}).
\newblock


\bibitem[Ma et~al\mbox{.}(2024)]%
        {ma2024coco_agent}
\bibfield{author}{\bibinfo{person}{Xinbei Ma}, \bibinfo{person}{Zhuosheng Zhang}, {and} \bibinfo{person}{Hai Zhao}.} \bibinfo{year}{2024}\natexlab{}.
\newblock \showarticletitle{CoCo-Agent: A Comprehensive Cognitive MLLM Agent for Smartphone GUI Automation}.
\newblock \bibinfo{journal}{\emph{arXiv preprint arXiv:2402.11941v3}} (\bibinfo{year}{2024}).
\newblock


\bibitem[Minsky(1988)]%
        {minsky1988society}
\bibfield{author}{\bibinfo{person}{Marvin Minsky}.} \bibinfo{year}{1988}\natexlab{}.
\newblock \bibinfo{booktitle}{\emph{Society of mind}}.
\newblock \bibinfo{publisher}{Simon and Schuster}.
\newblock


\bibitem[Mo et~al\mbox{.}(2024)]%
        {mo2024leverage}
\bibfield{author}{\bibinfo{person}{Fengran Mo}, \bibinfo{person}{Longxiang Zhao}, \bibinfo{person}{Kaiyu Huang}, \bibinfo{person}{Yue Dong}, \bibinfo{person}{Degen Huang}, {and} \bibinfo{person}{Jian-Yun Nie}.} \bibinfo{year}{2024}\natexlab{}.
\newblock \showarticletitle{How to leverage personal textual knowledge for personalized conversational information retrieval}. In \bibinfo{booktitle}{\emph{Proceedings of the 33rd ACM International Conference on Information and Knowledge Management}}. \bibinfo{pages}{3954--3958}.
\newblock


\bibitem[Nie et~al\mbox{.}(2024)]%
        {nie2024hybrid}
\bibfield{author}{\bibinfo{person}{Guangtao Nie}, \bibinfo{person}{Rong Zhi}, \bibinfo{person}{Xiaofan Yan}, \bibinfo{person}{Yufan Du}, \bibinfo{person}{Xiangyang Zhang}, \bibinfo{person}{Jianwei Chen}, \bibinfo{person}{Mi Zhou}, \bibinfo{person}{Hongshen Chen}, \bibinfo{person}{Tianhao Li}, \bibinfo{person}{Ziguang Cheng}, {et~al\mbox{.}}} \bibinfo{year}{2024}\natexlab{}.
\newblock \showarticletitle{A hybrid multi-agent conversational recommender system with llm and search engine in e-commerce}. In \bibinfo{booktitle}{\emph{Proceedings of the 18th ACM Conference on Recommender Systems}}. \bibinfo{pages}{745--747}.
\newblock


\bibitem[Ning et~al\mbox{.}(2024)]%
        {ning2024cheatagent}
\bibfield{author}{\bibinfo{person}{Liang-bo Ning}, \bibinfo{person}{Shijie Wang}, \bibinfo{person}{Wenqi Fan}, \bibinfo{person}{Qing Li}, \bibinfo{person}{Xin Xu}, \bibinfo{person}{Hao Chen}, {and} \bibinfo{person}{Feiran Huang}.} \bibinfo{year}{2024}\natexlab{}.
\newblock \showarticletitle{Cheatagent: Attacking llm-empowered recommender systems via llm agent}. In \bibinfo{booktitle}{\emph{Proceedings of the 30th ACM SIGKDD Conference on Knowledge Discovery and Data Mining}}. \bibinfo{pages}{2284--2295}.
\newblock


\bibitem[Nong et~al\mbox{.}(2024)]%
        {nong2024mobileflow}
\bibfield{author}{\bibinfo{person}{Songqin Nong}, \bibinfo{person}{Jiali Zhu}, \bibinfo{person}{Rui Wu}, \bibinfo{person}{Jiongchao Jin}, \bibinfo{person}{Shuo Shan}, \bibinfo{person}{Xiutian Huang}, {and} \bibinfo{person}{Wenhao Xu}.} \bibinfo{year}{2024}\natexlab{}.
\newblock \showarticletitle{Mobileflow: A multimodal llm for mobile gui agent}.
\newblock \bibinfo{journal}{\emph{arXiv preprint arXiv:2407.04346}} (\bibinfo{year}{2024}).
\newblock


\bibitem[O'Brien and Lewis(2023)]%
        {o2023contrastive}
\bibfield{author}{\bibinfo{person}{Sean O'Brien} {and} \bibinfo{person}{Mike Lewis}.} \bibinfo{year}{2023}\natexlab{}.
\newblock \showarticletitle{Contrastive decoding improves reasoning in large language models}.
\newblock \bibinfo{journal}{\emph{arXiv preprint arXiv:2309.09117}} (\bibinfo{year}{2023}).
\newblock


\bibitem[OpenAI(2022)]%
        {chatgpt}
\bibfield{author}{\bibinfo{person}{OpenAI}.} \bibinfo{year}{2022}\natexlab{}.
\newblock \bibinfo{title}{Introducing ChatGPT}.
\newblock \bibinfo{howpublished}{\url{https://openai.com/blog/chatgpt}}.
\newblock
\newblock
\shownote{(Accessed on 01/12/2023)}.


\bibitem[Park et~al\mbox{.}(2023)]%
        {park2023generative}
\bibfield{author}{\bibinfo{person}{Joon~Sung Park}, \bibinfo{person}{Joseph O'Brien}, \bibinfo{person}{Carrie~Jun Cai}, \bibinfo{person}{Meredith~Ringel Morris}, \bibinfo{person}{Percy Liang}, {and} \bibinfo{person}{Michael~S Bernstein}.} \bibinfo{year}{2023}\natexlab{}.
\newblock \showarticletitle{Generative agents: Interactive simulacra of human behavior}. In \bibinfo{booktitle}{\emph{Proceedings of the 36th Annual ACM Symposium on User Interface Software and Technology}}. \bibinfo{pages}{1--22}.
\newblock


\bibitem[Park et~al\mbox{.}(2024)]%
        {park2024generative}
\bibfield{author}{\bibinfo{person}{Joon~Sung Park}, \bibinfo{person}{Carolyn~Q Zou}, \bibinfo{person}{Aaron Shaw}, \bibinfo{person}{Benjamin~Mako Hill}, \bibinfo{person}{Carrie Cai}, \bibinfo{person}{Meredith~Ringel Morris}, \bibinfo{person}{Robb Willer}, \bibinfo{person}{Percy Liang}, {and} \bibinfo{person}{Michael~S Bernstein}.} \bibinfo{year}{2024}\natexlab{}.
\newblock \showarticletitle{Generative agent simulations of 1,000 people}.
\newblock \bibinfo{journal}{\emph{arXiv preprint arXiv:2411.10109}} (\bibinfo{year}{2024}).
\newblock


\bibitem[Pei et~al\mbox{.}(2019)]%
        {pei2019personalized}
\bibfield{author}{\bibinfo{person}{Changhua Pei}, \bibinfo{person}{Yi Zhang}, \bibinfo{person}{Yongfeng Zhang}, \bibinfo{person}{Fei Sun}, \bibinfo{person}{Xiao Lin}, \bibinfo{person}{Hanxiao Sun}, \bibinfo{person}{Jian Wu}, \bibinfo{person}{Peng Jiang}, \bibinfo{person}{Junfeng Ge}, \bibinfo{person}{Wenwu Ou}, {et~al\mbox{.}}} \bibinfo{year}{2019}\natexlab{}.
\newblock \showarticletitle{Personalized re-ranking for recommendation}. In \bibinfo{booktitle}{\emph{Proceedings of the 13th ACM conference on recommender systems}}. \bibinfo{pages}{3--11}.
\newblock


\bibitem[Peng et~al\mbox{.}(2025)]%
        {peng2025surveyllmpoweredagentsrecommender}
\bibfield{author}{\bibinfo{person}{Qiyao Peng}, \bibinfo{person}{Hongtao Liu}, \bibinfo{person}{Hua Huang}, \bibinfo{person}{Qing Yang}, {and} \bibinfo{person}{Minglai Shao}.} \bibinfo{year}{2025}\natexlab{}.
\newblock \bibinfo{title}{A Survey on LLM-powered Agents for Recommender Systems}.
\newblock
\newblock
\showeprint[arxiv]{2502.10050}~[cs.IR]
\urldef\tempurl%
\url{https://arxiv.org/abs/2502.10050}
\showURL{%
\tempurl}


\bibitem[Perez et~al\mbox{.}(2022)]%
        {perez2022discovering}
\bibfield{author}{\bibinfo{person}{Ethan Perez}, \bibinfo{person}{Sam Ringer}, \bibinfo{person}{Kamil{\.e} Luko{\v{s}}i{\=u}t{\.e}}, \bibinfo{person}{Karina Nguyen}, \bibinfo{person}{Edwin Chen}, \bibinfo{person}{Scott Heiner}, \bibinfo{person}{Craig Pettit}, \bibinfo{person}{Catherine Olsson}, \bibinfo{person}{Sandipan Kundu}, \bibinfo{person}{Saurav Kadavath}, {et~al\mbox{.}}} \bibinfo{year}{2022}\natexlab{}.
\newblock \showarticletitle{Discovering language model behaviors with model-written evaluations}.
\newblock \bibinfo{journal}{\emph{arXiv preprint arXiv:2212.09251}} (\bibinfo{year}{2022}).
\newblock


\bibitem[Putta et~al\mbox{.}(2024)]%
        {putta2024agent}
\bibfield{author}{\bibinfo{person}{Pranav Putta}, \bibinfo{person}{Edmund Mills}, \bibinfo{person}{Naman Garg}, \bibinfo{person}{Sumeet Motwani}, \bibinfo{person}{Chelsea Finn}, \bibinfo{person}{Divyansh Garg}, {and} \bibinfo{person}{Rafael Rafailov}.} \bibinfo{year}{2024}\natexlab{}.
\newblock \showarticletitle{Agent q: Advanced reasoning and learning for autonomous ai agents}.
\newblock \bibinfo{journal}{\emph{arXiv preprint arXiv:2408.07199}} (\bibinfo{year}{2024}).
\newblock


\bibitem[Qian et~al\mbox{.}(2023)]%
        {qian2023communicative}
\bibfield{author}{\bibinfo{person}{Chen Qian}, \bibinfo{person}{Xin Cong}, \bibinfo{person}{Cheng Yang}, \bibinfo{person}{Weize Chen}, \bibinfo{person}{Yusheng Su}, \bibinfo{person}{Juyuan Xu}, \bibinfo{person}{Zhiyuan Liu}, {and} \bibinfo{person}{Maosong Sun}.} \bibinfo{year}{2023}\natexlab{}.
\newblock \showarticletitle{Communicative agents for software development}.
\newblock \bibinfo{journal}{\emph{arXiv preprint arXiv:2307.07924}} (\bibinfo{year}{2023}).
\newblock


\bibitem[Qin et~al\mbox{.}(2023)]%
        {qin2023toolllm}
\bibfield{author}{\bibinfo{person}{Yujia Qin}, \bibinfo{person}{Shihao Liang}, \bibinfo{person}{Yining Ye}, \bibinfo{person}{Kunlun Zhu}, \bibinfo{person}{Lan Yan}, \bibinfo{person}{Yaxi Lu}, \bibinfo{person}{Yankai Lin}, \bibinfo{person}{Xin Cong}, \bibinfo{person}{Xiangru Tang}, \bibinfo{person}{Bill Qian}, {et~al\mbox{.}}} \bibinfo{year}{2023}\natexlab{}.
\newblock \showarticletitle{Toolllm: Facilitating large language models to master 16000+ real-world apis}.
\newblock \bibinfo{journal}{\emph{arXiv preprint arXiv:2307.16789}} (\bibinfo{year}{2023}).
\newblock


\bibitem[Ram et~al\mbox{.}(2023)]%
        {ram2023context}
\bibfield{author}{\bibinfo{person}{Ori Ram}, \bibinfo{person}{Yoav Levine}, \bibinfo{person}{Itay Dalmedigos}, \bibinfo{person}{Dor Muhlgay}, \bibinfo{person}{Amnon Shashua}, \bibinfo{person}{Kevin Leyton-Brown}, {and} \bibinfo{person}{Yoav Shoham}.} \bibinfo{year}{2023}\natexlab{}.
\newblock \showarticletitle{In-context retrieval-augmented language models}.
\newblock \bibinfo{journal}{\emph{Transactions of the Association for Computational Linguistics}}  \bibinfo{volume}{11} (\bibinfo{year}{2023}), \bibinfo{pages}{1316--1331}.
\newblock


\bibitem[Rawassizadeh and Rong(2023)]%
        {rawassizadeh2023odsearch}
\bibfield{author}{\bibinfo{person}{Reza Rawassizadeh} {and} \bibinfo{person}{Yi Rong}.} \bibinfo{year}{2023}\natexlab{}.
\newblock \showarticletitle{ODSearch: Fast and Resource Efficient On-device Natural Language Search for Fitness Trackers' Data}.
\newblock \bibinfo{journal}{\emph{Proceedings of the ACM on Interactive, Mobile, Wearable and Ubiquitous Technologies}} \bibinfo{volume}{6}, \bibinfo{number}{4} (\bibinfo{year}{2023}), \bibinfo{pages}{1--25}.
\newblock


\bibitem[Rawles et~al\mbox{.}(2024a)]%
        {rawles2024AndroidWorld}
\bibfield{author}{\bibinfo{person}{Christopher Rawles}, \bibinfo{person}{Sarah Clinckemaillie}, \bibinfo{person}{Yifan Chang}, \bibinfo{person}{Jonathan Waltz}, \bibinfo{person}{Gabrielle Lau}, \bibinfo{person}{Marybeth Fair}, \bibinfo{person}{Alice Li}, \bibinfo{person}{William Bishop}, \bibinfo{person}{Wei Li}, \bibinfo{person}{Folawiyo Campbell-Ajala}, {et~al\mbox{.}}} \bibinfo{year}{2024}\natexlab{a}.
\newblock \showarticletitle{AndroidWorld: A dynamic benchmarking environment for autonomous agents}.
\newblock \bibinfo{journal}{\emph{arXiv preprint arXiv:2405.14573}} (\bibinfo{year}{2024}).
\newblock


\bibitem[Rawles et~al\mbox{.}(2024b)]%
        {rawles2023AiTW}
\bibfield{author}{\bibinfo{person}{Christopher Rawles}, \bibinfo{person}{Alice Li}, \bibinfo{person}{Daniel Rodriguez}, \bibinfo{person}{Oriana Riva}, {and} \bibinfo{person}{Timothy Lillicrap}.} \bibinfo{year}{2024}\natexlab{b}.
\newblock \showarticletitle{Androidinthewild: A large-scale dataset for android device control}.
\newblock \bibinfo{journal}{\emph{Advances in Neural Information Processing Systems}}  \bibinfo{volume}{36} (\bibinfo{year}{2024}).
\newblock


\bibitem[Reddy et~al\mbox{.}(2024)]%
        {reddy2024infogent}
\bibfield{author}{\bibinfo{person}{Revanth~Gangi Reddy}, \bibinfo{person}{Sagnik Mukherjee}, \bibinfo{person}{Jeonghwan Kim}, \bibinfo{person}{Zhenhailong Wang}, \bibinfo{person}{Dilek Hakkani-Tur}, {and} \bibinfo{person}{Heng Ji}.} \bibinfo{year}{2024}\natexlab{}.
\newblock \showarticletitle{Infogent: An agent-based framework for web information aggregation}.
\newblock \bibinfo{journal}{\emph{arXiv preprint arXiv:2410.19054}} (\bibinfo{year}{2024}).
\newblock


\bibitem[Ren et~al\mbox{.}(2024)]%
        {ren2024bases}
\bibfield{author}{\bibinfo{person}{Ruiyang Ren}, \bibinfo{person}{Peng Qiu}, \bibinfo{person}{Yingqi Qu}, \bibinfo{person}{Jing Liu}, \bibinfo{person}{Wayne~Xin Zhao}, \bibinfo{person}{Hua Wu}, \bibinfo{person}{Ji-Rong Wen}, {and} \bibinfo{person}{Haifeng Wang}.} \bibinfo{year}{2024}\natexlab{}.
\newblock \showarticletitle{BASES: Large-scale Web Search User Simulation with Large Language Model based Agents}.
\newblock \bibinfo{journal}{\emph{arXiv preprint arXiv:2402.17505}} (\bibinfo{year}{2024}).
\newblock


\bibitem[Resnick and Varian(1997)]%
        {resnick1997recommender}
\bibfield{author}{\bibinfo{person}{Paul Resnick} {and} \bibinfo{person}{Hal~R Varian}.} \bibinfo{year}{1997}\natexlab{}.
\newblock \showarticletitle{Recommender systems}.
\newblock \bibinfo{journal}{\emph{Commun. ACM}} \bibinfo{volume}{40}, \bibinfo{number}{3} (\bibinfo{year}{1997}), \bibinfo{pages}{56--58}.
\newblock


\bibitem[Richardson et~al\mbox{.}(2023)]%
        {richardson2023integrating}
\bibfield{author}{\bibinfo{person}{Chris Richardson}, \bibinfo{person}{Yao Zhang}, \bibinfo{person}{Kellen Gillespie}, \bibinfo{person}{Sudipta Kar}, \bibinfo{person}{Arshdeep Singh}, \bibinfo{person}{Zeynab Raeesy}, \bibinfo{person}{Omar~Zia Khan}, {and} \bibinfo{person}{Abhinav Sethy}.} \bibinfo{year}{2023}\natexlab{}.
\newblock \showarticletitle{Integrating summarization and retrieval for enhanced personalization via large language models}.
\newblock \bibinfo{journal}{\emph{arXiv preprint arXiv:2310.20081}} (\bibinfo{year}{2023}).
\newblock


\bibitem[Russe et~al\mbox{.}(2024)]%
        {russe2024improving}
\bibfield{author}{\bibinfo{person}{Maximilian~Frederik Russe}, \bibinfo{person}{Marco Reisert}, \bibinfo{person}{Fabian Bamberg}, {and} \bibinfo{person}{Alexander Rau}.} \bibinfo{year}{2024}\natexlab{}.
\newblock \showarticletitle{Improving the use of LLMs in radiology through prompt engineering: from precision prompts to zero-shot learning}. In \bibinfo{booktitle}{\emph{R{\"o}Fo-Fortschritte auf dem Gebiet der R{\"o}ntgenstrahlen und der bildgebenden Verfahren}}. Georg Thieme Verlag KG.
\newblock


\bibitem[Schick et~al\mbox{.}(2024)]%
        {schick2024toolformer}
\bibfield{author}{\bibinfo{person}{Timo Schick}, \bibinfo{person}{Jane Dwivedi-Yu}, \bibinfo{person}{Roberto Dess{\`\i}}, \bibinfo{person}{Roberta Raileanu}, \bibinfo{person}{Maria Lomeli}, \bibinfo{person}{Eric Hambro}, \bibinfo{person}{Luke Zettlemoyer}, \bibinfo{person}{Nicola Cancedda}, {and} \bibinfo{person}{Thomas Scialom}.} \bibinfo{year}{2024}\natexlab{}.
\newblock \showarticletitle{Toolformer: Language models can teach themselves to use tools}.
\newblock \bibinfo{journal}{\emph{Advances in Neural Information Processing Systems}}  \bibinfo{volume}{36} (\bibinfo{year}{2024}).
\newblock


\bibitem[Schumann et~al\mbox{.}(2024)]%
        {schumann2024velma}
\bibfield{author}{\bibinfo{person}{Raphael Schumann}, \bibinfo{person}{Wanrong Zhu}, \bibinfo{person}{Weixi Feng}, \bibinfo{person}{Tsu-Jui Fu}, \bibinfo{person}{Stefan Riezler}, {and} \bibinfo{person}{William~Yang Wang}.} \bibinfo{year}{2024}\natexlab{}.
\newblock \showarticletitle{Velma: Verbalization embodiment of llm agents for vision and language navigation in street view}. In \bibinfo{booktitle}{\emph{Proceedings of the AAAI Conference on Artificial Intelligence}}, Vol.~\bibinfo{volume}{38}. \bibinfo{pages}{18924--18933}.
\newblock


\bibitem[Sch{\"u}tze et~al\mbox{.}(2008)]%
        {schutze2008introduction}
\bibfield{author}{\bibinfo{person}{Hinrich Sch{\"u}tze}, \bibinfo{person}{Christopher~D Manning}, {and} \bibinfo{person}{Prabhakar Raghavan}.} \bibinfo{year}{2008}\natexlab{}.
\newblock \bibinfo{booktitle}{\emph{Introduction to information retrieval}}. Vol.~\bibinfo{volume}{39}.
\newblock \bibinfo{publisher}{Cambridge University Press Cambridge}.
\newblock


\bibitem[Sekuli{\'c} et~al\mbox{.}(2024a)]%
        {sekulic2024analysing}
\bibfield{author}{\bibinfo{person}{Ivan Sekuli{\'c}}, \bibinfo{person}{Mohammad Alinannejadi}, {and} \bibinfo{person}{Fabio Crestani}.} \bibinfo{year}{2024}\natexlab{a}.
\newblock \showarticletitle{Analysing utterances in llm-based user simulation for conversational search}.
\newblock \bibinfo{journal}{\emph{ACM Transactions on Intelligent Systems and Technology}} \bibinfo{volume}{15}, \bibinfo{number}{3} (\bibinfo{year}{2024}), \bibinfo{pages}{1--22}.
\newblock


\bibitem[Sekuli{\'c} et~al\mbox{.}(2024b)]%
        {sekulic2024reliable}
\bibfield{author}{\bibinfo{person}{Ivan Sekuli{\'c}}, \bibinfo{person}{Silvia Terragni}, \bibinfo{person}{Victor Guimar{\~a}es}, \bibinfo{person}{Nghia Khau}, \bibinfo{person}{Bruna Guedes}, \bibinfo{person}{Modestas Filipavicius}, \bibinfo{person}{Andr{\'e}~Ferreira Manso}, {and} \bibinfo{person}{Roland Mathis}.} \bibinfo{year}{2024}\natexlab{b}.
\newblock \showarticletitle{Reliable LLM-based user simulator for task-oriented dialogue systems}.
\newblock \bibinfo{journal}{\emph{arXiv preprint arXiv:2402.13374}} (\bibinfo{year}{2024}).
\newblock


\bibitem[Sharma et~al\mbox{.}(2023)]%
        {sharma2023towards}
\bibfield{author}{\bibinfo{person}{Mrinank Sharma}, \bibinfo{person}{Meg Tong}, \bibinfo{person}{Tomasz Korbak}, \bibinfo{person}{David Duvenaud}, \bibinfo{person}{Amanda Askell}, \bibinfo{person}{Samuel~R Bowman}, \bibinfo{person}{Newton Cheng}, \bibinfo{person}{Esin Durmus}, \bibinfo{person}{Zac Hatfield-Dodds}, \bibinfo{person}{Scott~R Johnston}, {et~al\mbox{.}}} \bibinfo{year}{2023}\natexlab{}.
\newblock \showarticletitle{Towards understanding sycophancy in language models}.
\newblock \bibinfo{journal}{\emph{arXiv preprint arXiv:2310.13548}} (\bibinfo{year}{2023}).
\newblock


\bibitem[Sharma et~al\mbox{.}(2024)]%
        {sharma2024generative}
\bibfield{author}{\bibinfo{person}{Nikhil Sharma}, \bibinfo{person}{Q~Vera Liao}, {and} \bibinfo{person}{Ziang Xiao}.} \bibinfo{year}{2024}\natexlab{}.
\newblock \showarticletitle{Generative Echo Chamber? Effect of LLM-Powered Search Systems on Diverse Information Seeking}. In \bibinfo{booktitle}{\emph{Proceedings of the CHI Conference on Human Factors in Computing Systems}}. \bibinfo{pages}{1--17}.
\newblock


\bibitem[Sherstinsky(2020)]%
        {sherstinsky2020fundamentals}
\bibfield{author}{\bibinfo{person}{Alex Sherstinsky}.} \bibinfo{year}{2020}\natexlab{}.
\newblock \showarticletitle{Fundamentals of recurrent neural network (RNN) and long short-term memory (LSTM) network}.
\newblock \bibinfo{journal}{\emph{Physica D: Nonlinear Phenomena}}  \bibinfo{volume}{404} (\bibinfo{year}{2020}), \bibinfo{pages}{132306}.
\newblock


\bibitem[Shi et~al\mbox{.}(2023)]%
        {shi2023trusting}
\bibfield{author}{\bibinfo{person}{Weijia Shi}, \bibinfo{person}{Xiaochuang Han}, \bibinfo{person}{Mike Lewis}, \bibinfo{person}{Yulia Tsvetkov}, \bibinfo{person}{Luke Zettlemoyer}, {and} \bibinfo{person}{Scott Wen-tau Yih}.} \bibinfo{year}{2023}\natexlab{}.
\newblock \showarticletitle{Trusting your evidence: Hallucinate less with context-aware decoding}.
\newblock \bibinfo{journal}{\emph{arXiv preprint arXiv:2305.14739}} (\bibinfo{year}{2023}).
\newblock


\bibitem[Shi et~al\mbox{.}(2024)]%
        {shi2024enhancing}
\bibfield{author}{\bibinfo{person}{Wentao Shi}, \bibinfo{person}{Xiangnan He}, \bibinfo{person}{Yang Zhang}, \bibinfo{person}{Chongming Gao}, \bibinfo{person}{Xinyue Li}, \bibinfo{person}{Jizhi Zhang}, \bibinfo{person}{Qifan Wang}, {and} \bibinfo{person}{Fuli Feng}.} \bibinfo{year}{2024}\natexlab{}.
\newblock \showarticletitle{Enhancing Long-Term Recommendation with Bi-level Learnable Large Language Model Planning}.
\newblock \bibinfo{journal}{\emph{arXiv preprint arXiv:2403.00843}} (\bibinfo{year}{2024}).
\newblock


\bibitem[Shin et~al\mbox{.}(2021)]%
        {shin2021one4all}
\bibfield{author}{\bibinfo{person}{Kyuyong Shin}, \bibinfo{person}{Hanock Kwak}, \bibinfo{person}{Kyung-Min Kim}, \bibinfo{person}{Minkyu Kim}, \bibinfo{person}{Young-Jin Park}, \bibinfo{person}{Jisu Jeong}, {and} \bibinfo{person}{Seungjae Jung}.} \bibinfo{year}{2021}\natexlab{}.
\newblock \showarticletitle{One4all user representation for recommender systems in e-commerce}.
\newblock \bibinfo{journal}{\emph{arXiv preprint arXiv:2106.00573}} (\bibinfo{year}{2021}).
\newblock


\bibitem[Shu et~al\mbox{.}(2024)]%
        {shu2024rah}
\bibfield{author}{\bibinfo{person}{Yubo Shu}, \bibinfo{person}{Haonan Zhang}, \bibinfo{person}{Hansu Gu}, \bibinfo{person}{Peng Zhang}, \bibinfo{person}{Tun Lu}, \bibinfo{person}{Dongsheng Li}, {and} \bibinfo{person}{Ning Gu}.} \bibinfo{year}{2024}\natexlab{}.
\newblock \showarticletitle{RAH! RecSys--Assistant--Human: A Human-Centered Recommendation Framework With LLM Agents}.
\newblock \bibinfo{journal}{\emph{IEEE Transactions on Computational Social Systems}} (\bibinfo{year}{2024}).
\newblock


\bibitem[Sodhi et~al\mbox{.}(2023)]%
        {sodhi2023step}
\bibfield{author}{\bibinfo{person}{Paloma Sodhi}, \bibinfo{person}{SRK Branavan}, \bibinfo{person}{Yoav Artzi}, {and} \bibinfo{person}{Ryan McDonald}.} \bibinfo{year}{2023}\natexlab{}.
\newblock \showarticletitle{Step: Stacked llm policies for web actions}.
\newblock \bibinfo{journal}{\emph{arXiv preprint arXiv:2310.03720}} (\bibinfo{year}{2023}).
\newblock


\bibitem[Speretta and Gauch(2005)]%
        {speretta2005personalized}
\bibfield{author}{\bibinfo{person}{Mirco Speretta} {and} \bibinfo{person}{Susan Gauch}.} \bibinfo{year}{2005}\natexlab{}.
\newblock \showarticletitle{Personalized search based on user search histories}. In \bibinfo{booktitle}{\emph{The 2005 IEEE/WIC/ACM International Conference on Web Intelligence (WI'05)}}. IEEE, \bibinfo{pages}{622--628}.
\newblock


\bibitem[Sun et~al\mbox{.}(2024)]%
        {sun2024lawluo}
\bibfield{author}{\bibinfo{person}{Jingyun Sun}, \bibinfo{person}{Chengxiao Dai}, \bibinfo{person}{Zhongze Luo}, \bibinfo{person}{Yangbo Chang}, {and} \bibinfo{person}{Yang Li}.} \bibinfo{year}{2024}\natexlab{}.
\newblock \showarticletitle{Lawluo: A chinese law firm co-run by llm agents}.
\newblock \bibinfo{journal}{\emph{arXiv preprint arXiv:2407.16252}} (\bibinfo{year}{2024}).
\newblock


\bibitem[Takata et~al\mbox{.}(2024)]%
        {takata2024spontaneous}
\bibfield{author}{\bibinfo{person}{Ryosuke Takata}, \bibinfo{person}{Atsushi Masumori}, {and} \bibinfo{person}{Takashi Ikegami}.} \bibinfo{year}{2024}\natexlab{}.
\newblock \showarticletitle{Spontaneous Emergence of Agent Individuality through Social Interactions in LLM-Based Communities}.
\newblock \bibinfo{journal}{\emph{arXiv preprint arXiv:2411.03252}} (\bibinfo{year}{2024}).
\newblock


\bibitem[Team(2023)]%
        {team2023xagent}
\bibfield{author}{\bibinfo{person}{X Team}.} \bibinfo{year}{2023}\natexlab{}.
\newblock \showarticletitle{Xagent: An autonomous agent for complex task solving}.
\newblock \bibinfo{journal}{\emph{XAgent blog}} (\bibinfo{year}{2023}).
\newblock


\bibitem[Thakkar and Yadav(2024)]%
        {thakkar2024personalized}
\bibfield{author}{\bibinfo{person}{Param Thakkar} {and} \bibinfo{person}{Anushka Yadav}.} \bibinfo{year}{2024}\natexlab{}.
\newblock \showarticletitle{Personalized Recommendation Systems using Multimodal, Autonomous, Multi Agent Systems}.
\newblock \bibinfo{journal}{\emph{arXiv preprint arXiv:2410.19855}} (\bibinfo{year}{2024}).
\newblock


\bibitem[Van~Veen et~al\mbox{.}(2024)]%
        {van2024adapted}
\bibfield{author}{\bibinfo{person}{Dave Van~Veen}, \bibinfo{person}{Cara Van~Uden}, \bibinfo{person}{Louis Blankemeier}, \bibinfo{person}{Jean-Benoit Delbrouck}, \bibinfo{person}{Asad Aali}, \bibinfo{person}{Christian Bluethgen}, \bibinfo{person}{Anuj Pareek}, \bibinfo{person}{Malgorzata Polacin}, \bibinfo{person}{Eduardo~Pontes Reis}, \bibinfo{person}{Anna Seehofnerov{\'a}}, {et~al\mbox{.}}} \bibinfo{year}{2024}\natexlab{}.
\newblock \showarticletitle{Adapted large language models can outperform medical experts in clinical text summarization}.
\newblock \bibinfo{journal}{\emph{Nature medicine}} \bibinfo{volume}{30}, \bibinfo{number}{4} (\bibinfo{year}{2024}), \bibinfo{pages}{1134--1142}.
\newblock


\bibitem[Vaswani et~al\mbox{.}(2017)]%
        {vaswani2017attention}
\bibfield{author}{\bibinfo{person}{Ashish Vaswani}, \bibinfo{person}{Noam Shazeer}, \bibinfo{person}{Niki Parmar}, \bibinfo{person}{Jakob Uszkoreit}, \bibinfo{person}{Llion Jones}, \bibinfo{person}{Aidan~N Gomez}, \bibinfo{person}{{\L}ukasz Kaiser}, {and} \bibinfo{person}{Illia Polosukhin}.} \bibinfo{year}{2017}\natexlab{}.
\newblock \showarticletitle{Attention is all you need}.
\newblock \bibinfo{journal}{\emph{Advances in neural information processing systems}}  \bibinfo{volume}{30} (\bibinfo{year}{2017}).
\newblock


\bibitem[Wang et~al\mbox{.}(2023c)]%
        {wang2023voyager}
\bibfield{author}{\bibinfo{person}{Guanzhi Wang}, \bibinfo{person}{Yuqi Xie}, \bibinfo{person}{Yunfan Jiang}, \bibinfo{person}{Ajay Mandlekar}, \bibinfo{person}{Chaowei Xiao}, \bibinfo{person}{Yuke Zhu}, \bibinfo{person}{Linxi Fan}, {and} \bibinfo{person}{Anima Anandkumar}.} \bibinfo{year}{2023}\natexlab{c}.
\newblock \showarticletitle{Voyager: An open-ended embodied agent with large language models}.
\newblock \bibinfo{journal}{\emph{arXiv preprint arXiv:2305.16291}} (\bibinfo{year}{2023}).
\newblock


\bibitem[Wang et~al\mbox{.}(2024c)]%
        {wang2024Mobileagent}
\bibfield{author}{\bibinfo{person}{Junyang Wang}, \bibinfo{person}{Haiyang Xu}, \bibinfo{person}{Jiabo Ye}, \bibinfo{person}{Ming Yan}, \bibinfo{person}{Weizhou Shen}, \bibinfo{person}{Ji Zhang}, \bibinfo{person}{Fei Huang}, {and} \bibinfo{person}{Jitao Sang}.} \bibinfo{year}{2024}\natexlab{c}.
\newblock \showarticletitle{Mobile-agent: Autonomous multi-modal mobile device agent with visual perception}.
\newblock \bibinfo{journal}{\emph{arXiv preprint arXiv:2401.16158}} (\bibinfo{year}{2024}).
\newblock


\bibitem[Wang et~al\mbox{.}(2016)]%
        {wang2016comprehensive}
\bibfield{author}{\bibinfo{person}{Kaiye Wang}, \bibinfo{person}{Qiyue Yin}, \bibinfo{person}{Wei Wang}, \bibinfo{person}{Shu Wu}, {and} \bibinfo{person}{Liang Wang}.} \bibinfo{year}{2016}\natexlab{}.
\newblock \showarticletitle{A comprehensive survey on cross-modal retrieval}.
\newblock \bibinfo{journal}{\emph{arXiv preprint arXiv:1607.06215}} (\bibinfo{year}{2016}).
\newblock


\bibitem[Wang et~al\mbox{.}(2023d)]%
        {wang2023recagent}
\bibfield{author}{\bibinfo{person}{Lei Wang}, \bibinfo{person}{Jingsen Zhang}, \bibinfo{person}{Xu Chen}, \bibinfo{person}{Yankai Lin}, \bibinfo{person}{Ruihua Song}, \bibinfo{person}{Wayne~Xin Zhao}, {and} \bibinfo{person}{Ji-Rong Wen}.} \bibinfo{year}{2023}\natexlab{d}.
\newblock \showarticletitle{RecAgent: A Novel Simulation Paradigm for Recommender Systems}.
\newblock \bibinfo{journal}{\emph{arXiv preprint arXiv:2306.02552}} (\bibinfo{year}{2023}).
\newblock


\bibitem[Wang et~al\mbox{.}(2024a)]%
        {wang2024executable}
\bibfield{author}{\bibinfo{person}{Xingyao Wang}, \bibinfo{person}{Yangyi Chen}, \bibinfo{person}{Lifan Yuan}, \bibinfo{person}{Yizhe Zhang}, \bibinfo{person}{Yunzhu Li}, \bibinfo{person}{Hao Peng}, {and} \bibinfo{person}{Heng Ji}.} \bibinfo{year}{2024}\natexlab{a}.
\newblock \showarticletitle{Executable code actions elicit better llm agents}. In \bibinfo{booktitle}{\emph{Forty-first International Conference on Machine Learning}}.
\newblock


\bibitem[Wang et~al\mbox{.}(2023b)]%
        {wang2023rethinking}
\bibfield{author}{\bibinfo{person}{Xiaolei Wang}, \bibinfo{person}{Xinyu Tang}, \bibinfo{person}{Wayne~Xin Zhao}, \bibinfo{person}{Jingyuan Wang}, {and} \bibinfo{person}{Ji-Rong Wen}.} \bibinfo{year}{2023}\natexlab{b}.
\newblock \showarticletitle{Rethinking the evaluation for conversational recommendation in the era of large language models}.
\newblock \bibinfo{journal}{\emph{arXiv preprint arXiv:2305.13112}} (\bibinfo{year}{2023}).
\newblock


\bibitem[Wang et~al\mbox{.}(2022)]%
        {wang2022self}
\bibfield{author}{\bibinfo{person}{Xuezhi Wang}, \bibinfo{person}{Jason Wei}, \bibinfo{person}{Dale Schuurmans}, \bibinfo{person}{Quoc Le}, \bibinfo{person}{Ed Chi}, \bibinfo{person}{Sharan Narang}, \bibinfo{person}{Aakanksha Chowdhery}, {and} \bibinfo{person}{Denny Zhou}.} \bibinfo{year}{2022}\natexlab{}.
\newblock \showarticletitle{Self-consistency improves chain of thought reasoning in language models}.
\newblock \bibinfo{journal}{\emph{arXiv preprint arXiv:2203.11171}} (\bibinfo{year}{2022}).
\newblock


\bibitem[Wang et~al\mbox{.}(2023a)]%
        {wang2023recmind}
\bibfield{author}{\bibinfo{person}{Yancheng Wang}, \bibinfo{person}{Ziyan Jiang}, \bibinfo{person}{Zheng Chen}, \bibinfo{person}{Fan Yang}, \bibinfo{person}{Yingxue Zhou}, \bibinfo{person}{Eunah Cho}, \bibinfo{person}{Xing Fan}, \bibinfo{person}{Xiaojiang Huang}, \bibinfo{person}{Yanbin Lu}, {and} \bibinfo{person}{Yingzhen Yang}.} \bibinfo{year}{2023}\natexlab{a}.
\newblock \showarticletitle{Recmind: Large language model powered agent for recommendation}.
\newblock \bibinfo{journal}{\emph{arXiv preprint arXiv:2308.14296}} (\bibinfo{year}{2023}).
\newblock


\bibitem[Wang et~al\mbox{.}(2024b)]%
        {wang2024poisoning}
\bibfield{author}{\bibinfo{person}{Zongwei Wang}, \bibinfo{person}{Min Gao}, \bibinfo{person}{Junliang Yu}, \bibinfo{person}{Hao Ma}, \bibinfo{person}{Hongzhi Yin}, {and} \bibinfo{person}{Shazia Sadiq}.} \bibinfo{year}{2024}\natexlab{b}.
\newblock \showarticletitle{Poisoning attacks against recommender systems: A survey}.
\newblock \bibinfo{journal}{\emph{arXiv preprint arXiv:2401.01527}} (\bibinfo{year}{2024}).
\newblock


\bibitem[Wang et~al\mbox{.}(2024d)]%
        {wang2024multi}
\bibfield{author}{\bibinfo{person}{Zhefan Wang}, \bibinfo{person}{Yuanqing Yu}, \bibinfo{person}{Wendi Zheng}, \bibinfo{person}{Weizhi Ma}, {and} \bibinfo{person}{Min Zhang}.} \bibinfo{year}{2024}\natexlab{d}.
\newblock \showarticletitle{Multi-Agent Collaboration Framework for Recommender Systems}.
\newblock \bibinfo{journal}{\emph{arXiv preprint arXiv:2402.15235}} (\bibinfo{year}{2024}).
\newblock


\bibitem[Waterman(1985)]%
        {waterman1985guide}
\bibfield{author}{\bibinfo{person}{Donald~A Waterman}.} \bibinfo{year}{1985}\natexlab{}.
\newblock \bibinfo{booktitle}{\emph{A guide to expert systems}}.
\newblock \bibinfo{publisher}{Addison-Wesley Longman Publishing Co., Inc.}
\newblock


\bibitem[Wei et~al\mbox{.}(2023)]%
        {wei2023simple}
\bibfield{author}{\bibinfo{person}{Jerry Wei}, \bibinfo{person}{Da Huang}, \bibinfo{person}{Yifeng Lu}, \bibinfo{person}{Denny Zhou}, {and} \bibinfo{person}{Quoc~V Le}.} \bibinfo{year}{2023}\natexlab{}.
\newblock \showarticletitle{Simple synthetic data reduces sycophancy in large language models}.
\newblock \bibinfo{journal}{\emph{arXiv preprint arXiv:2308.03958}} (\bibinfo{year}{2023}).
\newblock


\bibitem[Wei et~al\mbox{.}(2022)]%
        {wei2022chain}
\bibfield{author}{\bibinfo{person}{Jason Wei}, \bibinfo{person}{Xuezhi Wang}, \bibinfo{person}{Dale Schuurmans}, \bibinfo{person}{Maarten Bosma}, \bibinfo{person}{Fei Xia}, \bibinfo{person}{Ed Chi}, \bibinfo{person}{Quoc~V Le}, \bibinfo{person}{Denny Zhou}, {et~al\mbox{.}}} \bibinfo{year}{2022}\natexlab{}.
\newblock \showarticletitle{Chain-of-thought prompting elicits reasoning in large language models}.
\newblock \bibinfo{journal}{\emph{Advances in Neural Information Processing Systems}}  \bibinfo{volume}{35} (\bibinfo{year}{2022}), \bibinfo{pages}{24824--24837}.
\newblock


\bibitem[White(2024)]%
        {white2024advancing}
\bibfield{author}{\bibinfo{person}{Ryen~W White}.} \bibinfo{year}{2024}\natexlab{}.
\newblock \showarticletitle{Advancing the search frontier with AI agents}.
\newblock \bibinfo{journal}{\emph{Commun. ACM}} (\bibinfo{year}{2024}).
\newblock


\bibitem[Wilkins et~al\mbox{.}(2024)]%
        {wilkins2024hybrid}
\bibfield{author}{\bibinfo{person}{Grant Wilkins}, \bibinfo{person}{Srinivasan Keshav}, {and} \bibinfo{person}{Richard Mortier}.} \bibinfo{year}{2024}\natexlab{}.
\newblock \showarticletitle{Hybrid Heterogeneous Clusters Can Lower the Energy Consumption of LLM Inference Workloads}. In \bibinfo{booktitle}{\emph{Proceedings of the 15th ACM International Conference on Future and Sustainable Energy Systems}}. \bibinfo{pages}{506--513}.
\newblock


\bibitem[Wondergem et~al\mbox{.}(1998)]%
        {wondergem1998agents}
\bibfield{author}{\bibinfo{person}{BCM Wondergem}, \bibinfo{person}{P~van Bommel}, \bibinfo{person}{Theo~WC Huibers}, {and} \bibinfo{person}{Th~P Weide}.} \bibinfo{year}{1998}\natexlab{}.
\newblock \showarticletitle{Agents in Cyberspace--Towards a Framework for Multi-Agent Systems in Information Discovery}. In \bibinfo{booktitle}{\emph{20th Annual BCS-IRSG Colloquium on IR}}. BCS Learning \& Development.
\newblock


\bibitem[Wu et~al\mbox{.}(2024b)]%
        {wu2024survey}
\bibfield{author}{\bibinfo{person}{Likang Wu}, \bibinfo{person}{Zhi Zheng}, \bibinfo{person}{Zhaopeng Qiu}, \bibinfo{person}{Hao Wang}, \bibinfo{person}{Hongchao Gu}, \bibinfo{person}{Tingjia Shen}, \bibinfo{person}{Chuan Qin}, \bibinfo{person}{Chen Zhu}, \bibinfo{person}{Hengshu Zhu}, \bibinfo{person}{Qi Liu}, {et~al\mbox{.}}} \bibinfo{year}{2024}\natexlab{b}.
\newblock \showarticletitle{A survey on large language models for recommendation}.
\newblock \bibinfo{journal}{\emph{World Wide Web}} \bibinfo{volume}{27}, \bibinfo{number}{5} (\bibinfo{year}{2024}), \bibinfo{pages}{60}.
\newblock


\bibitem[Wu et~al\mbox{.}(2025)]%
        {wu2025avatar}
\bibfield{author}{\bibinfo{person}{Shirley Wu}, \bibinfo{person}{Shiyu Zhao}, \bibinfo{person}{Qian Huang}, \bibinfo{person}{Kexin Huang}, \bibinfo{person}{Michihiro Yasunaga}, \bibinfo{person}{Kaidi Cao}, \bibinfo{person}{Vassilis Ioannidis}, \bibinfo{person}{Karthik Subbian}, \bibinfo{person}{Jure Leskovec}, {and} \bibinfo{person}{James~Y Zou}.} \bibinfo{year}{2025}\natexlab{}.
\newblock \showarticletitle{AvaTaR: Optimizing LLM Agents for Tool Usage via Contrastive Reasoning}.
\newblock \bibinfo{journal}{\emph{Advances in Neural Information Processing Systems}}  \bibinfo{volume}{37} (\bibinfo{year}{2025}), \bibinfo{pages}{25981--26010}.
\newblock


\bibitem[Wu et~al\mbox{.}(2024a)]%
        {wu2024avatar}
\bibfield{author}{\bibinfo{person}{Shirley Wu}, \bibinfo{person}{Shiyu Zhao}, \bibinfo{person}{Qian Huang}, \bibinfo{person}{Kexin Huang}, \bibinfo{person}{Michihiro Yasunaga}, \bibinfo{person}{Kaidi Cao}, \bibinfo{person}{Vassilis~N Ioannidis}, \bibinfo{person}{Karthik Subbian}, \bibinfo{person}{Jure Leskovec}, {and} \bibinfo{person}{James Zou}.} \bibinfo{year}{2024}\natexlab{a}.
\newblock \showarticletitle{AvaTaR: Optimizing LLM Agents for Tool-Assisted Knowledge Retrieval}.
\newblock \bibinfo{journal}{\emph{arXiv preprint arXiv:2406.11200}} (\bibinfo{year}{2024}).
\newblock


\bibitem[Xi et~al\mbox{.}(2023)]%
        {xi2023rise}
\bibfield{author}{\bibinfo{person}{Zhiheng Xi}, \bibinfo{person}{Wenxiang Chen}, \bibinfo{person}{Xin Guo}, \bibinfo{person}{Wei He}, \bibinfo{person}{Yiwen Ding}, \bibinfo{person}{Boyang Hong}, \bibinfo{person}{Ming Zhang}, \bibinfo{person}{Junzhe Wang}, \bibinfo{person}{Senjie Jin}, \bibinfo{person}{Enyu Zhou}, {et~al\mbox{.}}} \bibinfo{year}{2023}\natexlab{}.
\newblock \showarticletitle{The Rise and Potential of Large Language Model Based Agents: A Survey}.
\newblock \bibinfo{journal}{\emph{arXiv preprint arXiv:2309.07864}} (\bibinfo{year}{2023}).
\newblock


\bibitem[Xiao et~al\mbox{.}(2024)]%
        {xiao2024cellagent}
\bibfield{author}{\bibinfo{person}{Yihang Xiao}, \bibinfo{person}{Jinyi Liu}, \bibinfo{person}{Yan Zheng}, \bibinfo{person}{Xiaohan Xie}, \bibinfo{person}{Jianye Hao}, \bibinfo{person}{Mingzhi Li}, \bibinfo{person}{Ruitao Wang}, \bibinfo{person}{Fei Ni}, \bibinfo{person}{Yuxiao Li}, \bibinfo{person}{Jintian Luo}, {et~al\mbox{.}}} \bibinfo{year}{2024}\natexlab{}.
\newblock \showarticletitle{CellAgent: An LLM-driven Multi-Agent Framework for Automated Single-cell Data Analysis}.
\newblock \bibinfo{journal}{\emph{bioRxiv}} (\bibinfo{year}{2024}), \bibinfo{pages}{2024--05}.
\newblock


\bibitem[Xie and Zou(2024)]%
        {xie2024human}
\bibfield{author}{\bibinfo{person}{Chengxing Xie} {and} \bibinfo{person}{Difan Zou}.} \bibinfo{year}{2024}\natexlab{}.
\newblock \showarticletitle{A Human-Like Reasoning Framework for Multi-Phases Planning Task with Large Language Models}.
\newblock \bibinfo{journal}{\emph{arXiv preprint arXiv:2405.18208}} (\bibinfo{year}{2024}).
\newblock


\bibitem[Xie et~al\mbox{.}(2024)]%
        {xie2024large}
\bibfield{author}{\bibinfo{person}{Junlin Xie}, \bibinfo{person}{Zhihong Chen}, \bibinfo{person}{Ruifei Zhang}, \bibinfo{person}{Xiang Wan}, {and} \bibinfo{person}{Guanbin Li}.} \bibinfo{year}{2024}\natexlab{}.
\newblock \showarticletitle{Large multimodal agents: A survey}.
\newblock \bibinfo{journal}{\emph{arXiv preprint arXiv:2402.15116}} (\bibinfo{year}{2024}).
\newblock


\bibitem[Xu et~al\mbox{.}(2018)]%
        {xu2018deep}
\bibfield{author}{\bibinfo{person}{Jun Xu}, \bibinfo{person}{Xiangnan He}, {and} \bibinfo{person}{Hang Li}.} \bibinfo{year}{2018}\natexlab{}.
\newblock \showarticletitle{Deep learning for matching in search and recommendation}. In \bibinfo{booktitle}{\emph{The 41st International ACM SIGIR Conference on Research \& Development in Information Retrieval}}. \bibinfo{pages}{1365--1368}.
\newblock


\bibitem[Xu et~al\mbox{.}(2024a)]%
        {xu2024tad}
\bibfield{author}{\bibinfo{person}{Xinhao Xu}, \bibinfo{person}{Hui Chen}, \bibinfo{person}{Zijia Lin}, \bibinfo{person}{Jungong Han}, \bibinfo{person}{Lixing Gong}, \bibinfo{person}{Guoxin Wang}, \bibinfo{person}{Yongjun Bao}, {and} \bibinfo{person}{Guiguang Ding}.} \bibinfo{year}{2024}\natexlab{a}.
\newblock \showarticletitle{Tad: A plug-and-play task-aware decoding method to better adapt llms on downstream tasks}. In \bibinfo{booktitle}{\emph{Proceedings of the Thirty-Third International Joint Conference on Artificial Intelligence, IJCAI}}.
\newblock


\bibitem[Xu et~al\mbox{.}(2024b)]%
        {xu2024can}
\bibfield{author}{\bibinfo{person}{Zhenyu Xu}, \bibinfo{person}{Hailin Xu}, \bibinfo{person}{Zhouyang Lu}, \bibinfo{person}{Yingying Zhao}, \bibinfo{person}{Rui Zhu}, \bibinfo{person}{Yujiang Wang}, \bibinfo{person}{Mingzhi Dong}, \bibinfo{person}{Yuhu Chang}, \bibinfo{person}{Qin Lv}, \bibinfo{person}{Robert~P Dick}, {et~al\mbox{.}}} \bibinfo{year}{2024}\natexlab{b}.
\newblock \showarticletitle{Can Large Language Models Be Good Companions? An LLM-Based Eyewear System with Conversational Common Ground}.
\newblock \bibinfo{journal}{\emph{Proceedings of the ACM on Interactive, Mobile, Wearable and Ubiquitous Technologies}} \bibinfo{volume}{8}, \bibinfo{number}{2} (\bibinfo{year}{2024}), \bibinfo{pages}{1--41}.
\newblock


\bibitem[Yan et~al\mbox{.}(2024)]%
        {yan2024clinicallab}
\bibfield{author}{\bibinfo{person}{Weixiang Yan}, \bibinfo{person}{Haitian Liu}, \bibinfo{person}{Tengxiao Wu}, \bibinfo{person}{Qian Chen}, \bibinfo{person}{Wen Wang}, \bibinfo{person}{Haoyuan Chai}, \bibinfo{person}{Jiayi Wang}, \bibinfo{person}{Weishan Zhao}, \bibinfo{person}{Yixin Zhang}, \bibinfo{person}{Renjun Zhang}, {et~al\mbox{.}}} \bibinfo{year}{2024}\natexlab{}.
\newblock \showarticletitle{ClinicalLab: Aligning Agents for Multi-Departmental Clinical Diagnostics in the Real World}.
\newblock \bibinfo{journal}{\emph{arXiv preprint arXiv:2406.13890}} (\bibinfo{year}{2024}).
\newblock


\bibitem[Yin et~al\mbox{.}(2024)]%
        {yin2024device}
\bibfield{author}{\bibinfo{person}{Hongzhi Yin}, \bibinfo{person}{Liang Qu}, \bibinfo{person}{Tong Chen}, \bibinfo{person}{Wei Yuan}, \bibinfo{person}{Ruiqi Zheng}, \bibinfo{person}{Jing Long}, \bibinfo{person}{Xin Xia}, \bibinfo{person}{Yuhui Shi}, {and} \bibinfo{person}{Chengqi Zhang}.} \bibinfo{year}{2024}\natexlab{}.
\newblock \showarticletitle{On-device recommender systems: A comprehensive survey}.
\newblock \bibinfo{journal}{\emph{arXiv preprint arXiv:2401.11441}} (\bibinfo{year}{2024}).
\newblock


\bibitem[Yoon et~al\mbox{.}(2024)]%
        {yoon2024evaluating}
\bibfield{author}{\bibinfo{person}{Se-eun Yoon}, \bibinfo{person}{Zhankui He}, \bibinfo{person}{Jessica~Maria Echterhoff}, {and} \bibinfo{person}{Julian McAuley}.} \bibinfo{year}{2024}\natexlab{}.
\newblock \showarticletitle{Evaluating Large Language Models as Generative User Simulators for Conversational Recommendation}.
\newblock \bibinfo{journal}{\emph{arXiv preprint arXiv:2403.09738}} (\bibinfo{year}{2024}).
\newblock


\bibitem[Yuan et~al\mbox{.}(2021)]%
        {yuan2021Conure}
\bibfield{author}{\bibinfo{person}{Fajie Yuan}, \bibinfo{person}{Guoxiao Zhang}, \bibinfo{person}{Alexandros Karatzoglou}, \bibinfo{person}{Joemon Jose}, \bibinfo{person}{Beibei Kong}, {and} \bibinfo{person}{Yudong Li}.} \bibinfo{year}{2021}\natexlab{}.
\newblock \showarticletitle{One person, one model, one world: Learning continual user representation without forgetting}. In \bibinfo{booktitle}{\emph{Proceedings of the 44th International ACM SIGIR Conference on Research and Development in Information Retrieval}}. \bibinfo{pages}{696--705}.
\newblock


\bibitem[Yuan et~al\mbox{.}(2024)]%
        {yuan2024easytool}
\bibfield{author}{\bibinfo{person}{Siyu Yuan}, \bibinfo{person}{Kaitao Song}, \bibinfo{person}{Jiangjie Chen}, \bibinfo{person}{Xu Tan}, \bibinfo{person}{Yongliang Shen}, \bibinfo{person}{Ren Kan}, \bibinfo{person}{Dongsheng Li}, {and} \bibinfo{person}{Deqing Yang}.} \bibinfo{year}{2024}\natexlab{}.
\newblock \showarticletitle{Easytool: Enhancing llm-based agents with concise tool instruction}.
\newblock \bibinfo{journal}{\emph{arXiv preprint arXiv:2401.06201}} (\bibinfo{year}{2024}).
\newblock


\bibitem[Yuan et~al\mbox{.}(2023)]%
        {yuan2023federated}
\bibfield{author}{\bibinfo{person}{Wei Yuan}, \bibinfo{person}{Hongzhi Yin}, \bibinfo{person}{Fangzhao Wu}, \bibinfo{person}{Shijie Zhang}, \bibinfo{person}{Tieke He}, {and} \bibinfo{person}{Hao Wang}.} \bibinfo{year}{2023}\natexlab{}.
\newblock \showarticletitle{Federated unlearning for on-device recommendation}. In \bibinfo{booktitle}{\emph{Proceedings of the sixteenth ACM international conference on web search and data mining}}. \bibinfo{pages}{393--401}.
\newblock


\bibitem[Zeng et~al\mbox{.}(2024)]%
        {zeng2024automated}
\bibfield{author}{\bibinfo{person}{Yankai Zeng}, \bibinfo{person}{Abhiramon Rajasekharan}, \bibinfo{person}{Parth Padalkar}, \bibinfo{person}{Kinjal Basu}, \bibinfo{person}{Joaqu{\'\i}n Arias}, {and} \bibinfo{person}{Gopal Gupta}.} \bibinfo{year}{2024}\natexlab{}.
\newblock \showarticletitle{Automated interactive domain-specific conversational agents that understand human dialogs}. In \bibinfo{booktitle}{\emph{International Symposium on Practical Aspects of Declarative Languages}}. Springer, \bibinfo{pages}{204--222}.
\newblock


\bibitem[Zerhoudi and Granitzer(2024)]%
        {zerhoudi2024personarag}
\bibfield{author}{\bibinfo{person}{Saber Zerhoudi} {and} \bibinfo{person}{Michael Granitzer}.} \bibinfo{year}{2024}\natexlab{}.
\newblock \showarticletitle{PersonaRAG: Enhancing Retrieval-Augmented Generation Systems with User-Centric Agents}.
\newblock \bibinfo{journal}{\emph{arXiv preprint arXiv:2407.09394}} (\bibinfo{year}{2024}).
\newblock


\bibitem[Zhan and Zhang(2023)]%
        {zhang2024AutoGUI}
\bibfield{author}{\bibinfo{person}{Zhuosheng Zhan} {and} \bibinfo{person}{Aston Zhang}.} \bibinfo{year}{2023}\natexlab{}.
\newblock \showarticletitle{You only look at screens: Multimodal chain-of-action agents}.
\newblock \bibinfo{journal}{\emph{arXiv preprint arXiv:2309.11436}} (\bibinfo{year}{2023}).
\newblock


\bibitem[Zhang et~al\mbox{.}(2024b)]%
        {zhang2024generative}
\bibfield{author}{\bibinfo{person}{An Zhang}, \bibinfo{person}{Yuxin Chen}, \bibinfo{person}{Leheng Sheng}, \bibinfo{person}{Xiang Wang}, {and} \bibinfo{person}{Tat-Seng Chua}.} \bibinfo{year}{2024}\natexlab{b}.
\newblock \showarticletitle{On generative agents in recommendation}. In \bibinfo{booktitle}{\emph{Proceedings of the 47th International ACM SIGIR Conference on Research and Development in Information Retrieval}}. \bibinfo{pages}{1807--1817}.
\newblock


\bibitem[Zhang et~al\mbox{.}(2023)]%
        {zhang2023generative}
\bibfield{author}{\bibinfo{person}{An Zhang}, \bibinfo{person}{Leheng Sheng}, \bibinfo{person}{Yuxin Chen}, \bibinfo{person}{Hao Li}, \bibinfo{person}{Yang Deng}, \bibinfo{person}{Xiang Wang}, {and} \bibinfo{person}{Tat-Seng Chua}.} \bibinfo{year}{2023}\natexlab{}.
\newblock \showarticletitle{On Generative Agents in Recommendation}.
\newblock \bibinfo{journal}{\emph{arXiv preprint arXiv:2310.10108}} (\bibinfo{year}{2023}).
\newblock


\bibitem[Zhang et~al\mbox{.}(2024g)]%
        {zhang2024usimagent}
\bibfield{author}{\bibinfo{person}{Erhan Zhang}, \bibinfo{person}{Xingzhu Wang}, \bibinfo{person}{Peiyuan Gong}, \bibinfo{person}{Yankai Lin}, {and} \bibinfo{person}{Jiaxin Mao}.} \bibinfo{year}{2024}\natexlab{g}.
\newblock \showarticletitle{Usimagent: Large language models for simulating search users}. In \bibinfo{booktitle}{\emph{Proceedings of the 47th International ACM SIGIR Conference on Research and Development in Information Retrieval}}. \bibinfo{pages}{2687--2692}.
\newblock


\bibitem[Zhang et~al\mbox{.}(2024a)]%
        {zhang2024prospect}
\bibfield{author}{\bibinfo{person}{Jizhi Zhang}, \bibinfo{person}{Keqin Bao}, \bibinfo{person}{Wenjie Wang}, \bibinfo{person}{Yang Zhang}, \bibinfo{person}{Wentao Shi}, \bibinfo{person}{Wanhong Xu}, \bibinfo{person}{Fuli Feng}, {and} \bibinfo{person}{Tat-Seng Chua}.} \bibinfo{year}{2024}\natexlab{a}.
\newblock \showarticletitle{Prospect Personalized Recommendation on Large Language Model-based Agent Platform}.
\newblock \bibinfo{journal}{\emph{arXiv preprint arXiv:2402.18240}} (\bibinfo{year}{2024}).
\newblock


\bibitem[Zhang et~al\mbox{.}(2024c)]%
        {zhang2024NineRec}
\bibfield{author}{\bibinfo{person}{Jiaqi Zhang}, \bibinfo{person}{Yu Cheng}, \bibinfo{person}{Yongxin Ni}, \bibinfo{person}{Yunzhu Pan}, \bibinfo{person}{Zheng Yuan}, \bibinfo{person}{Junchen Fu}, \bibinfo{person}{Youhua Li}, \bibinfo{person}{Jie Wang}, {and} \bibinfo{person}{Fajie Yuan}.} \bibinfo{year}{2024}\natexlab{c}.
\newblock \showarticletitle{Ninerec: A benchmark dataset suite for evaluating transferable recommendation}.
\newblock \bibinfo{journal}{\emph{IEEE Transactions on Pattern Analysis and Machine Intelligence}} (\bibinfo{year}{2024}).
\newblock


\bibitem[Zhang et~al\mbox{.}(2024d)]%
        {zhang2024smartagent}
\bibfield{author}{\bibinfo{person}{Jiaqi Zhang}, \bibinfo{person}{Chen Gao}, \bibinfo{person}{Liyuan Zhang}, \bibinfo{person}{Yong Li}, {and} \bibinfo{person}{Hongzhi Yin}.} \bibinfo{year}{2024}\natexlab{d}.
\newblock \showarticletitle{SmartAgent: Chain-of-User-Thought for Embodied Personalized Agent in Cyber World}.
\newblock \bibinfo{journal}{\emph{arXiv preprint arXiv:2412.07472}} (\bibinfo{year}{2024}).
\newblock


\bibitem[Zhang et~al\mbox{.}(2024e)]%
        {zhang2024agentcf}
\bibfield{author}{\bibinfo{person}{Junjie Zhang}, \bibinfo{person}{Yupeng Hou}, \bibinfo{person}{Ruobing Xie}, \bibinfo{person}{Wenqi Sun}, \bibinfo{person}{Julian McAuley}, \bibinfo{person}{Wayne~Xin Zhao}, \bibinfo{person}{Leyu Lin}, {and} \bibinfo{person}{Ji-Rong Wen}.} \bibinfo{year}{2024}\natexlab{e}.
\newblock \showarticletitle{Agentcf: Collaborative learning with autonomous language agents for recommender systems}. In \bibinfo{booktitle}{\emph{Proceedings of the ACM on Web Conference 2024}}. \bibinfo{pages}{3679--3689}.
\newblock


\bibitem[Zhang et~al\mbox{.}(2019)]%
        {zhang2019deep}
\bibfield{author}{\bibinfo{person}{Shuai Zhang}, \bibinfo{person}{Lina Yao}, \bibinfo{person}{Aixin Sun}, {and} \bibinfo{person}{Yi Tay}.} \bibinfo{year}{2019}\natexlab{}.
\newblock \showarticletitle{Deep learning based recommender system: A survey and new perspectives}.
\newblock \bibinfo{journal}{\emph{ACM computing surveys (CSUR)}} \bibinfo{volume}{52}, \bibinfo{number}{1} (\bibinfo{year}{2019}), \bibinfo{pages}{1--38}.
\newblock


\bibitem[Zhang et~al\mbox{.}(2024h)]%
        {zhang2024multimodal}
\bibfield{author}{\bibinfo{person}{Wentao Zhang}, \bibinfo{person}{Lingxuan Zhao}, \bibinfo{person}{Haochong Xia}, \bibinfo{person}{Shuo Sun}, \bibinfo{person}{Jiaze Sun}, \bibinfo{person}{Molei Qin}, \bibinfo{person}{Xinyi Li}, \bibinfo{person}{Yuqing Zhao}, \bibinfo{person}{Yilei Zhao}, \bibinfo{person}{Xinyu Cai}, {et~al\mbox{.}}} \bibinfo{year}{2024}\natexlab{h}.
\newblock \showarticletitle{A multimodal foundation agent for financial trading: Tool-augmented, diversified, and generalist}. In \bibinfo{booktitle}{\emph{Proceedings of the 30th ACM SIGKDD Conference on Knowledge Discovery and Data Mining}}. \bibinfo{pages}{4314--4325}.
\newblock


\bibitem[Zhang et~al\mbox{.}(2018)]%
        {zhang2018towards}
\bibfield{author}{\bibinfo{person}{Yongfeng Zhang}, \bibinfo{person}{Xu Chen}, \bibinfo{person}{Qingyao Ai}, \bibinfo{person}{Liu Yang}, {and} \bibinfo{person}{W~Bruce Croft}.} \bibinfo{year}{2018}\natexlab{}.
\newblock \showarticletitle{Towards conversational search and recommendation: System ask, user respond}. In \bibinfo{booktitle}{\emph{Proceedings of the 27th acm international conference on information and knowledge management}}. \bibinfo{pages}{177--186}.
\newblock


\bibitem[Zhang et~al\mbox{.}(2024f)]%
        {zhang2024llm}
\bibfield{author}{\bibinfo{person}{Zijian Zhang}, \bibinfo{person}{Shuchang Liu}, \bibinfo{person}{Ziru Liu}, \bibinfo{person}{Rui Zhong}, \bibinfo{person}{Qingpeng Cai}, \bibinfo{person}{Xiangyu Zhao}, \bibinfo{person}{Chunxu Zhang}, \bibinfo{person}{Qidong Liu}, {and} \bibinfo{person}{Peng Jiang}.} \bibinfo{year}{2024}\natexlab{f}.
\newblock \showarticletitle{LLM-Powered User Simulator for Recommender System}.
\newblock \bibinfo{journal}{\emph{arXiv preprint arXiv:2412.16984}} (\bibinfo{year}{2024}).
\newblock


\bibitem[Zhao et~al\mbox{.}(2024a)]%
        {zhao2024expel}
\bibfield{author}{\bibinfo{person}{Andrew Zhao}, \bibinfo{person}{Daniel Huang}, \bibinfo{person}{Quentin Xu}, \bibinfo{person}{Matthieu Lin}, \bibinfo{person}{Yong-Jin Liu}, {and} \bibinfo{person}{Gao Huang}.} \bibinfo{year}{2024}\natexlab{a}.
\newblock \showarticletitle{Expel: Llm agents are experiential learners}. In \bibinfo{booktitle}{\emph{Proceedings of the AAAI Conference on Artificial Intelligence}}, Vol.~\bibinfo{volume}{38}. \bibinfo{pages}{19632--19642}.
\newblock


\bibitem[Zhao et~al\mbox{.}(2023b)]%
        {zhao2023survey}
\bibfield{author}{\bibinfo{person}{Wayne~Xin Zhao}, \bibinfo{person}{Kun Zhou}, \bibinfo{person}{Junyi Li}, \bibinfo{person}{Tianyi Tang}, \bibinfo{person}{Xiaolei Wang}, \bibinfo{person}{Yupeng Hou}, \bibinfo{person}{Yingqian Min}, \bibinfo{person}{Beichen Zhang}, \bibinfo{person}{Junjie Zhang}, \bibinfo{person}{Zican Dong}, {et~al\mbox{.}}} \bibinfo{year}{2023}\natexlab{b}.
\newblock \showarticletitle{A survey of large language models}.
\newblock \bibinfo{journal}{\emph{arXiv preprint arXiv:2303.18223}} (\bibinfo{year}{2023}).
\newblock


\bibitem[Zhao et~al\mbox{.}(2024b)]%
        {zhao2024let}
\bibfield{author}{\bibinfo{person}{Yuyue Zhao}, \bibinfo{person}{Jiancan Wu}, \bibinfo{person}{Xiang Wang}, \bibinfo{person}{Wei Tang}, \bibinfo{person}{Dingxian Wang}, {and} \bibinfo{person}{Maarten de Rijke}.} \bibinfo{year}{2024}\natexlab{b}.
\newblock \showarticletitle{Let Me Do It For You: Towards LLM Empowered Recommendation via Tool Learning}. In \bibinfo{booktitle}{\emph{Proceedings of the 47th International ACM SIGIR Conference on Research and Development in Information Retrieval}}. \bibinfo{pages}{1796--1806}.
\newblock


\bibitem[Zhao et~al\mbox{.}(2023a)]%
        {zhao2024See_and_Think}
\bibfield{author}{\bibinfo{person}{Zhonghan Zhao}, \bibinfo{person}{Wenhao Chai}, \bibinfo{person}{Xuan Wang}, \bibinfo{person}{Li Boyi}, \bibinfo{person}{Shengyu Hao}, \bibinfo{person}{Shidong Cao}, \bibinfo{person}{Tian Ye}, \bibinfo{person}{Jenq-Neng Hwang}, {and} \bibinfo{person}{Gaoang Wang}.} \bibinfo{year}{2023}\natexlab{a}.
\newblock \showarticletitle{See and think: Embodied agent in virtual environment}.
\newblock \bibinfo{journal}{\emph{arXiv preprint arXiv:2311.15209}} (\bibinfo{year}{2023}).
\newblock


\bibitem[Zheng et~al\mbox{.}(2024a)]%
        {zheng2024SEEACT}
\bibfield{author}{\bibinfo{person}{Boyuan Zheng}, \bibinfo{person}{Boyu Gou}, \bibinfo{person}{Jihyung Kil}, \bibinfo{person}{Huan Sun}, {and} \bibinfo{person}{Yu Su}.} \bibinfo{year}{2024}\natexlab{a}.
\newblock \showarticletitle{Gpt-4v (ision) is a generalist web agent, if grounded}.
\newblock \bibinfo{journal}{\emph{arXiv preprint arXiv:2401.01614}} (\bibinfo{year}{2024}).
\newblock


\bibitem[Zheng et~al\mbox{.}(2024b)]%
        {zheng2024gpt}
\bibfield{author}{\bibinfo{person}{Boyuan Zheng}, \bibinfo{person}{Boyu Gou}, \bibinfo{person}{Jihyung Kil}, \bibinfo{person}{Huan Sun}, {and} \bibinfo{person}{Yu Su}.} \bibinfo{year}{2024}\natexlab{b}.
\newblock \showarticletitle{Gpt-4v (ision) is a generalist web agent, if grounded}.
\newblock \bibinfo{journal}{\emph{arXiv preprint arXiv:2401.01614}} (\bibinfo{year}{2024}).
\newblock


\bibitem[Zheng et~al\mbox{.}(2023)]%
        {zheng2023large}
\bibfield{author}{\bibinfo{person}{Chujie Zheng}, \bibinfo{person}{Hao Zhou}, \bibinfo{person}{Fandong Meng}, \bibinfo{person}{Jie Zhou}, {and} \bibinfo{person}{Minlie Huang}.} \bibinfo{year}{2023}\natexlab{}.
\newblock \showarticletitle{Large language models are not robust multiple choice selectors}. In \bibinfo{booktitle}{\emph{The Twelfth International Conference on Learning Representations}}.
\newblock


\bibitem[Zhou et~al\mbox{.}(2023c)]%
        {zhou2023language}
\bibfield{author}{\bibinfo{person}{Andy Zhou}, \bibinfo{person}{Kai Yan}, \bibinfo{person}{Michal Shlapentokh-Rothman}, \bibinfo{person}{Haohan Wang}, {and} \bibinfo{person}{Yu-Xiong Wang}.} \bibinfo{year}{2023}\natexlab{c}.
\newblock \showarticletitle{Language agent tree search unifies reasoning acting and planning in language models}.
\newblock \bibinfo{journal}{\emph{arXiv preprint arXiv:2310.04406}} (\bibinfo{year}{2023}).
\newblock


\bibitem[Zhou et~al\mbox{.}(2023b)]%
        {zhou2023webarena}
\bibfield{author}{\bibinfo{person}{Shuyan Zhou}, \bibinfo{person}{Frank~F Xu}, \bibinfo{person}{Hao Zhu}, \bibinfo{person}{Xuhui Zhou}, \bibinfo{person}{Robert Lo}, \bibinfo{person}{Abishek Sridhar}, \bibinfo{person}{Xianyi Cheng}, \bibinfo{person}{Yonatan Bisk}, \bibinfo{person}{Daniel Fried}, \bibinfo{person}{Uri Alon}, {et~al\mbox{.}}} \bibinfo{year}{2023}\natexlab{b}.
\newblock \showarticletitle{Webarena: A realistic web environment for building autonomous agents}.
\newblock \bibinfo{journal}{\emph{arXiv preprint arXiv:2307.13854}} (\bibinfo{year}{2023}).
\newblock


\bibitem[Zhou et~al\mbox{.}(2023a)]%
        {zhou2023agents}
\bibfield{author}{\bibinfo{person}{Wangchunshu Zhou}, \bibinfo{person}{Yuchen~Eleanor Jiang}, \bibinfo{person}{Long Li}, \bibinfo{person}{Jialong Wu}, \bibinfo{person}{Tiannan Wang}, \bibinfo{person}{Shi Qiu}, \bibinfo{person}{Jintian Zhang}, \bibinfo{person}{Jing Chen}, \bibinfo{person}{Ruipu Wu}, \bibinfo{person}{Shuai Wang}, {et~al\mbox{.}}} \bibinfo{year}{2023}\natexlab{a}.
\newblock \showarticletitle{Agents: An open-source framework for autonomous language agents}.
\newblock \bibinfo{journal}{\emph{arXiv preprint arXiv:2309.07870}} (\bibinfo{year}{2023}).
\newblock


\bibitem[Zhu et~al\mbox{.}(2024a)]%
        {zhu2024reliable}
\bibfield{author}{\bibinfo{person}{Lixi Zhu}, \bibinfo{person}{Xiaowen Huang}, {and} \bibinfo{person}{Jitao Sang}.} \bibinfo{year}{2024}\natexlab{a}.
\newblock \showarticletitle{How Reliable is Your Simulator? Analysis on the Limitations of Current LLM-based User Simulators for Conversational Recommendation}. In \bibinfo{booktitle}{\emph{Companion Proceedings of the ACM on Web Conference 2024}}. \bibinfo{pages}{1726--1732}.
\newblock


\bibitem[Zhu et~al\mbox{.}(2024b)]%
        {zhu2024llm}
\bibfield{author}{\bibinfo{person}{Lixi Zhu}, \bibinfo{person}{Xiaowen Huang}, {and} \bibinfo{person}{Jitao Sang}.} \bibinfo{year}{2024}\natexlab{b}.
\newblock \showarticletitle{A LLM-based Controllable, Scalable, Human-Involved User Simulator Framework for Conversational Recommender Systems}.
\newblock \bibinfo{journal}{\emph{arXiv preprint arXiv:2405.08035}} (\bibinfo{year}{2024}).
\newblock


\bibitem[Zhu et~al\mbox{.}(2024c)]%
        {zhu2024knowagent}
\bibfield{author}{\bibinfo{person}{Yuqi Zhu}, \bibinfo{person}{Shuofei Qiao}, \bibinfo{person}{Yixin Ou}, \bibinfo{person}{Shumin Deng}, \bibinfo{person}{Ningyu Zhang}, \bibinfo{person}{Shiwei Lyu}, \bibinfo{person}{Yue Shen}, \bibinfo{person}{Lei Liang}, \bibinfo{person}{Jinjie Gu}, {and} \bibinfo{person}{Huajun Chen}.} \bibinfo{year}{2024}\natexlab{c}.
\newblock \showarticletitle{Knowagent: Knowledge-augmented planning for llm-based agents}.
\newblock \bibinfo{journal}{\emph{arXiv preprint arXiv:2403.03101}} (\bibinfo{year}{2024}).
\newblock


\bibitem[Zhu et~al\mbox{.}(2023)]%
        {zhu2023large}
\bibfield{author}{\bibinfo{person}{Yutao Zhu}, \bibinfo{person}{Huaying Yuan}, \bibinfo{person}{Shuting Wang}, \bibinfo{person}{Jiongnan Liu}, \bibinfo{person}{Wenhan Liu}, \bibinfo{person}{Chenlong Deng}, \bibinfo{person}{Zhicheng Dou}, {and} \bibinfo{person}{Ji-Rong Wen}.} \bibinfo{year}{2023}\natexlab{}.
\newblock \showarticletitle{Large language models for information retrieval: A survey}.
\newblock \bibinfo{journal}{\emph{arXiv preprint arXiv:2308.07107}} (\bibinfo{year}{2023}).
\newblock


\end{thebibliography}
